\documentclass[apj]{emulateapj}
\usepackage{natbib,xspace,color}
\usepackage{verbatim}
\usepackage{multirow}
\usepackage{rotating}
\usepackage{gensymb}
\usepackage{mathrsfs}  
\usepackage{amsmath}
\usepackage{amssymb}
\usepackage[backref,breaklinks,colorlinks,citecolor=blue]{hyperref}
\usepackage[all]{hypcap} 

\usepackage{url}

\newcommand{\lir}{$L_{\rm{IR}}$}

\shorttitle{Continuous Mid-Infrared Star Formation Rate Indicators}
\shortauthors{Battisti et al.}

\begin{document}

\title{Continuous Mid-Infrared Star Formation Rate Indicators: Diagnostics for $0<\lowercase{z}<3$ Star Forming Galaxies}
\author{A. J. Battisti\altaffilmark{1}, D. Calzetti\altaffilmark{1}, B. D. Johnson\altaffilmark{2}, D. Elbaz\altaffilmark{3} }

\altaffiltext{1}{Department of Astronomy, University of Massachusetts, Amherst, MA 01003, USA; abattist@astro.umass.edu}
\altaffiltext{2}{Institut d'Astrophysique de Paris, UMR 7095, 75014, Paris, France}
\altaffiltext{3}{Laboratoire AIM-Paris-Saclay, CEA/DSM/Irfu, CNRS, Universit\'e
Paris Diderot, Saclay, pt courrier 131, 91191 Gif-sur-Yvette, France}

\begin{abstract}
We present continuous, monochromatic star formation rate (SFR) indicators over the mid-infrared wavelength range of $6-70~\micron$. We use a sample of 58 star forming galaxies (SFGs) in the \textit{Spitzer}-SDSS-\textit{GALEX} Spectroscopic Survey (SSGSS) at $z<0.2$, for which there is a rich suite of multi-wavelength photometry and spectroscopy from the ultraviolet through to the infrared. The data from the \textit{Spitzer} infrared spectrograph (IRS) of these galaxies, which spans $5-40~\micron$, is anchored to their photometric counterparts. The spectral region between $40-70~\micron$ is interpolated using dust model fits to the IRS spectrum and \textit{Spitzer} 70 and $160~\micron$ photometry. Since there are no sharp spectral features in this region, we expect these interpolations to be robust. This spectral range is calibrated as a SFR diagnostic using several reference SFR indicators to mitigate potential bias. Our band-specific continuous SFR indicators are found to be consistent with monochromatic calibrations in the local universe, as derived from \textit{Spitzer}, WISE, and Herschel photometry. Our local composite template and continuous SFR diagnostics are made available for public use through the NASA/IPAC Infrared Science Archive (IRSA) and have typical dispersions of 30\% or less. We discuss the validity and range of applicability for our SFR indicators in the context of unveiling the formation and evolution of galaxies. Additionally, in the era of the James Webb Space Telescope this will become a flexible tool, applicable to any SFG up to $z\sim3$.
\end{abstract}

\section{Introduction}
Star formation is a fundamental parameter of galaxies that describes how galaxies evolve, when used in conjunction with mass. As the process of star formation depletes a galaxy of its gas, it must be continuously replenished by infall from the intergalactic medium to be supported for an extended time. When massive stars die, they enrich the surrounding interstellar medium with heavy metals, thus altering a galaxy's chemical composition. Therefore, accurately tracing star formation though cosmic time gives key constraints on how galaxies are able to form and evolve \citep[e.g.,][and references therein]{tinsley68,somerville12,madau&dickinson14}.     

For these reasons, great efforts have been made to calibrate a wide range of the electromagnetic spectrum that can be linked to processes involved with recent star formation \citep[see review by][]{kennicutt12}. In particular, infrared (IR) wavelength calibrations are proving to be critical to understanding galaxies in the early Universe. Deep IR surveys with the \textit{Spitzer} and \textit{Herschel} Space Telescopes have revealed that the majority of star formation that occurs at redshift $z\sim1-3$ is enshrouded by dust \citep[e.g.,][]{murphy11a, elbaz11}, making it very difficult to measure accurate SFRs at optical wavelengths. In addition, IR-bright galaxies ($L\gtrsim\times10^{11} L_\odot$) are much more prevalent during that time than today \citep[e.g.,][and references therein]{chary&elbaz01,lefloch05,magnelli09,murphy11a, elbaz11, lutz14}. Furthermore, observations suggest that $\sim$$85\%$ of today's stars were formed at redshift  $0<z<2.5$ \citep{marchesini09,muzzin13,tomczak14}. Together these results have renewed interest in monochromatic (i.e., single-band) mid-IR (MIR) star formation rate (SFR) indicators, as distant galaxies can easily be observed in the MIR. 

Dust emission in the MIR is more closely related to star formation than longer IR wavelengths, where heating by low mass (i.e., long-living) stars becomes important, which has allowed several wavelength bands in the MIR to be well calibrated locally as SFR diagnostics \citep{zhu08,rieke09,calzetti10}. However, difficulties arise in utilizing local calibrations because the regions of rest-frame wavelengths probed by a given band will vary with redshift. As a reference, the \textit{Spitzer} $24~\micron$ and the \textit{Herschel} $70~\micron$ bands target the rest-frame $8~\micron$ and $23~\micron$ emission, respectively, for a galaxy at redshift $z=2$. Correcting for this effect is most commonly achieved through $k$-corrections which depend heavily on the assumed galaxy spectral energy distribution (SED) template. Such templates \citep[e.g.,][]{chary&elbaz01,dale&helou02,polletta07,rieke09,brown14} typically require many photometric bands for accurate matching. Therefore, in order to fully utilize current and future deep IR imaging surveys for a greater understanding the formation and evolution of galaxies without a reliance on extensive multi-band imaging, continuous single-band SFR indicators will be imperative.

In the near future, the Mid-Infrared Instrument (MIRI; 5--28~$\mu$m) on board the James Webb Space Telescope (JWST) will expand our ability to probe galaxies in the MIR, detecting down to the regime of normal star-forming disk galaxies ($L\lesssim3\times10^{11} L_\odot$) out to $z=3$ and representing an order of magnitude improvement in sensitivity over \textit{Spitzer} bands of similar wavelength\footnote{\url{http://www.stsci.edu/jwst/instruments/miri/instrumentdesign/filters/}}. Thus, current and future cosmological surveys are highlighting the need for continuous monochromatic SFR indicators that cover, without breaks, the MIR wavelength range of $6-70~\micron$. This will provide a flexible tool that can be applied to any galaxy up to redshift $z\approx3$. With the release of the Wide-field Infrared Survey Explorer (WISE) All-Sky Survey \citep{wright10}, times are ripe for consolidating all these data into a coherent picture. In this study, we use \textit{GALEX}, Sloan Digital Sky Survey (SDSS), WISE, and \textit{Spitzer} data of a large sample of local galaxies to perform the calibration of SFR($\lambda$) in the $6-70~\micron$ range.

Throughout this work we adopt the WMAP five-year cosmological parameters, $H_0=70.5$~km/s/Mpc, $\Omega_M=0.27$, $\Omega_{\Lambda}=0.73$ \citep{komatsu09}. We assume a  \citet{kroupa01} initial mass function (IMF) for all SFR calibrations. The infrared luminosity, \lir, of a galaxy refers to the integrated luminosity over the region from 8 to $1000~\micron$ ($L_{\rm{IR}}=\int_{8\mu\rm{m}}^{1000\mu\rm{m}} L_\nu d\nu$). 

\section{Data}
\subsection{The SSGSS Sample}
The \textit{Spitzer}--SDSS--\textit{GALEX} Spectroscopic Survey (SSGSS) is a sample of 101 galaxies located within the \textit{Spitzer} Wide-Area Infrared Extragalactic (SWIRE) Survey/Lockman Hole area at $0.03 < z < 0.22$ \citep{treyer10,odowd11}. These galaxies represent a subset of the 912 galaxies within the \citet{johnson06} sample, which has extensive multi-wavelength coverage from \textit{Spitzer}, SDSS, and \textit{GALEX}, and for which \textit{Spitzer} infrared spectrograph (IRS) measurements have also been obtained. The UV data is from pipeline-processed \textit{GALEX} observations of this regions with average exposures of $\sim$1.5~ks. The optical photometry comes from the seventh data release of the SDSS main galaxy sample \citep[DR7;][]{abazajian09}. The optical spectroscopic measurements are from the Max Planck Institute for Astrophysics and Johns Hopkins University (MPA/JHU) group\footnote{\url{http://www.mpa-garching.mpg.de/SDSS/DR7/}}, which is based on the method presented in \citet{tremonti04}. The infrared photometry comes from the SWIRE survey observations \citep{lonsdale03}. 

Each galaxy in this sample has been observed with the \textit{Spitzer} IRAC and MIPS bands, in addition to observations using the blue filter of the IRS peak-up facility. These blue filter peak-ups have spectral coverage from $13.3-18.7~\micron$ and give an additional photometric point at $16~\micron$, between the IRAC $8~\micron$ and MIPS $24~\micron$ bands. Aperture photometry was performed in 7\arcsec\ and 12\arcsec\ radius apertures for the $3.6-8~\micron$ IRAC and $24~\micron$ MIPS, respectively, and then aperture-corrected to 12.2\arcsec\ and $>35$\arcsec, respectively. For the MIPS 70 and $160~\micron$ bands, nearly all the galaxies can be treated as point sources, and aperture corrections were taken from the MIPS handbook. A full description of these aperture corrections is described in \citet{johnson07}. For a more detailed description of the SSGSS dataset, we refer the reader to \citet{odowd11}. 

The IRS spectroscopy for the SSGSS sample is obtained through the NASA/IPAC Infrared Science Archive (IRSA) website\footnote{\url{http://irsa.ipac.caltech.edu/data/SPITZER/SSGSS/}} and is from the work of \citet{odowd11}.  This study utilizes the lower resolution Short-Low (SL) and Long-Low (LL) IRS modules, as these have been obtained for the entire SSGSS sample. The SL module spans $5.2-14.5~\micron$ with resolving power $R = 60-125$ and has a slit width of $3.6-3.7$\arcsec. The LL module spans $14-38~\micron$ with resolving power $R = 57-126$ and has a slit width of $10.5-10.7$\arcsec. The spectra from the two modules were combined by weighted mean and a detailed description of the method can be found in \citet{odowd11}. At $z\sim0.1$ these galaxies are sufficiently distant such that the IRS slit encompasses a significant fraction of each galaxy ($r$-band Petrosian diameters are $\sim$$10$\arcsec), providing some of the best MIR SEDs for a continuous SFR($\lambda$) determination.

In order to accurately determine a diagnostic for star formation across the MIR, it is necessary to only consider cases where the majority of light is being contributed from stars (i.e., star-forming galaxies; SFGs) and not from an active galactic nucleus (AGN). The galaxy type is traditionally determined according to their location on the Baldwin-Phillips-Terlevich (BPT) diagram \citep{baldwin81,kewley01,kauffmann03}. By adopting the DR7 values of emission line measurements, we find that 64 of the 101 SSGSS galaxies are classified as SFGs. We note that the initial classification of SSGSS galaxies by \citet{odowd11} utilized SDSS DR4 measurements, which results in a few BPT designations to differ between these works. We further exclude 6 of the SSGSS galaxies classified as SFGs from our analysis for the following reasons: SSGSS~18 appears to be a merger, SSGSS~19, 22, and 96 have significant breaks in their IRS spectra due to low signal-to-noise (S/N), and SSGSS~35 and 51 suffer from problems with IRS confusion. This leaves 58 galaxies to be used in our calibration of a monochromatic MIR SFR indicator. Table~\ref{Tab:IRS_correct} shows the SSGSS IDs of the galaxies used for this study along with some of their properties (additional parameters in the table are introduced in later sections).

Our sample of 58 galaxies span a redshift range of $0.03\le z \le0.22$ with a median redshift of 0.075. The range of infrared luminosity is $9.53\le \log(L_{\rm{IR}}/L_{\odot})\le 11.37$, with a median of $10.55$. All measurements of \lir\ for these galaxies are taken from the original SSGSS dataset \citep[presented in][]{treyer10}. The selection criteria for the SSGSS sample was based on $5.8~\micron$ surface brightness and $24~\micron$ flux density, and this restricts our sample of 58 galaxies to relatively high stellar masses ($1.6\times10^9\le M/M_{\odot}\le 1.7\times10^{11}$) and metallicities ($8.7\le 12+\log(\rm{O/H})\le 9.2$). These stellar mass and metallicity estimates are updated from the SSGSS dataset values (based on DR4) to the MPA-JHU DR7 estimates. 

\begin{table*}
\begin{center}
\caption{Summary of Galaxy Properties and IRS Correction Terms \label{Tab:IRS_correct}}
\begin{tabular}{ccccccccc}
\hline\hline 
SSGSS& R.A. & Decl. & $z$   & log(\lir)& $\langle SFR \rangle$ &$c_{\rm{phot}}$ & $k$ & \multirow{2}{*}{$\frac{(16+k)}{(8+k)}$} \\  
   ID & (J2000) & (J2000) &       & $(L_{\odot})$ &$(M_{\odot}/\rm{yr})$&  & &  \\ \hline
  1 & 160.34398 & 58.89201 & 0.066 & 10.55 &  4.22 & 0.963 &  14.10 & 1.362 \\
  2 & 159.86748 & 58.79165 & 0.045 &  9.83 &  0.65 & 1.062 &  23.64 & 1.253 \\
  3 & 162.41000 & 59.58426 & 0.117 & 10.59 &  3.41 & 1.091 &  54.09 & 1.129 \\
  4 & 162.54131 & 59.50806 & 0.066 & 10.22 &  1.29 & 0.871 &--4.466 & 3.264 \\
  5 & 162.36443 & 59.54812 & 0.217 & 11.37 & 19.55 & 1.088 &  24.48 & 1.246 \\
  6 & 162.52991 & 59.54828 & 0.115 & 10.99 &  7.41 & 0.851 &  1.310 & 1.859 \\
  8 & 161.48123 & 59.15443 & 0.044 &  9.97 &  0.79 & 0.843 &--5.224 & 3.882 \\
 14 & 161.92709 & 56.31395 & 0.153 & 11.06 &  9.60 & 0.918 &  10.69 & 1.428 \\
 16 & 162.04231 & 56.38041 & 0.072 & 10.42 &  2.15 & 0.943 &  15.73 & 1.337 \\
 17 & 161.76901 & 56.34029 & 0.047 & 10.83 &  5.58 & 0.922 &  21.94 & 1.267 \\
 24 & 163.53931 & 56.82104 & 0.046 & 10.59 &  3.43 & 1.077 &  0.240 & 1.971 \\
 25 & 158.22482 & 58.10917 & 0.073 & 10.30 &  1.89 & 1.061 &  2.363 & 1.772 \\
 27 & 159.34668 & 57.52069 & 0.072 & 11.01 &  7.70 & 0.950 &--0.933 & 2.132 \\
 30 & 159.73558 & 57.26361 & 0.046 & 10.11 &  1.06 & 1.040 &  63.01 & 1.113 \\
 32 & 161.48724 & 57.45520 & 0.117 & 10.82 &  6.30 & 1.121 &  278.9 & 1.028 \\
 34 & 160.30701 & 57.08246 & 0.046 &  9.96 &  0.79 & 0.987 &  46.47 & 1.147 \\
 36 & 159.98523 & 57.40522 & 0.072 & 10.39 &  2.01 & 0.996 &--1.629 & 2.256 \\
 38 & 160.20963 & 57.39475 & 0.118 & 10.92 &  6.43 & 0.905 &  1.989 & 1.801 \\
 39 & 159.38356 & 57.38491 & 0.074 & 10.14 &  1.28 & 0.684 &  1.615 & 1.832 \\
 41 & 158.99098 & 57.41671 & 0.102 & 10.34 &  1.68 & 0.818 &  2.159 & 1.787 \\
 42 & 158.97563 & 58.31007 & 0.155 & 11.03 &  9.32 & 1.170 &  14.84 & 1.350 \\
 46 & 159.02698 & 57.78402 & 0.044 & 10.02 &  0.99 & 0.932 &  3.183 & 1.715 \\
 47 & 159.22287 & 57.91185 & 0.102 & 10.68 &  4.34 & 1.331 &  61.86 & 1.115 \\
 48 & 159.98817 & 58.65948 & 0.200 & 11.24 & 15.53 & 0.869 &  11.59 & 1.408 \\
 49 & 159.51942 & 58.04882 & 0.091 & 10.49 &  3.13 & 1.107 &  0.495 & 1.942 \\
 52 & 160.54201 & 58.66098 & 0.031 &  9.53 &  0.29 & 1.129 &--2.795 & 2.537 \\
 54 & 160.41264 & 58.58743 & 0.115 & 11.20 & 11.53 & 0.901 &  6.117 & 1.567 \\
 55 & 160.29353 & 58.25641 & 0.121 & 10.54 &  2.95 & 0.865 &  3.155 & 1.717 \\
 56 & 160.41617 & 58.31722 & 0.072 & 10.01 &  0.85 & 0.886 &  4.645 & 1.633 \\
 57 & 160.12233 & 58.16783 & 0.073 &  9.92 &  0.75 & 0.689 &  15.28 & 1.344 \\
 59 & 159.89861 & 57.98557 & 0.075 & 10.36 &  2.00 & 0.928 &  15.09 & 1.346 \\
 60 & 160.51027 & 57.89706 & 0.116 & 10.48 &  2.89 & 0.910 &  6.530 & 1.551 \\
 62 & 160.91280 & 58.04736 & 0.133 & 11.08 &  9.85 & 0.978 &  0.688 & 1.921 \\
 64 & 161.00317 & 58.76030 & 0.073 & 10.88 &  5.44 & 0.913 &  3.276 & 1.709 \\
 65 & 161.37666 & 58.20886 & 0.118 & 11.18 & 11.85 & 0.965 &  12.55 & 1.389 \\
 66 & 161.25533 & 57.77575 & 0.113 & 10.86 &  6.75 & 0.937 &  13.86 & 1.366 \\
 67 & 161.18829 & 58.45495 & 0.031 & 10.09 &  1.21 & 1.581 &  8.790 & 1.476 \\
 68 & 163.63458 & 57.15902 & 0.068 & 10.54 &  3.37 & 0.975 &  25.56 & 1.238 \\
 70 & 163.17673 & 57.32074 & 0.090 & 10.49 &  2.81 & 1.037 &  255.8 & 1.030 \\
 71 & 163.21991 & 57.13160 & 0.163 & 10.98 &  8.32 & 1.271 &  285.0 & 1.027 \\
 72 & 163.25565 & 57.09528 & 0.080 & 10.79 &  5.01 & 0.951 &  14.00 & 1.364 \\
 74 & 161.95050 & 57.57723 & 0.118 & 10.80 &  5.07 & 0.878 &--2.430 & 2.436 \\
 76 & 162.02142 & 57.81512 & 0.074 & 10.55 &  2.62 & 0.938 &  6.128 & 1.566 \\
 77 & 162.10524 & 57.66665 & 0.044 &  9.69 &  0.62 & 1.130 &  66.79 & 1.107 \\
 78 & 162.12204 & 57.89890 & 0.074 & 10.56 &  3.17 & 1.027 &--4.810 & 3.508 \\
 79 & 161.25693 & 57.66116 & 0.045 &  9.82 &  0.81 & 1.023 &--0.238 & 2.031 \\
 80 & 162.07401 & 57.40280 & 0.075 & 10.35 &  2.14 & 0.992 &--1.949 & 2.322 \\
 81 & 162.04674 & 57.40856 & 0.075 & 10.24 &  1.69 & 0.909 &  1.709 & 1.824 \\
 82 & 161.03609 & 57.86136 & 0.121 & 10.77 &  4.98 & 0.854 &  2.444 & 1.766 \\
 83 & 160.77402 & 58.69774 & 0.119 & 10.92 &  6.56 & 0.919 &--3.474 & 2.768 \\
 88 & 161.38522 & 58.50156 & 0.116 & 10.51 &  2.54 & 1.106 &  42.43 & 1.159 \\
 90 & 162.64168 & 59.37266 & 0.153 & 11.02 &  8.66 & 0.796 &  8.033 & 1.499 \\
 91 & 162.53705 & 58.92866 & 0.117 & 10.78 &  5.31 & 0.782 &--3.869 & 2.937 \\
 92 & 162.65512 & 59.09582 & 0.032 & 10.18 &  1.42 & 0.966 &  6.675 & 1.545 \\
 94 & 161.80573 & 58.17759 & 0.061 & 10.13 &  0.90 & 0.919 &  9.281 & 1.463 \\
 95 & 163.71245 & 58.39082 & 0.115 & 10.82 &  5.74 & 0.956 &  22.68 & 1.261 \\
 98 & 164.14571 & 58.79676 & 0.050 & 10.59 &  3.70 & 0.982 &  5.797 & 1.580 \\
 99 & 164.33247 & 57.95170 & 0.077 & 10.82 &  5.58 & 0.903 &--1.944 & 2.321 \\ \hline
\end{tabular}
\end{center}
\textbf{Notes.} Columns list the (1) galaxy ID number, (2) redshift, (3) integrated infrared luminosity from $8-1000~\micron$, (4) average SFR from the diagnostics in Table~\ref{Tab:calibrations}, (5) offset between IRS spectra and global photometry above $16~\micron$ (6) correction parameter for wavelength-dependent aperture loss of IRS spectrum below $16~\micron$, (7) correction factor of the spectrum at $8~\micron$ due to wavelength-dependent aperture loss.
\end{table*}

\subsection{WISE Data}
The WISE All-Sky Survey provides photometry at 3.4, 4.6, 12, and $22~\micron$ \citep{wright10} which complements the wealth of IR data available for the SSGSS sample. Most importantly for this study, the WISE $12~\micron$ band provides a crucial photometric point that bridges the gap between the \textit{Spitzer} $8~\micron$ and $24~\micron$ bands, a section of the MIR SED that experiences a transition from polycyclic aromatic hydrocarbon (PAH) emission features and dust continuum. The WISE photometry for the SSGSS sample is obtained through the NASA/IPAC Infrared Science Archive (IRSA) website\footnote{\url{http://irsa.ipac.caltech.edu/Missions/wise.html}}. Using an approach similar to that of \citet{johnson07} for the \textit{Spitzer} bands, we utilize the 13.75\arcsec\ radius aperture measurements for $12~\micron$ and then apply an aperture correction of 1.20. This correction term was found using sources in our sample with no obvious contamination from neighbors and measuring the flux density out to 24.75\arcsec\ to determine their total flux density. The WISE photometry at $22~\micron$ is less accurate than the \textit{Spitzer} $24~\micron$, owing to it having two orders of magnitude lower sensitivity \citep{dole04,wright10}, and is not used for our analysis. In addition, the $22~\micron$ observations suffer from a effective wavelength error, which systematically brightens the photometry of star forming galaxies \citep{wright10,brown14}.

\section{Analysis}

\subsection{Anchoring the IRS Spectra to Global Photometry} \label{IRS_anchor}
For this study, we focus on utilizing the \textit{Spitzer} $5.8-24~\micron$ and WISE $12~\micron$ photometry to anchor the \textit{Spitzer} IRS $5-40~\micron$ spectroscopy available for the entire sample. The reasoning for this approach is to have spectroscopy that is representative of the global flux density of each galaxy, which is required to create a calibrated continuous, monochromatic SFR($\lambda$) indicator. Offsets between the IRS spectrum and the photometry can occur from differences in data reduction methods, or from the width of the IRS slit being smaller than the size of the galaxy. The former effect results in a uniform offset across the entire spectra and can be corrected with a normalization factor. The behavior of the latter effect will be dependent on whether the galaxy observed is an unresolved point-like source. The default \textit{Spitzer} IRS custom extraction (SPICE) does include a correction for light lost from the slit due to the changing angular resolution as a function of wavelength but assumes the object to be a point source. To correct for both of these effects, we utilize photometry from the \textit{Spitzer} 8, 16, and $24~\micron$ and WISE $12~\micron$ bands as a reference. The end-of-channel transmission drop of the SL module below $\sim$5.8$~\micron$, combined with a typical redshift of $z\sim0.1$, makes the $5.8~\micron$ band region unreliable for use in most cases, and so it is not used as an anchor. However, we do make use of the $5.8~\micron$ band to inspect our photometric matching in the lowest redshift galaxies (see below). Here we outline our approach to correct for offset effects so that these spectra are well representative of global photometric measurements. 

\citet{odowd11} found that IRS measurements of SSGSS galaxies from the LL module did not show evidence for significant aperture loss when compared to the \textit{Spitzer} 16 and $24~\micron$ photometry. This is attributed to the fact that the lower resolution (larger PSF) of sources in this longer wavelength module, coupled with the larger slit width of $10.6$\arcsec\, allows for the standard SPICE algorithm to accurately recover the total flux density, since objects are close to point sources \citep[see][]{odowd11}. This would suggest that any offsets between the photometry and spectroscopy beyond $16~\micron$ should be uniform across the module (i.e., a global loss in flux density). Therefore, each spectrum is first fit to match the $16~\micron$ and $24~\micron$ photometric points using a constant offset, $c_{\rm{phot}}$, found using chi-squared minimization, 
\begin{equation}
c_{\mathrm{phot}}=\frac{\sum_i (S_{\mathrm{phot},i}\,S_{\mathrm{IRS},i})/\sigma(S_{\mathrm{phot},i})^2}{\sum_i (S_{\mathrm{phot},i}/\sigma(S_{\mathrm{phot},i}))^2}
\label{cphot}
\end{equation}
\begin{equation}
\chi^2=\sum_i \left(\frac{S_{\mathrm{phot},i}-c_{\mathrm{phot}} S_{\mathrm{IRS},i}}{\sigma(S_{\mathrm{phot},i})}\right)^2\,,
\label{chisq}
\end{equation}
where $S_{\mathrm{phot},i}$ is the \textit{Spitzer} photometric flux density of band $i$, $\sigma(S_{\mathrm{phot},i})$ is the uncertainty of the \textit{Spitzer} photometric flux density, and $S_{\mathrm{IRS},i}$ is the effective IRS photometric flux density found using the transmission curve for each band, $T_i(\lambda)$,
\begin{equation}
S_{\mathrm{IRS},i}= \frac{\int S_{\rm{IRS}}(\lambda)T_i(\lambda)\,d\lambda}{\int T_i(\lambda)\,d\lambda}\,.
\label{convolve}
\end{equation}
This method ignores the method of calibration that was utilized for each specific bandpass (i.e., a conversion of number of electrons measured by the detector into a flux density in Jy requires knowing the shape of the incoming flux of an object, which varies as a function of wavelength). However, discrepancies between the adopted method and correcting for calibration effects amount to $\sim$$1\%$, and is not important for this study. The offset required to match photometric values is typically small, with values of $c_{\rm{phot}}$ being between $0.7-1.6$.

\begin{figure*}
\begin{center}$
\begin{array}{cc}
\includegraphics[width=3.5in]{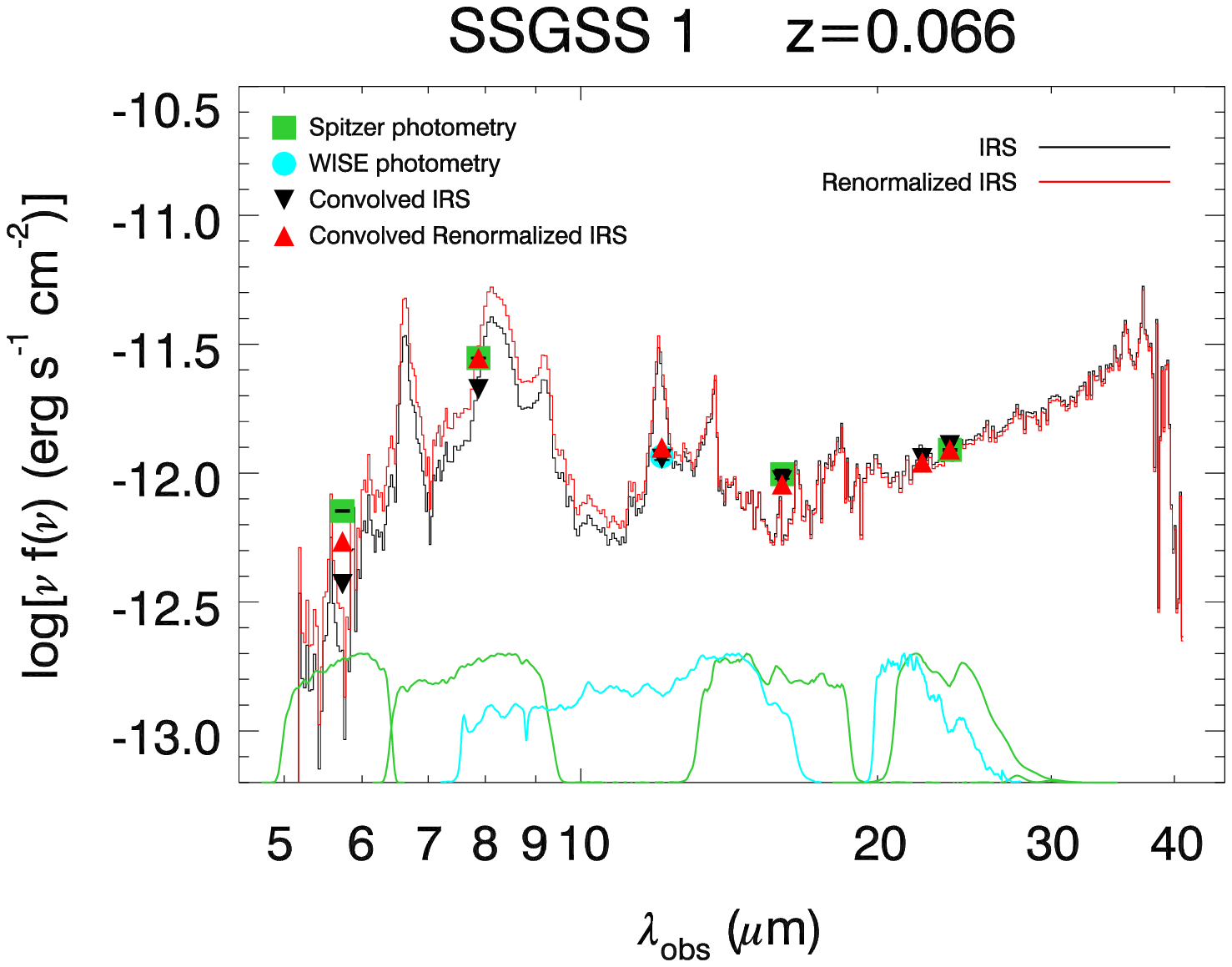} &
\includegraphics[width=3.5in]{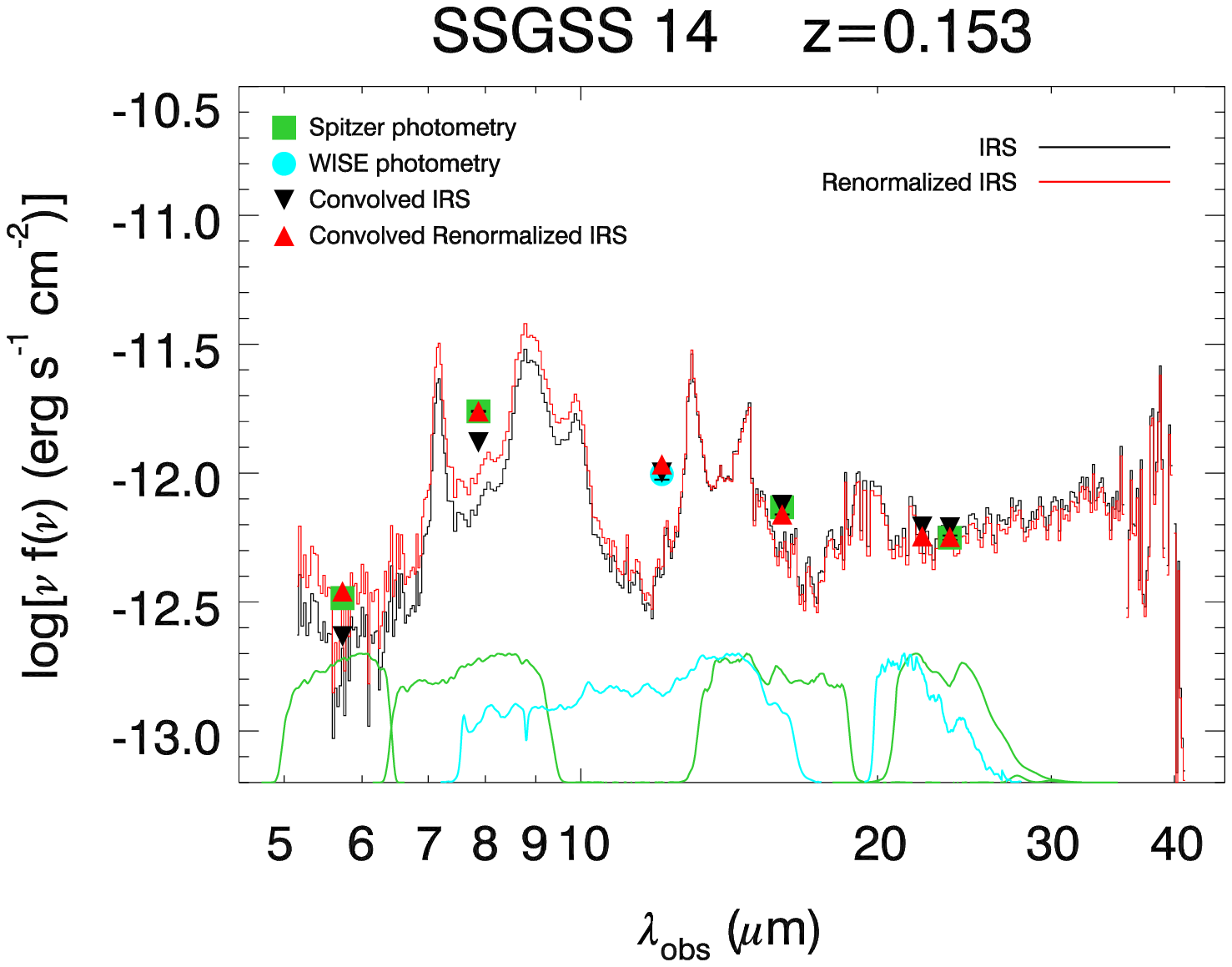} \\ 
\includegraphics[width=3.5in]{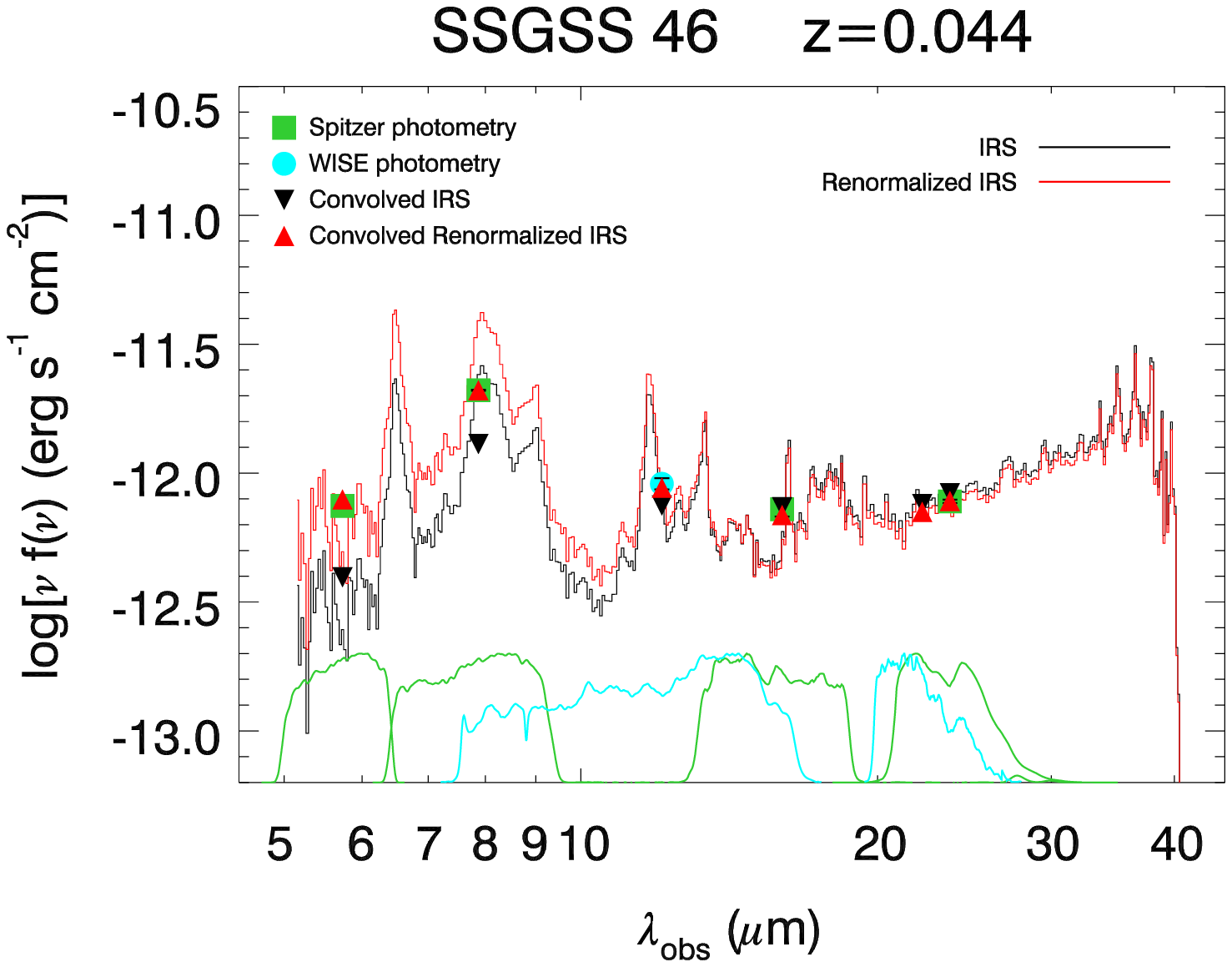} &
\includegraphics[width=3.5in]{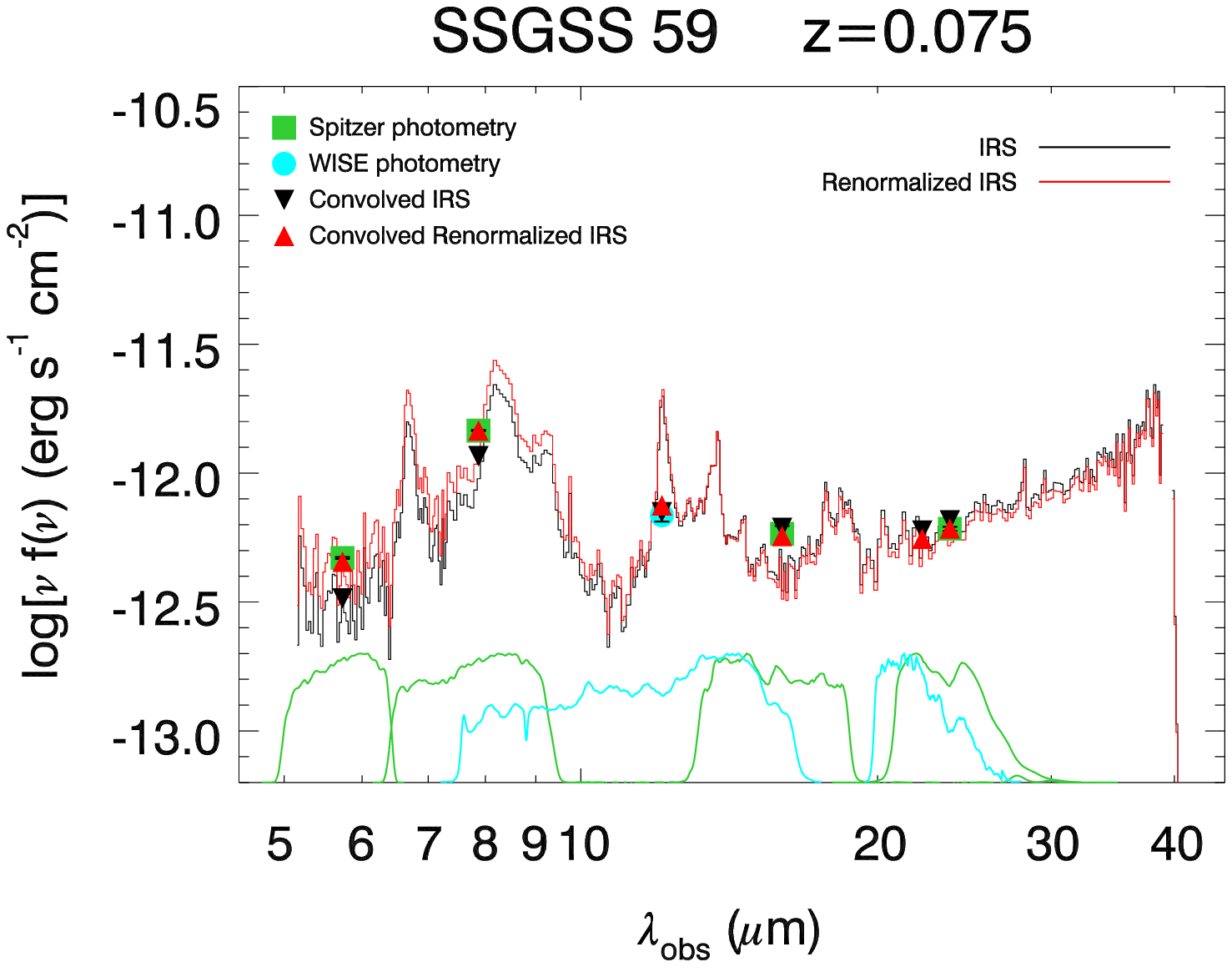} 
\end{array}$
\end{center}
\caption{IRS spectroscopy (black line), \textit{Spitzer} photometry (green squares), and WISE photometry (cyan circle) for some SSGSS galaxies. The IRS spectrum is normalized to the \textit{Spitzer} $8~\micron$, $16~\micron$, and $24~\micron$ and WISE $12~\micron$ photometric flux densities according to the method described in \S~\ref{IRS_anchor} (red line). The effective IRS photometry, found using the transmission curve for each bandpass filter, are shown as triangles. The normalized transmission curves for the \textit{Spitzer} and WISE bands in this region are shown as green and cyan lines, respectively. \label{fig:SSGSS_spectra}}
\end{figure*}

In contrast to the LL module, \citet{odowd11} found that IRS measurements short-ward of $16~\micron$ from the SL module did show evidence for aperture loss when compared to the \textit{Spitzer} $8~\micron$ photometry. In this case, the increasing resolution of the SL module at shorter wavelengths results in many of the galaxies in this sample being resolved in this module.  Also taking into account that the SL slit is 3.6\arcsec, which is smaller than the average extent of $\sim$$10$\arcsec\ ($r$-band Petrosian diameter) for SSGSS galaxies, implies that flux density loss in this wavelength region is more pronounced for more extended objects. For this reason, an additional correction term must be introduced below $16~\micron$ which has a $1/\lambda$ dependency to reflect the additional losses as resolution increases at shorter wavelengths (i.e., the PSF is decreasing at shorter wavelengths, resulting in less correction of light outside the slit). The correction terms adopted are summarized in the following equations,
 \begin{equation}
   S_{\rm{IRS,corr}}(\lambda) = \left\{
     \begin{array}{lr}
       S_{\rm{IRS}}(\lambda)/c_{\rm{phot}}\times \left( \frac{16+k}{\lambda+k} \right) & : \lambda <16~\mu m \\
       S_{\rm{IRS}}(\lambda)/c_{\rm{phot}} & : \lambda \ge 16~\mu m
     \end{array}
   \right.
\end{equation}
where $k$ is a constant found by performing a Levenberg-Marquardt least-squares fit of this function, using the IDL code \texttt{MPFITFUN} \citep{markwardt09}, such that the IRS spectrum matches the 8 and $12~\micron$ photometric flux density. Smaller values of $k$ correspond to larger correction factors in the spectrum. Examples of normalizing the IRS spectroscopy to the photometry are shown in Figure~\ref{fig:SSGSS_spectra}. A list of the normalization parameters is shown in Table~\ref{Tab:IRS_correct}.

As a consistency check on the $1/\lambda$ dependency, a more accurate check is made on the few cases where spectra have high S/N and low redshifts, such that the $5.8~\micron$ band region of the spectrum is reliable to use for convolution. In nearly all these cases the correction using a $1/\lambda$ dependency matches the observed photometric point, as demonstrated by SSGSS 46 in Figure~\ref{fig:SSGSS_spectra}.  In addition, the agreement of the \textit{Spitzer} $8~\micron$ and WISE $12~\micron$ data with this approach suggests these normalized spectra are well representative of photometric values.

\begin{figure*}
\begin{center}$
\begin{array}{cc}
\includegraphics[width=3.5in]{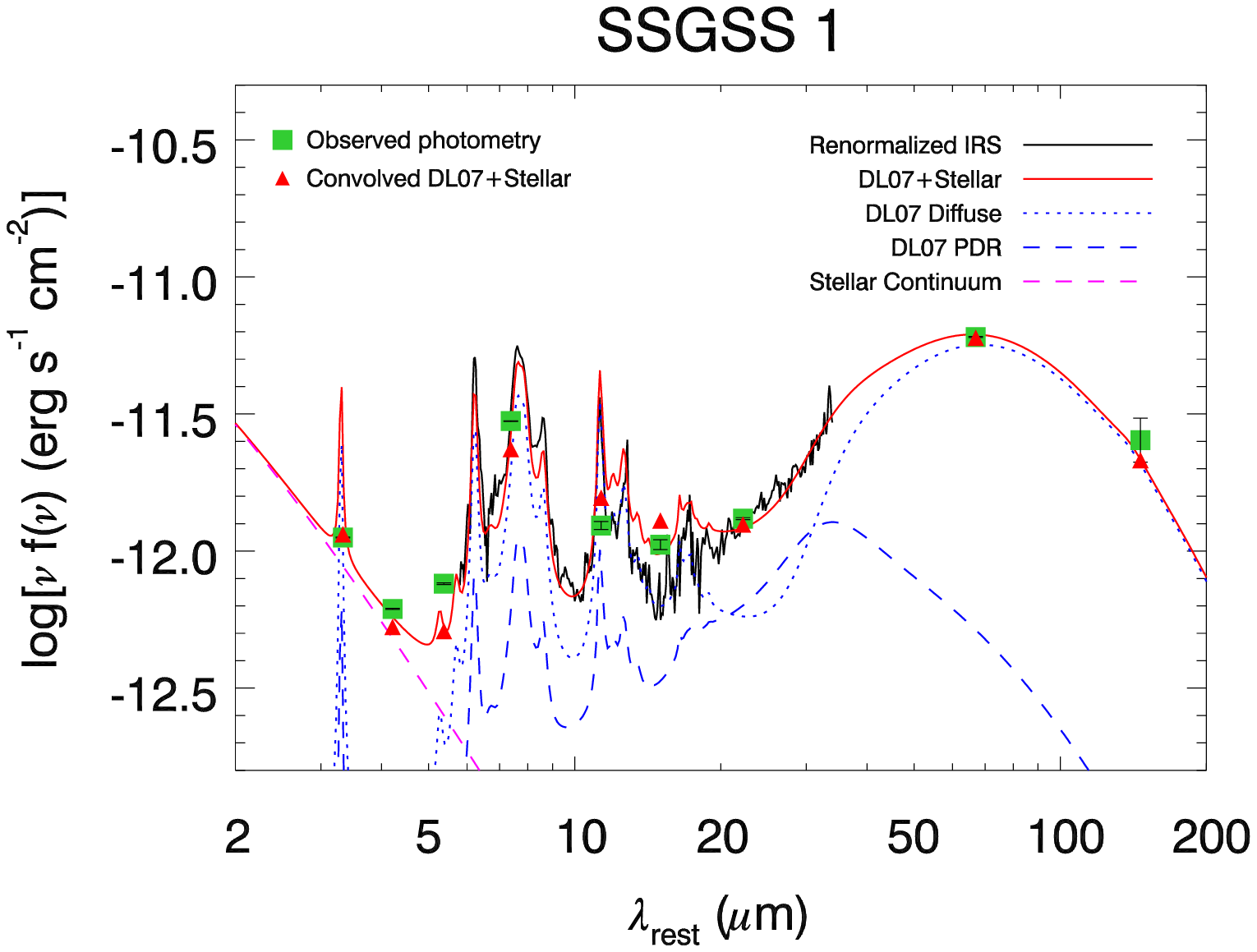} &
\includegraphics[width=3.5in]{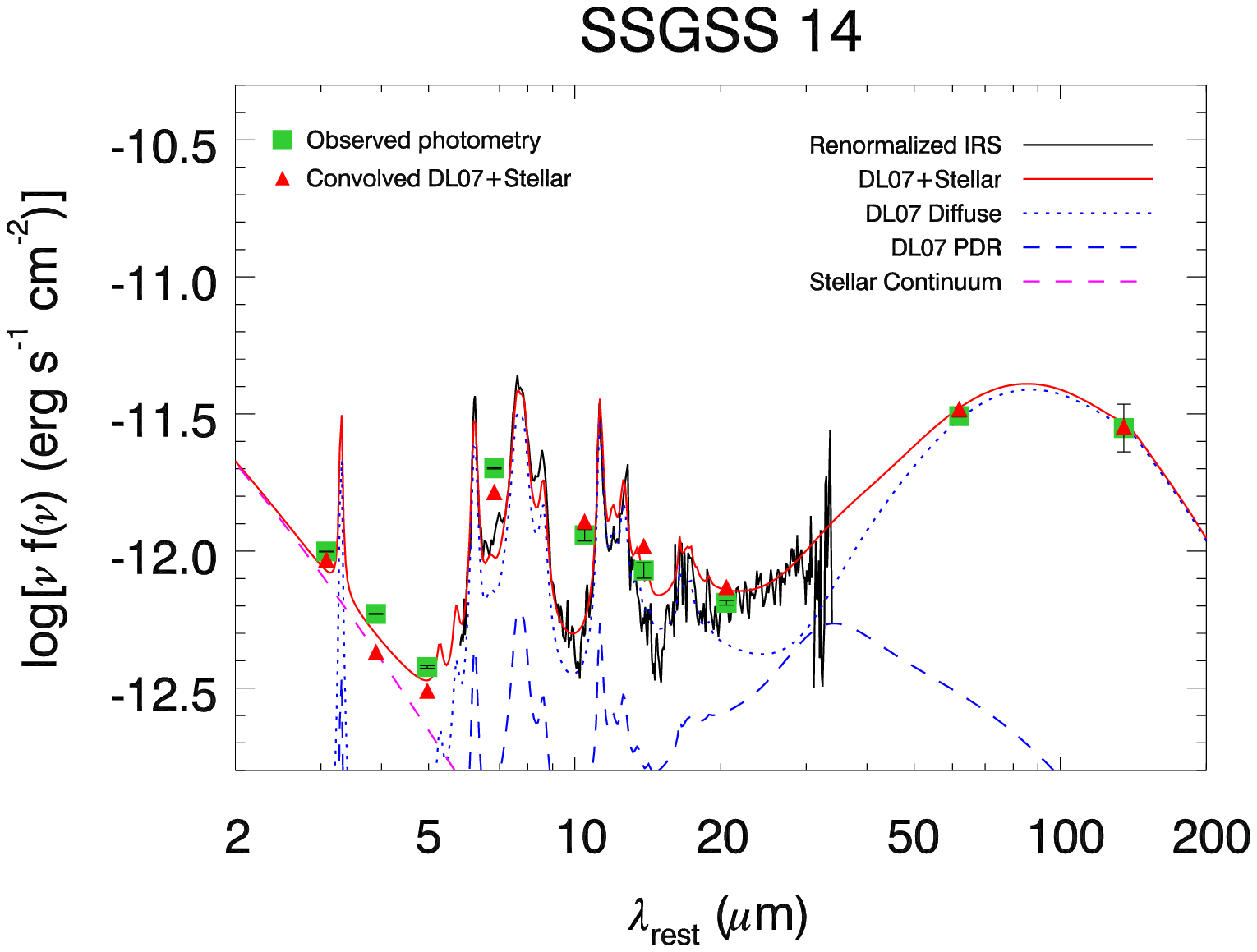} \\ 
\includegraphics[width=3.5in]{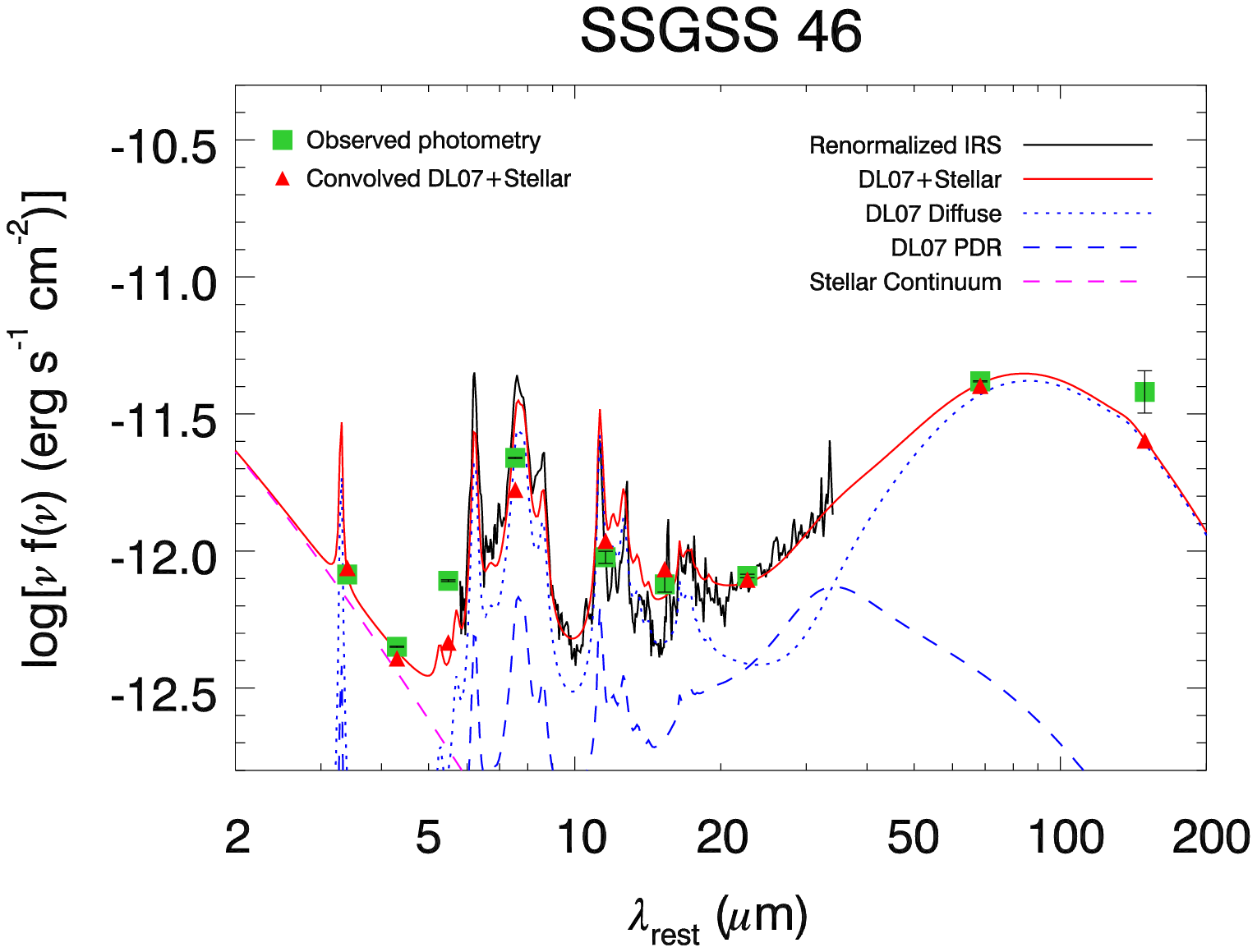} &
\includegraphics[width=3.5in]{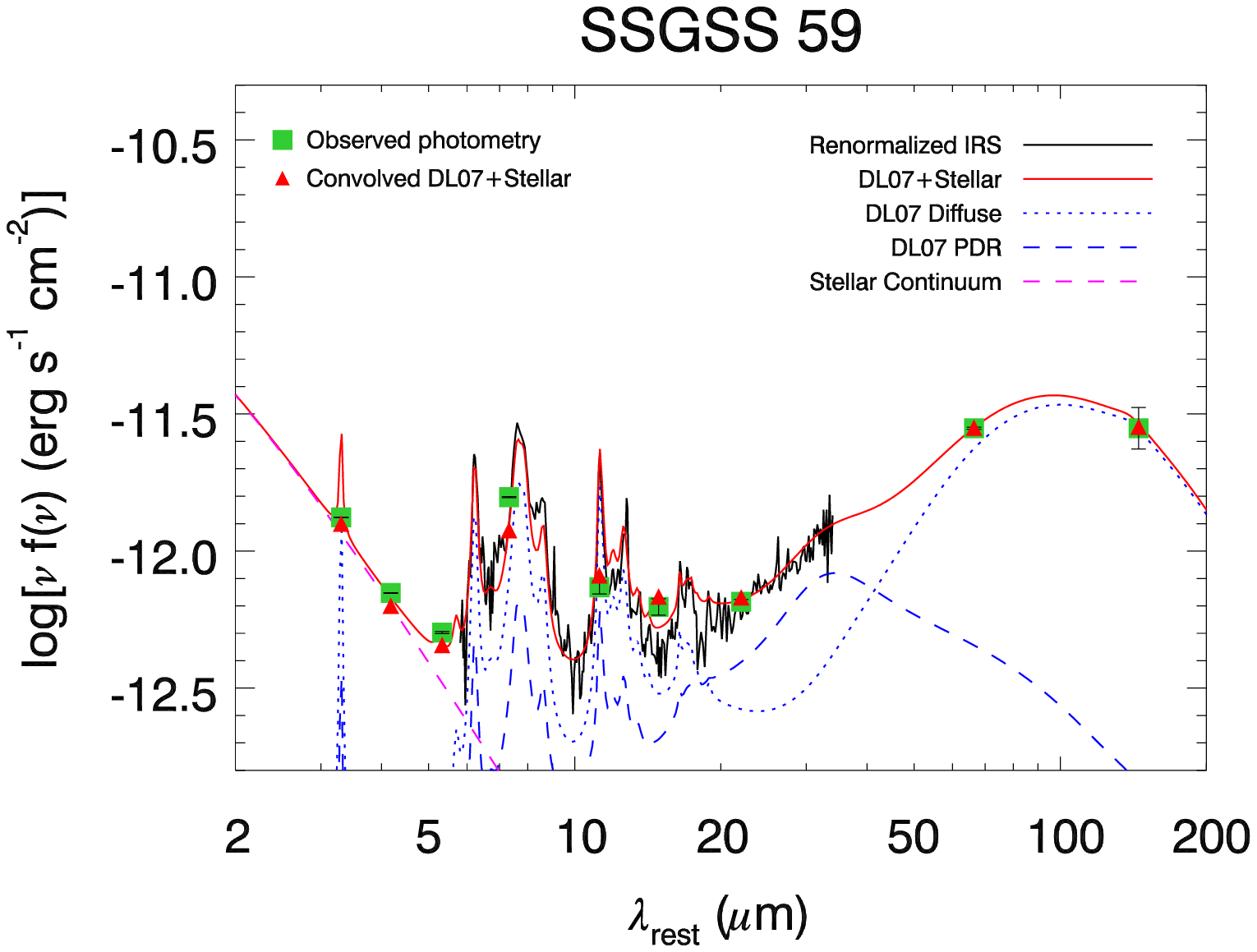} 
\end{array}$
\end{center}
\caption{A fit to the spectra of some SSGSS galaxies using the dust models of \citet{draine&li07}, shown as the solid red line. The IRAC 3.6, 4.5, and $5.8~\micron$ along with the MIPS 70 and $160~\micron$ photometry is used to fit the dust continuum in the absence of the IRS spectrum. The bands within the IRS region are shown for comparison and not used directly for the fit. The regions of the IRS spectrum associated with transmission drops in the instrument are not shown for clarity. We attribute the relatively poor match in the $6-20~\micron$ region to using a simple three component model. However, the purpose of these fits is only to determine the shape of emission in the $34-70~\micron$ region, for which the data is found to be in good agreement with the models. \label{fig:DLFIT_SSGSS}}
\end{figure*}

\subsection{Extending out to 70~\texorpdfstring{$\micron$}{um}} \label{DL07_method}
The IRS spectrum extends out to $40~\micron$, however, the combination of end-of-channel transmission drop around $\lambda_{\rm{obs}}\sim37~\micron$ and redshift effects makes these spectra unreliable for $\lambda_{\rm{rest}}\gtrsim37/(1+z)~\micron$. For the mean redshift of this sample ($z\sim0.1)$, this corresponds to roughly $\lambda_{\rm{rest}}\sim34~\micron$. In order to utilize the wavelength region between $34$ and $70~\micron$, which contains poor quality or no spectral data, we interpolate with dust models. This interpolation is expected to be robust since there are no sharp emission features in this wavelength range. The shape of the emission in this region is dependent on the temperature distribution of of the dust, the grain size distribution, as well as the relative importance of stochastic versus thermal equilibrium heating (the former gives an almost-constant continuum and the latter is responsible to the Wien-like rise of the spectrum). 

To extend the wavelength region of our study out to $70~\micron$, we fit the dust models of \citet{draine&li07}, combined with an additional stellar continuum component, to our IR photometry and IRS spectroscopy. For these models, the emission spectrum is given by \citet{draine07} as,
\begin{multline}
S_{\nu,\rm{model}} = \Omega_{*}B_\nu(T_*) + \frac{M_{\rm{dust}}}{4\pi D_{\rm{lum}}^{2}}[(1-\gamma)p_\nu^{(0)}(j_M,U_{\rm{min}}) \\ +\gamma p_\nu(j_M,U_{\rm{min}},U_{\rm{max}},\alpha),
\end{multline}
where $\Omega_{*}$ is the solid angle subtended by stars, $T_*$ is the effective temperature of the stellar contribution, $M_{\rm{dust}}$ is the total dust mass, $D_{\rm{lum}}$ is the distance to the galaxy, $p_\nu$ is the specific power per unit dust mass, $U_{\rm{min}}$ ($U_{\rm{max}}$) is the minimum (maximum) interstellar radiation, $\gamma$ is the fraction of the dust mass exposed to radiation with intensity $U>U_{\rm{min}}$, $j_M$ corresponds to the dust model \citep[i.e., the PAH abundance relative to dust, $q_{\rm{PAH}}$; shown in table 3 of][]{draine&li07}, and $\alpha$ is the power-law factor for the starlight intensity. In summary, this emission spectrum is a linear combination of three components: (1) a stellar continuum with effective temperature $T_*$ which dominates at $\lambda\lesssim5~\micron$; (2) a diffuse ISM component with an intensity factor $U=U_{\rm{min}}$; and (3) a component arising from photo-dissociation regions (PDRs). Typically, component (2) comprises a much larger amount of the total dust mass and, as such, is dominant over component (3) in the emission spectrum \citep{draine07, draine14}.

To fit this model we follow the approach outlined in \citet{draine07}, which found that the SEDs of galaxies in the \textit{Spitzer} Infrared Nearby Galaxies Survey (SINGS) were well reproduced with fixed values of $\alpha=2$, $U_{\rm{max}}=10^6$, and $T_*=5000$~K. Holding these parameters fixed, $q_{\rm{PAH}}$, $U_{\rm{min}}$, $\gamma$, $M_{\rm{dust}}$, and $\Omega_{*}$ are varied to find the dust model that comes closest to reproducing the photometry and spectroscopy. For this work, a grid of $\gamma$ values is constructed for all $q_{\rm{PAH}}$ (MW, LMC, SMC) and $U_{\rm{min}}$ values. The value of $M_{\rm{dust}}$ for each grid point is determined by minimizing the $\chi^2$ parameter in a similar manner to eq. (\ref{cphot}) and (\ref{chisq}), as $M_{\rm{dust}}$ represents a constant offset value. The goodness-of-fit for each case is assessed using the $\chi^2$ parameter,
\begin{equation}
\chi^2\equiv\sum_i\frac{S_{\mathrm{obs},i}-S_{\mathrm{model},i}}{\sigma_{\mathrm{obs},i}^2+\sigma_{\mathrm{model},i}^2}\,,
\end{equation}
where the sum is over observed bands and spectroscopic channels, $S_{\mathrm{model},i}$ is the model spectrum (for band comparison, the model spectrum is convolved with the response function of that band), $\sigma_{\mathrm{obs},i}$ is the observational uncertainty in the observed flux density $S_{\mathrm{obs},i}$, and $\sigma_{\mathrm{model},i}=0.1S_{\mathrm{model},i}$ as adopted by \citet{draine07}. The observed flux densities used in determining the best fit is comprised of the IRS spectroscopy in addition to the IRAC 3.6, 4.5, and $5.8~\micron$ and MIPS 70, and $160~\micron$ photometry. The model which minimizes the value of $\chi^2$ is adopted for use in representing the region $\lambda_{\rm{rest}}\gtrsim37/(1+z)~\micron$. Examples of the best fitting model for SSGSS galaxies are shown in Figure~\ref{fig:DLFIT_SSGSS}. The bands within the IRS region are shown only for comparison and are not used directly for the fit.

Our choice to adopt the model which minimizes the value of $\chi^2$ is not necessarily the most accurate representation of the spectra, as the degeneracy of the model parameters can allow for multiple fits to have similar $\chi^2$ values while having different FIR SED shapes. However, we do not consider this to be of great significance for this study for two reasons. First, the flux density variation due to changes in the SED shape for cases with $(\bar\chi^2-\bar\chi^2_{\rm{min}})<1$, where $\bar\chi^2_{\rm{min}}$ is the minimum value of the reduced $\chi^2$, is typically less than $25\%$ over the $30-70~\micron$ region, which is lower than the scatter among individual galaxy templates. As we will be utilizing an average of our galaxy templates for our diagnostic, the uncertainty from model SED variations will not be the dominant source of uncertainty. Second, the parameters of these fits are not used to determine the properties of these galaxies, which are more sensitive to these degeneracy effects than the total flux density.    

All cases are best fit by Milky Way dust models with $q_{\rm{PAH}}\geq2.50\%$. There is a systematic trend among most fits to underestimate the flux density around the $8~\micron$ PAH feature and overestimate the $10-20~\micron$ region. This is most likely due to the limitations of fitting only three components to the data. However, since the main focus of these fits is to provide a description of the region from $\sim$$34-70~\micron$, these deviations are not considered to be significant, as they should have little effect on matching the shape of the emission beyond $\sim$$34~\micron$. As will be discussed in \S~\ref{template_variation}, these fits are consistent with other $z=0$ SFG templates found in the literature that make use of \citet{draine&li07} models, suggesting that these deviations could be a common problem. Investigation into the cause of these discrepancies warrants additional study, as it will improve our understanding of dust properties in galaxies. We do not make use of these fits to determine \lir\ values, instead using the values provided in the SSGSS catalog, which have been extensively checked \citep{treyer10} across a variety of diagnostic methods.

\subsection{Determining Rest-Frame Luminosities}
As these galaxies span a redshift range of $0.03\le z\le 0.22$, the IRS spectrum and photometry of each galaxy span slightly different regions in rest-frame wavelength. This offset causes observed photometric values to vary by up to 30\% from the rest-frame values. This would affect our determination of SFR if not accounted for and introduce additional scatter. Since previous MIR calibrations have been performed for local samples of galaxies ($z\sim0$) to accuracies around 30\% \citep{rieke09,kennicutt09,calzetti10,hao11}, this is a non-negligible effect. 

To correct for redshift effects, photometric values for each band are determined by convolving the spectrum at the rest-frame filter postions for each band according to eq (\ref{convolve}), only now using the corrected spectrum, $S_{\rm{IRS,corr}}(\lambda)$, instead of the original IRS spectrum. This is performed for the \textit{Spitzer} 8, 24, and $70~\micron$ and WISE 12 and $22~\micron$ bands. These corrected flux densities are used to calculate the rest-frame luminosity (erg~s$^{-1}$) of each band,
\begin{equation} \label{Lrest}
L_{\mathrm{rest}}= (\nu L_{\nu})_{\mathrm{rest}}=(\nu S_{\mathrm{IRS,corr}})_{\mathrm{obs}}\,4\pi D_{\mathrm{lum}}^2 \,,
\end{equation}
where $\nu_{\mathrm{rest}}$ and $\nu_{\mathrm{obs}}$ are the effective rest-frame and observer-frame frequency of each band, respectively, and $D_{\mathrm{lum}}$ is the luminosity distance for the galaxy, calculated from its redshift. These rest-frame luminosities are used to determine SFRs for each of the galaxies in our sample. In a similar manner, each IRS spectrum is expressed as a wavelength dependent rest-frame luminosity, $L(\lambda)_{\mathrm{rest}}$, using the continuous spectrum, $S_{\mathrm{IRS,corr}}(\lambda)$. This is used later for calibrating our wavelength continuous SFR-luminosity conversion factors, $C(\lambda)$. 

To correct the \textit{Spitzer} 3.6 and $4.5~\micron$ bands, which lie outside of the IRS spectral coverage, a correction is applied assuming these bands encompass the Rayleigh-Jeans tail of the stellar continuum emission,
\begin{equation}
S'_{\mathrm{obs}}= S_{\mathrm{obs}}\left(\frac{\lambda_{\mathrm{rest}}}{\lambda_{\mathrm{obs}}}\right)^{-2}=S_{\mathrm{obs}}\times(1+z)^{-2}\,.
\end{equation} 
The luminosity is then found following eq (\ref{Lrest}) using $S'_{\mathrm{obs}}$. The rest-frame \textit{Spitzer} 3.6 and $4.5~\micron$ luminosities will only be used to examine the origins of scatter within our conversion factors in \S~\ref{scatter}, and has no influence on our estimates for the MIR conversion factors. The fitting results in \S~\ref{DL07_method} suggest that this simple approach is reasonable for our sample of SFGs.

\begin{table*}
\begin{center}
\caption{Reference Star Formation Rate Calibrations \label{Tab:calibrations}}
\begin{tabular}{ccccc}
\hline\hline 
Band(s) & $L_x$ Range & $SFR$ & $\log C_x$ & Reference \\
     & (erg s$^{-1}$)&       &            &           \\ \hline
FUV+TIR & \nodata & $[L(\mathrm{FUV})_{\mathrm{obs}}+0.46\,L(\mathrm{TIR})]/C_x$ & 43.35 & \citet{hao11} \\
FUV+$24~\micron$ & \nodata & $[L(\mathrm{FUV})_{\mathrm{obs}}+3.89\,L(24)]/C_x$ & 43.35 & \citet{hao11} \\
NUV+TIR & \nodata & $[L(\mathrm{NUV})_{\mathrm{obs}}+0.27\,L(\mathrm{TIR})]/C_x$ & 43.17 & \citet{hao11} \\
NUV+$24~\micron$ & \nodata & $[L(\mathrm{NUV})_{\mathrm{obs}}+2.26\,L(24)]/C_x$ & 43.17 & \citet{hao11} \\
H$\alpha$+$8~\micron$ & \nodata & $[L(\mathrm{H}\alpha)_{\mathrm{obs}}+0.011\,L(8)]/C_x$ & 41.27 & \citet{kennicutt09,hao11} \\
H$\alpha$+$24~\micron$ & \nodata & $[L(\mathrm{H}\alpha)_{\mathrm{obs}}+0.020\,L(24)]/C_x$ & 41.27 & \citet{kennicutt09,hao11} \\
H$\alpha$+$24~\micron$ & $L(24)< 4\times10^{42}$ & $[L(\mathrm{H}\alpha)_{\mathrm{obs}}+0.020\,L(24)]/C_x$ & 41.26 & \citet{calzetti10} \\
 & $4\times10^{42}\le L(24)< 5\times10^{43}$ & $[L(\mathrm{H}\alpha)_{\mathrm{obs}}+0.031\,L(24)]/C_x$ & 41.26 & \citet{calzetti10} \\
 & $ L(24)\ge 5\times10^{43}$ & $L(24)\times[2.03\times10^{-44}\,L(24)]^{0.048}/C_x$ & 42.77 & \citet{calzetti10} \\
H$\alpha$+TIR & \nodata & $[L(\mathrm{H}\alpha)_{\mathrm{obs}}+0.0024\,L(\mathrm{TIR})]/C_x$ & 41.27 & \citet{kennicutt09,hao11} \\
$24~\micron$ & $2.3\times10^{42}\le L(24)\le 5\times10^{43}$ & $L(24)/C_x$ & 42.69 & \citet{rieke09} \\
 & $L(24)>5\times10^{43}$ & $L(24)\times(2.03\times10^{-44}L(24))^{0.048}/C_x$ & 42.69 & \citet{rieke09} \\
$70~\micron$ & $L(70)\gtrsim 1.4\times10^{42}$ & $L(70)/C_x$ & 43.23 & \citet{calzetti10} \\ \hline
\end{tabular}
\end{center}
\textbf{Notes.} Columns list the (1) bands used in the calibration, (2) luminosity range over which the calibration can be used; empty fields denote an unspecified range, (3) $SFR$ conversion formula, (4) conversion constant, and (5) reference for calibration.
\end{table*}

\subsection{Reference Monochromatic SFR Indicators}
In order to perform any calibration of luminosity as a SFR indicator, it is necessary to rely on previous, well-calibrated SFR indicators. In this work, we utilize the calibrations of \citet{kennicutt09}, \citet{rieke09}, \citet{calzetti10}, and \citet{hao11}, which incorporate the full suite of data available for this sample. This list is shown in Table~\ref{Tab:calibrations}. The reference SFR, $\langle SFR \rangle$, for each galaxy is taken to be the average of the SFRs from these calibrations. By utilizing the average of a large number of diagnostics, we limit the risk of potential biases that any single diagnostic can be subject to. Several of these reference diagnostics make use of the total infrared luminosity, $L_{\rm{TIR}}$, which refers to the integrated luminosity over the region from 3 to $1100~\micron$. All measurements of $L_{\rm{TIR}}$ for these galaxies have been obtained from the original SSGSS dataset \citep{treyer10}. For the SSGSS sample, \citet{treyer10} find that $L_{\rm{TIR}}$ is larger than \lir\ by $\sim$$0.04$~dex.

To utilize SDSS measurements of H$\alpha$ for a SFR estimation, it is necessary to apply an aperture correction. The diameter of the SDSS spectroscopic fiber spans 3\arcsec, which is a factor of $\sim$$3$ smaller than the typical size of the SSGSS galaxies and results in only a fraction of the light being measured. We correct for this aperture effect using the prescription from \citet{hopkins03}, which uses the difference between the $r$-band Petrosian magnitude and the $r$-band fiber magnitude \citep[see also][]{treyer10}. The values of these corrections range from $1.9-8.3$ for our sample, with the exception of SSGSS~67 with a correction of $21.4$ due to its much larger size. 

\begin{figure*}
\plotone{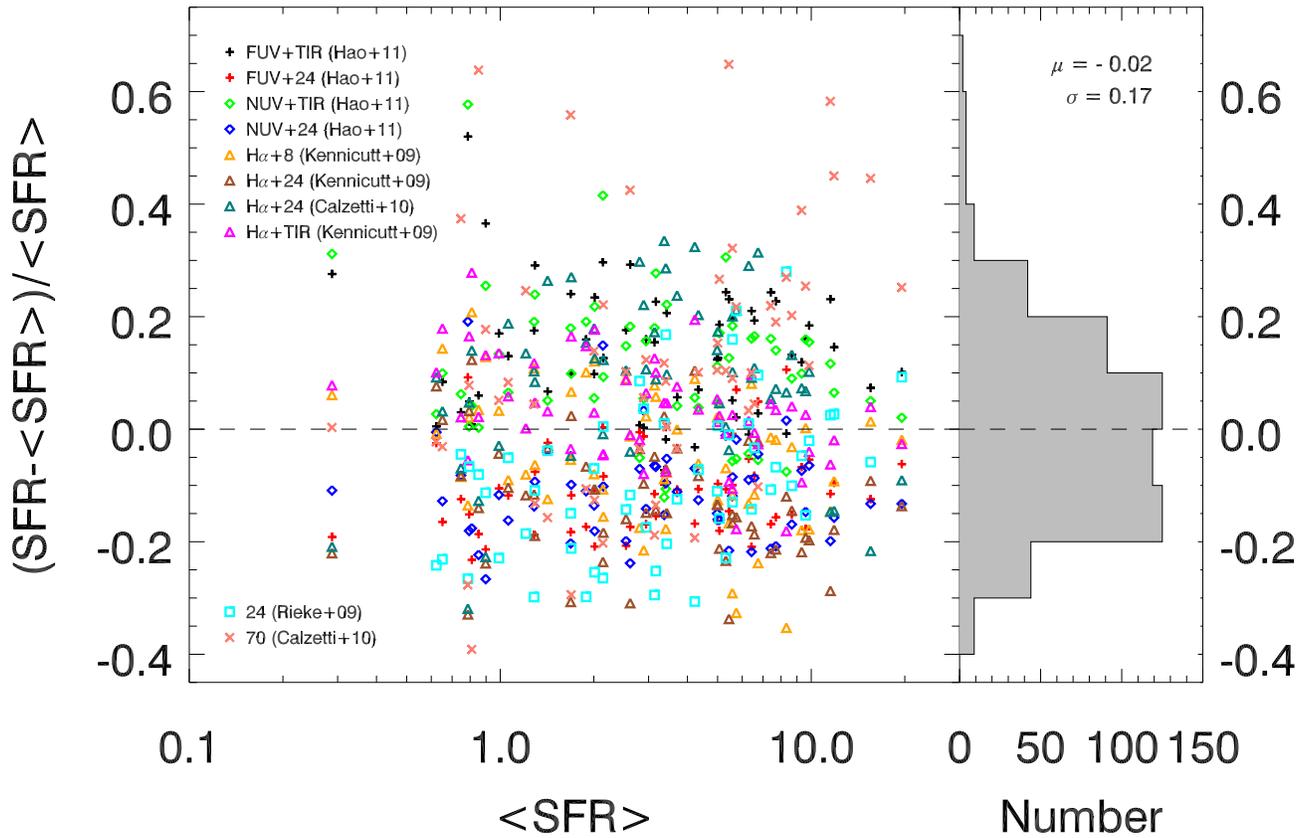}
\caption{{\it Left:} Comparison of SFRs determined from the calibrations listed in Table~\ref{Tab:calibrations} for the SFGs in the SSGSS. Each vertical strip of values shows the SFR values for each method on a single galaxy. The reference SFR, $\langle SFR \rangle$, for these galaxies is taken to be the average of the SFR values from these calibrations. {\it Right:} Histogram showing the distribution of SFR offsets relative to the reference value. This distribution is well fit by a Gaussian with $\mu=-0.02$ and $\sigma=0.17$, suggesting that the majority of these values agree within the uncertainties associated with the individual calibrations. \label{fig:SFR_compare}}
\end{figure*}

\begin{figure*}
\plotone{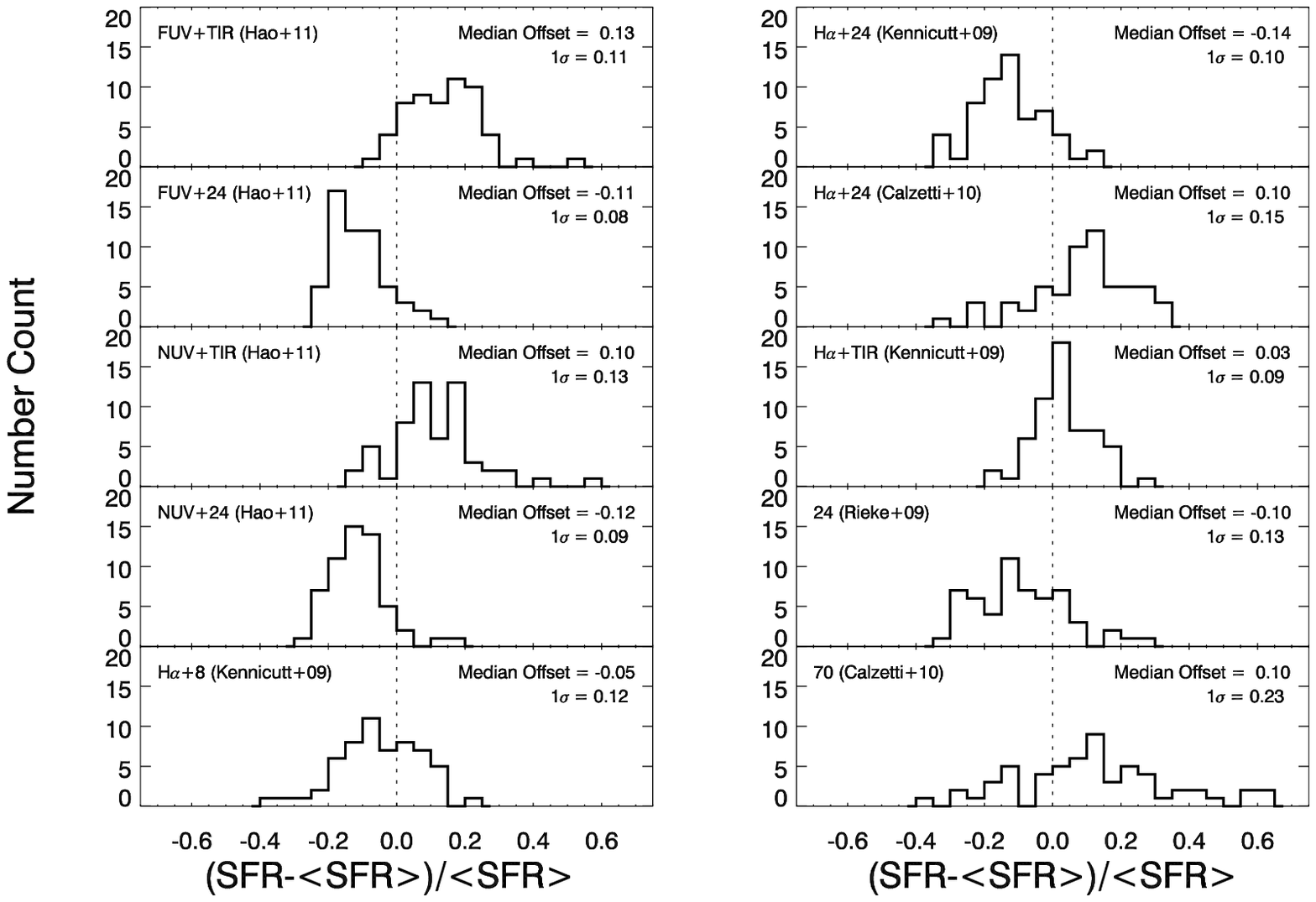}
\caption{Histograms showing the distribution of the individual SFR estimates relative to the reference value, $\langle SFR \rangle$. The calibrations being considered are listed in Table~\ref{Tab:calibrations}. The median offset and $1\sigma$ dispersion are shown in each panel. \label{fig:SFR_compare_hist}}
\end{figure*}

As a consistency check, the SFRs inferred from each indicator for our galaxies is compared in Figure~\ref{fig:SFR_compare}. It is seen from the distribution that the majority of these values agree within the $\sim$30\% uncertainty associated with individual calibrations \citep{rieke09,kennicutt09,calzetti10,hao11}. A formal fit of this distribution to a Gaussian profile gives values of $\mu=-0.02$ and $\sigma=0.17$. Our choice in using the average of the diagnostics, $\langle SFR \rangle$, for each galaxy instead of the median value appears to cause no significant differences, with typical offsets of only a few percent between the two, which are symmetric. A formal Gaussian fit of fractional difference between the mean and median gives $\mu=-0.01$ and $\sigma=0.04$. We illustrate the relative offsets for the individual calibrations in Figure~\ref{fig:SFR_compare_hist}. We note that the MPA/JHU group provides independent estimates for SFRs based on the technique discussed in \citet{brinchmann04} using extrapolated H$\alpha$ measurements. However, we find that the spread in values for these estimates relative to $\langle SFR \rangle$ are significantly larger than our other diagnostics ($1\sigma=0.48$), which we attribute to their larger SFR uncertainties ($\sim$$50\%$), and as such were excluded from our analysis (including Figures~\ref{fig:SFR_compare} and \ref{fig:SFR_compare_hist}).

\subsection{\texorpdfstring{\lir}{LIR} as a SFR Indicator} \label{lir}
A commonly utilized method to determine SFRs for galaxies relies on measuring the integrated luminosity over most of the IR wavelength range, \lir ($8-1000~\micron$).  However, physically understanding the conversion of \lir\ to a SFR is non-trivial and sensitive to many assumptions, such as the timescale of star formation, $\tau$, the star formation history (SFH), the metallicity, and the initial mass function \citep[IMF; see][]{murphy11b,calzetti13}.  For example, a galaxy with a constant SFH, a fixed metallicity, and a fixed IMF will have the calibration constant for \lir\ change by a factor of 1.75 between assuming $\tau=100$~Myr and $\tau=10$~Gyr \citep{calzetti13}.

We chose to avoid the use of SFRs based solely on \lir\ for reference because of the sensitivity to these assumptions. However, in order to compare the accuracy of our calibration on higher redshift samples (in \S~\ref{Cx_test}) for which \lir\ is the only technique available to estimate SFRs, we use a SFR-\lir\ conversion which reproduces the values of $\langle SFR \rangle$ seen for the SSGSS sample. This occurs for a conversion factor of $\log[C(L_{\mathrm{IR}})]=43.64$~erg~s$^{-1}/(M_\odot \rm{yr}^{-1}$). Utilizing Starburst99 \citep{leitherer99}, with a constant SFH, solar metallicity, a Kroupa IMF over $0.1-100~M_\odot$, and assuming all of the stellar light (UV+visible) is reradiated by dust, this corresponds to a timescale of $\tau\sim500$~Myr \citep[e.g.,][]{calzetti13}. 

This adopted conversion factor differs slightly from other commonly adopted values. In the case of \citet{murphy11b}, $\log[C(L_{\mathrm{IR}})]=43.41$~erg~s$^{-1}/(M_\odot \rm{yr}^{-1}$), our calibration is larger by 70\%. This large difference is due to two reasons: (1) \citet{murphy11b} assume that only UV light is being reradiated by the dust and does not account for the optical light that would also be reradited ($\sim$40\% of the discrepancy), and (2) they assume a 100~Myr constant star-forming population ($\sim$20\% of the discrepancy). In the case of \citet{kennicutt98} after converting from a \citet{salpeter55} IMF to a \citet{kroupa01} IMF, $\log[C(L_{\mathrm{IR}})]=43.53$~erg~s$^{-1}/(M_\odot \rm{yr}^{-1}$), our calibration is larger by 30\%. Most of this difference is due to them assuming a $100$~Myr constant star-forming population.

\section{A Calibrated Continuous, Monochromatic SFR(\texorpdfstring{$\lambda$}{lambda})}
\subsection{Composite IRS Spectrum}
The SFR of a galaxy, using a calibrated single-band luminosity, can be written as
\begin{equation}
\mathrm{SFR} (M_\odot \mathrm{yr}^{-1})= L_x/C_x\,,
\label{sfr}
\end{equation}
where $L_x$ is the monochromatic luminosity, in units of erg~s$^{-1}$, and $C_x$ is the conversion factor between SFR and luminosity for filter $x$ \citep[following convention of][]{kennicutt12}. In this respect, the appropriate conversion factor at a given band is found by normalizing the luminosity by the SFR determined independently from a reference calibration.

This same approach is taken to calibrate our continuous wavelength conversion factors, 
\begin{equation}
C(\lambda) (\mathrm{erg~s}^{-1}/(M_\odot \mathrm{yr}^{-1}))= L(\lambda)_{\mathrm{rest}}/\langle \mathrm{SFR}\rangle\,,
\end{equation}
where $L(\lambda)_{\mathrm{rest}}$ is the wavelength dependent IRS luminosity and $\langle SFR \rangle$ is the reference SFR. To achieve our calibration of $C(\lambda)$, the SFR-normalized IRS spectra are averaged together to create a composite spectrum for the group. As a result of shifting the spectra to the rest-frame, the wavelengths associated with each spectral channel no longer match exactly. Therefore, to perform this average, the channel wavelengths in the spectrum of the first galaxy in our group is taken to be the reference grid. Next, the normalized luminosity values of the other galaxies are re-gridded to this (i.e., each channel is associated to the nearest neighboring reference channel). In using this approach, the smoothing of sharp features that result from direct interpolation is avoided. The uncertainties associated with this re-griding to determine a composite spectrum are small relative to the channel flux density uncertainly. Furthermore, these uncertainties are much smaller than the scatter between spectra, which drives the uncertainty of our template, and can be considered negligible for the purposes of this study.

\begin{figure*}
\plotone{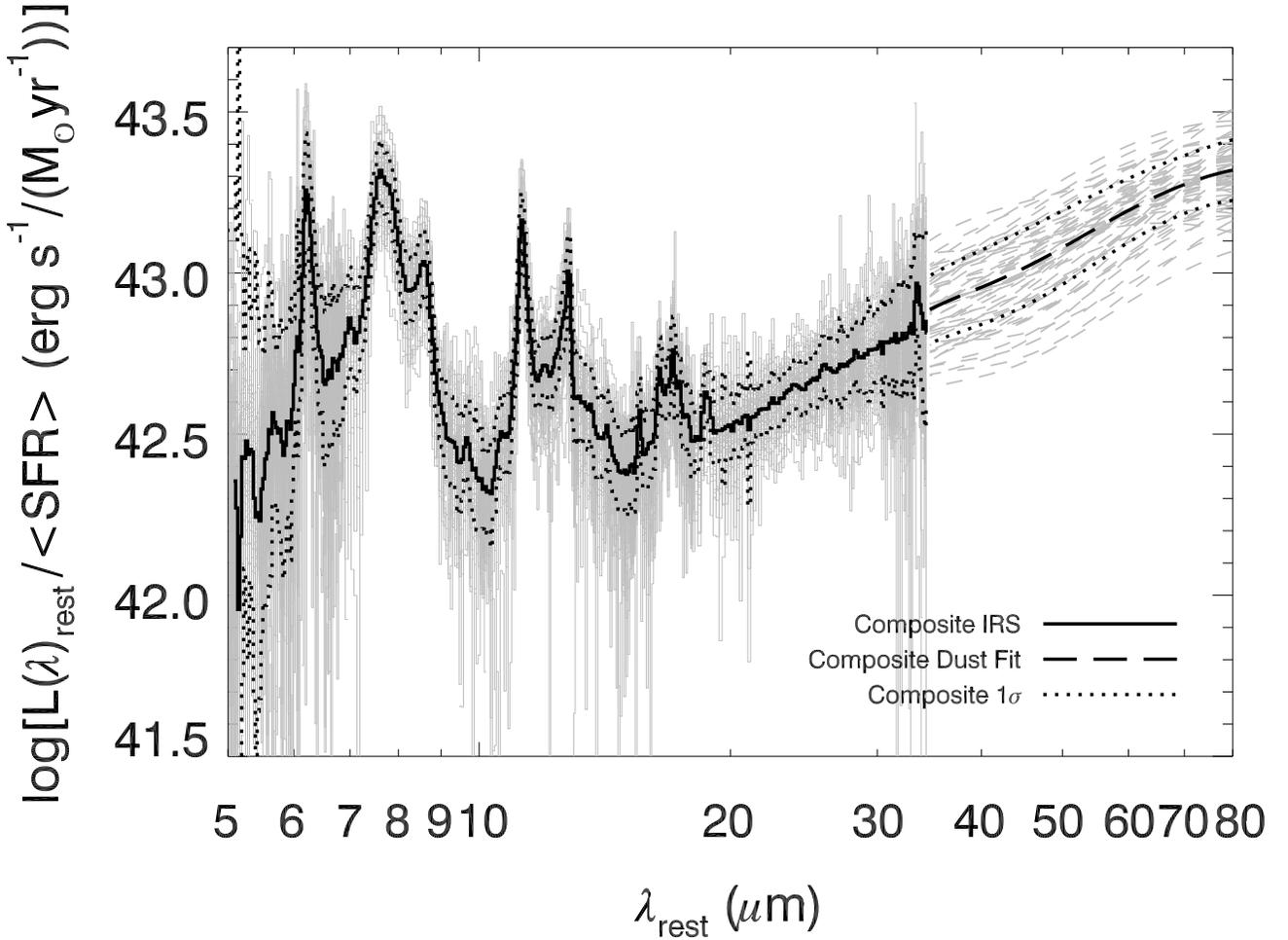}
\caption{Normalized IRS luminosity, $L(\lambda)_{\mathrm{rest}}/\langle SFR \rangle$, for all SFG galaxies (gray solid lines). The composite spectrum of this group is shown (thick black line) along with the standard deviation from this average (black dotted lines), which at most wavelengths is between $20-30\%$. The fits to the dust continuum for each galaxy (gray dashed lines; described in \S~\ref{DL07_method}) along with the average (black dashed line) are also shown. The low dispersion among normalized spectra suggests that the $6-70~\mu$m region can be utilized for SFR diagnostics. \label{fig:group_all}}
\end{figure*}

The result of an average for the entire sample of SFGs in the SSGSS sample is shown in Figure~\ref{fig:group_all}. The uncertainty of each channel in the composite spectrum is taken to be the standard deviation of the the group value for that channel. The sample standard deviation is the dominant source of uncertainty (typically between $20-30\%$ of the normalized luminosity value) and is larger than the flux density uncertainties of individual IRS channels (typically $\sim$2\%) by roughly an order of magnitude. This template can be used to determine the appropriate conversion factor for any luminosity within our wavelength coverage.

\subsection{Filter Smoothed Composite Spectrum} \label{filter_smooth}
In practice, observations of a galaxy are made using specific bandpass filters that encompass a portion of their SED. Therefore, it is more practical to utilize a composite spectrum that corresponds to photometric luminosities observed by various bands as functions of redshift. To accomplish this, the normalized IRS spectrum of each SFG in our sample is convolved with the filter response of specific bands as functions of redshift (i.e., the effective wavelength blue-shifts and the bandpass narrows, both by a factor of ($1+z$), as one goes to higher redshifts). Performing this convolution is similar to smoothing by the bandpass filter, only with the filter width changing with redshift. Throughout the rest of this paper, the term ``smoothed'' is used interchangeably to mean this convolution process. The redshift limit imposed for each band occurs at the shortest usable rest-frame wavelengths of the IRS spectrum for that band. The composite IRS template and the filter smoothed composites presented in this section are publicly available for download from the IRSA\footnote{\url{http://irsa.ipac.caltech.edu/data/SPITZER/MIR\_SFR}}.

Each normalized IRS spectra is smoothed using the \textit{Spitzer}\footnote{\url{http://irsa.ipac.caltech.edu/data/SPITZER/docs/irac/}}$^,$\footnote{\url{http://irsa.ipac.caltech.edu/data/SPITZER/docs/mips/}}, WISE\footnote{\url{http://www.astro.ucla.edu/~wright/WISE/passbands.html}}, and JWST/MIRI filters. The properties of these filters is listed in Table~\ref{Tab:filter_properties}. We note that the \textit{Herschel} PACS $70~\micron$ is close enough to \textit{Spitzer} $70~\micron$ that these can be interchanged for use with $C_{S70}(\lambda)$. We emphasize to the reader that care should be taken when considering the $22~\micron$ band, as it has been shown to suffer from an effective wavelength error \citep[see][]{wright10,brown14}. For the MIRI filters, we use the response functions of \citet{glasse15}. Since these curves do not take into account the wavelength dependent quantum efficiency, the instrument transmission, or the responsivity of the detector, they should be updated once better curves become available.

\begin{figure*}
\begin{center}$
\begin{array}{cc}
\includegraphics[width=3.0in]{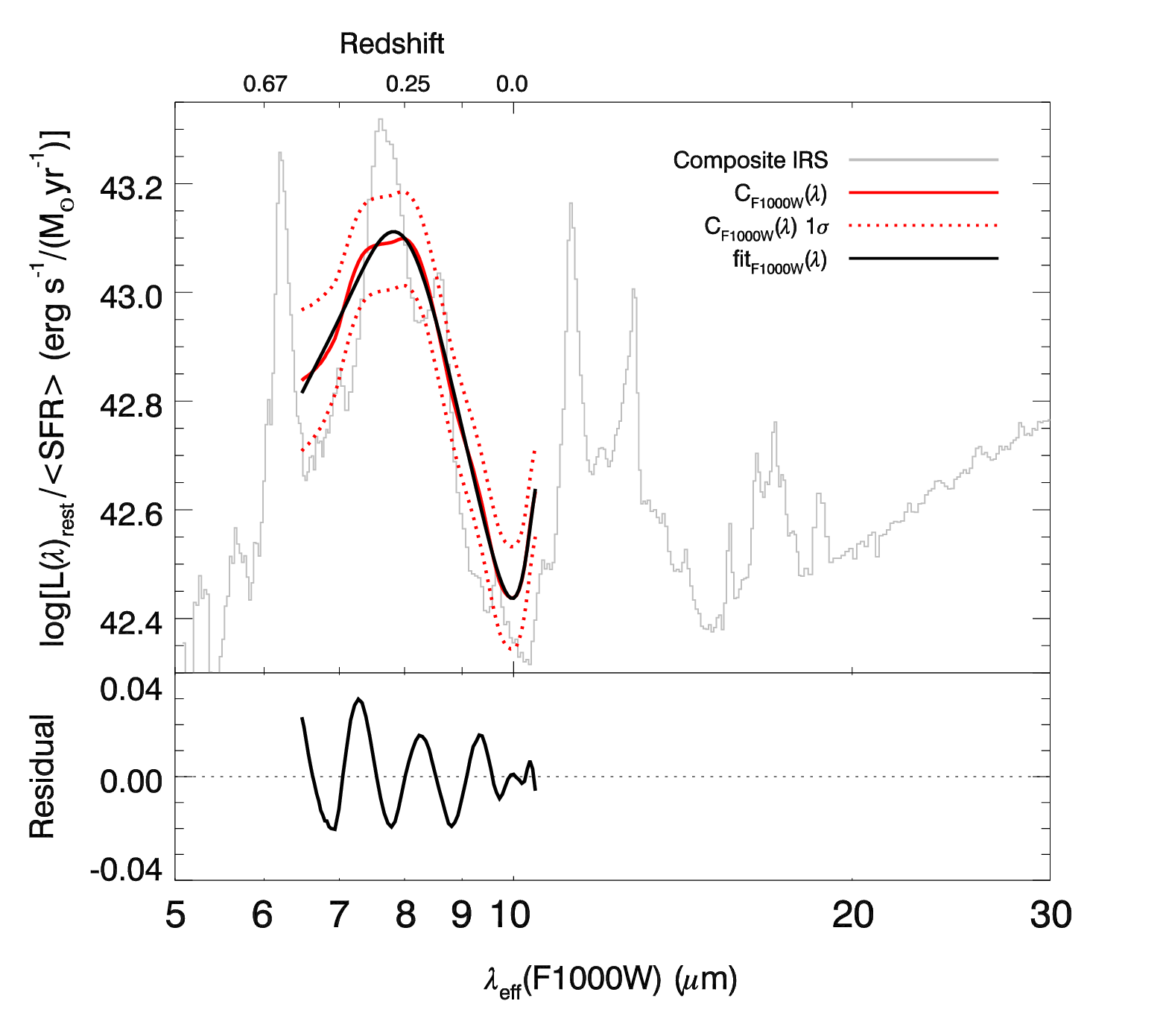} & 
\includegraphics[width=3.0in]{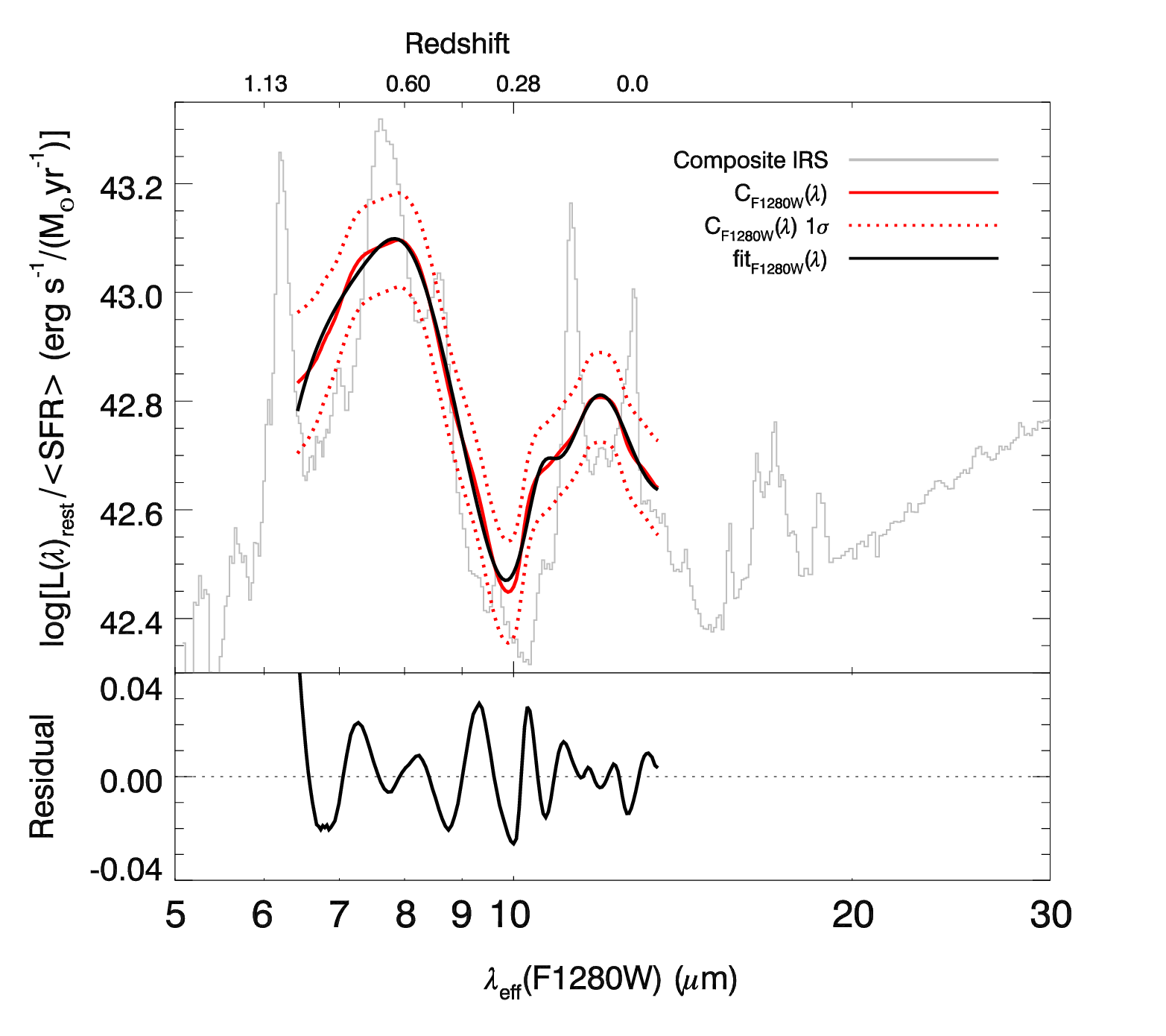} \\
\includegraphics[width=3.0in]{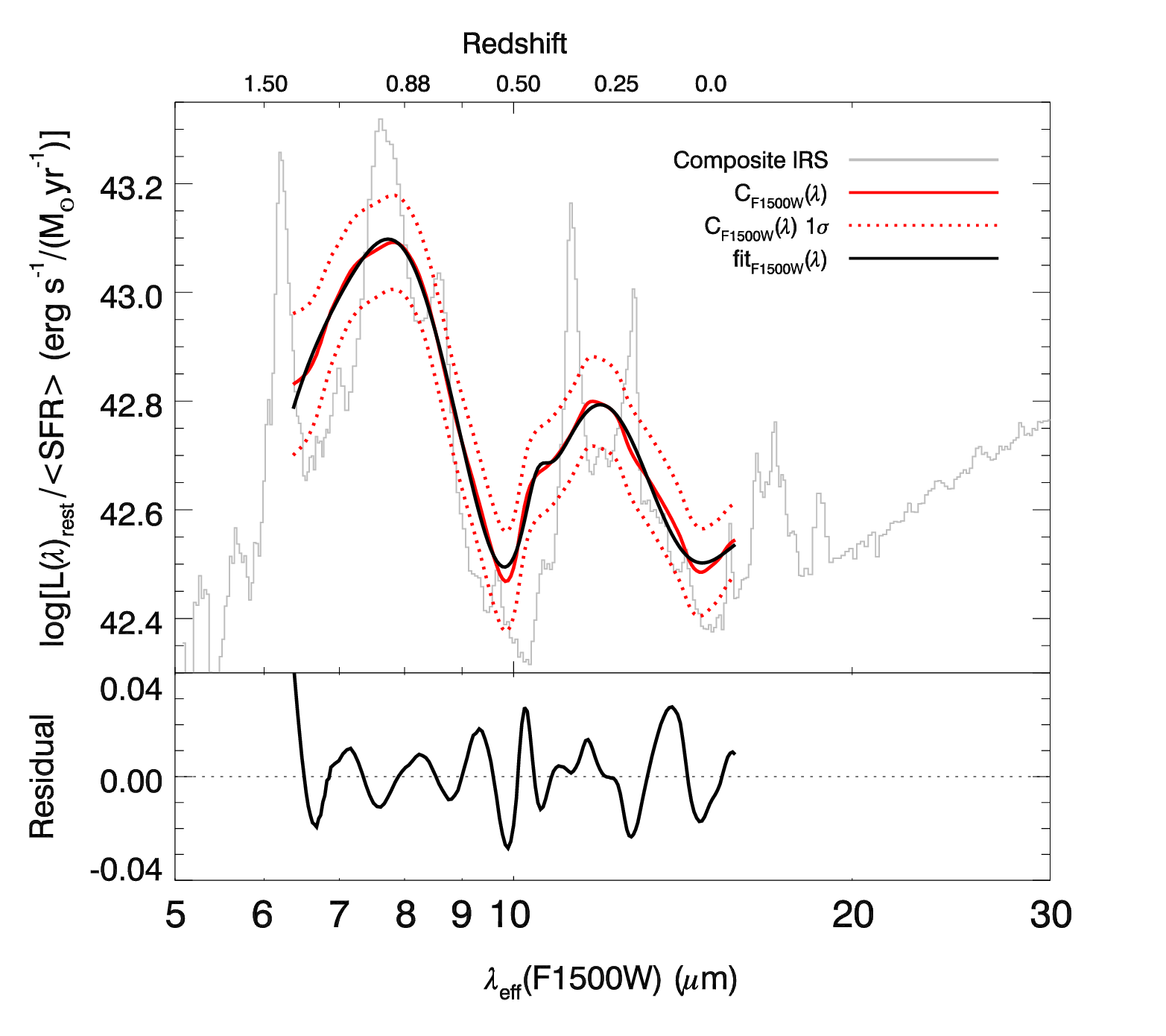} & 
\includegraphics[width=3.0in]{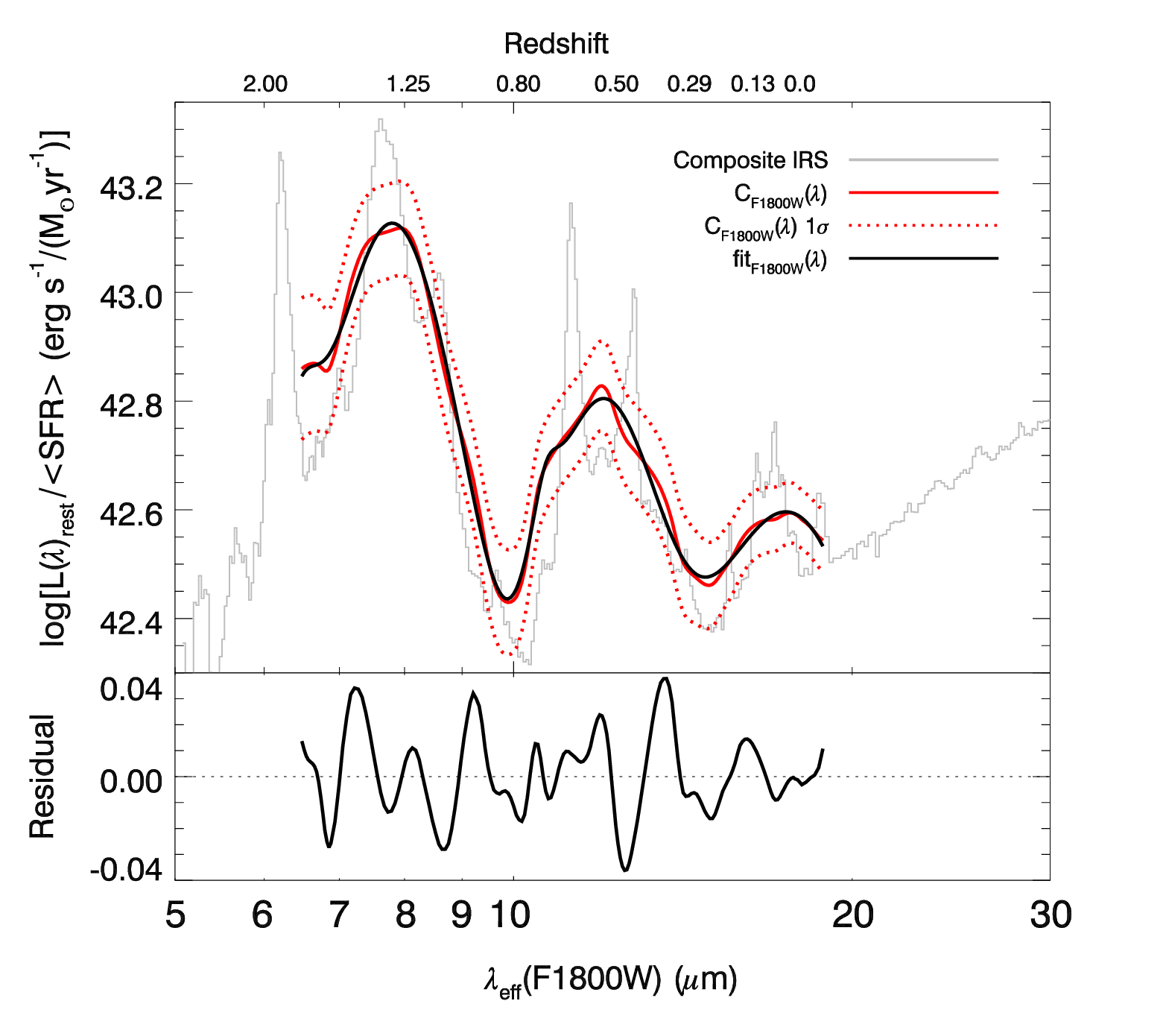} \\
\includegraphics[width=3.0in]{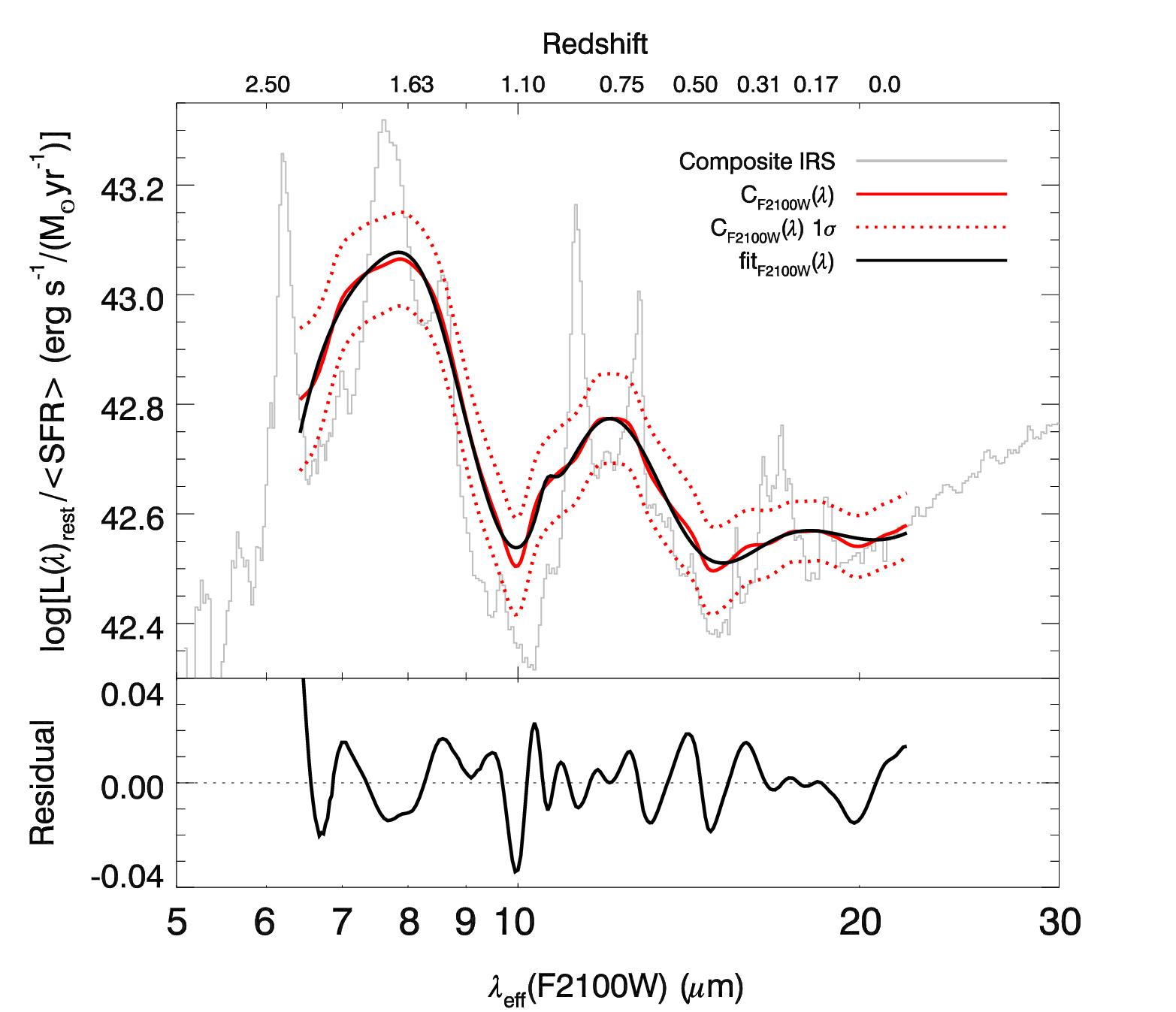} & 
\includegraphics[width=3.0in]{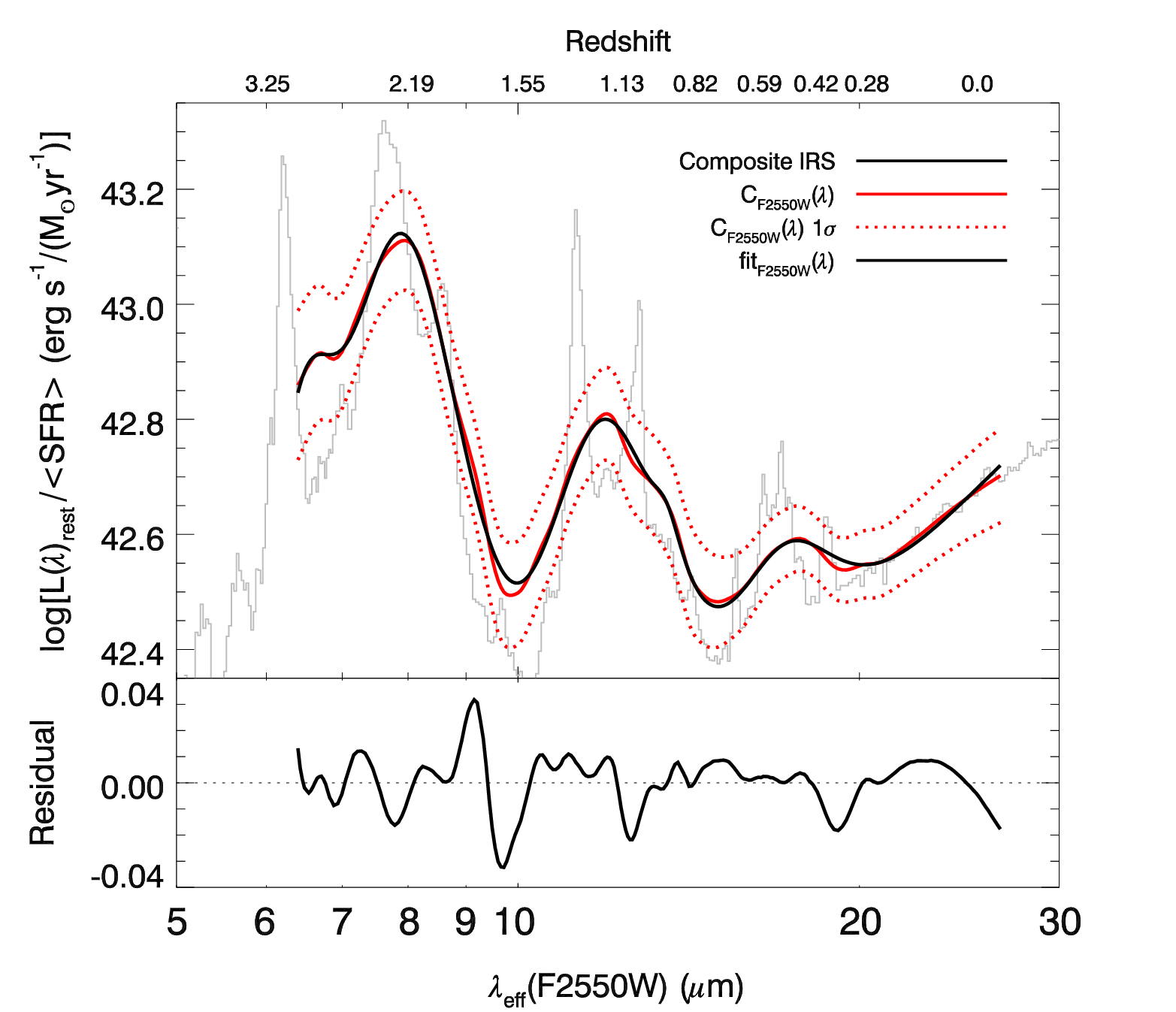} 
\end{array}$
\end{center}
\caption{The top of each panel shows the conversion factor for select JWST/MIRI bands using all SFG galaxies (solid red lines), along with their uncertainty (dotted red lines), which for most cases is between $15-20\%$. The region below $\sim$$6~\micron$ for each band is excluded due to significantly increased uncertainty in the composite spectrum (see \S~\ref{filter_smooth}). The bottom of each panel shows the residuals between the conversion factor and a fit to the curve ($\log[C_{x}(\lambda)/fit_{x}(\lambda)]$). \label{fig:JWST_calibration_all_SFG}}
\end{figure*}

 The composite spectra for each of these bands is created by averaging the smoothed spectra together, in the same manner as for the group average. The result of averaging the convolved spectra is shown in Figures~\ref{fig:band_calibration_all_SFG} \& \ref{fig:JWST_calibration_all_SFG}. The associated uncertainty with each channel (sample standard deviation) is slightly lower than the native composite spectrum owing to the smoothing from the convolution and is typically between $15-20\%$ of the normalized luminosity value (except for $70~\micron$ case, which is still around 30\%), making them comparable to accuracies achieved in many previous calibrations. Previously determined MIR conversion factors (from $z\sim0$ samples) are also shown and appear in good agreement. 

For the filter bands considered here, the smoothed IRS spectra show a very large increase in scatter below $\sim$$6~\micron$, which is due to a combination of the end-of-channel uncertainties being very high and also from variations in the old stellar populations of these galaxies. For these reasons, we only consider the regions for which the $1\sigma$ uncertainty is less than $30\%$ suitable for calibration. In the case of the WISE $12~\micron$ band, this region occurs below $\sim$$7~\micron$ because of the significantly wider filter bandwidth. The ranges chosen for the calibration of each band is shown in Table~\ref{Tab:calibration_parameters}. 

\begin{table}
\begin{center}
\caption{Filter Properties \label{Tab:filter_properties}}
\begin{tabular}{cccc}
\hline\hline 
Instrument & Band & $\lambda_{\rm{eff,0}}$ & FWHM \\
 &  & ($\micron$) & ($\micron$) \\ \hline
IRAC & $8~\micron$ & 7.87 & 2.8 \\
MIPS & $24~\micron$ & 23.68 & 5.3 \\
MIPS & $70~\micron$ & 71.42 & 19.0 \\
WISE & $12~\micron$ & 12.08 & 8.7 \\
WISE & $22~\micron$ & 22.19$^a$ & 3.5 \\
MIRI & F1000W & 10.00 & 2.0 \\
MIRI & F1280W & 12.80 & 2.4 \\
MIRI & F1500W & 15.00 & 3.0 \\
MIRI & F1800W & 18.00 & 3.0 \\
MIRI & F2100W & 21.00 & 5.0 \\
MIRI & F2550W & 25.50 & 4.0 \\ \hline
\end{tabular}
\end{center}
\textbf{Notes.} Columns list the (1) instrument, (2) passband name, (3) rest-frame effective wavelength, and (4) full width at half maximum. $^a$The $22~\micron$ observations suffer from a effective wavelength error \citep[see][]{wright10,brown14}.
\end{table}

\subsection{Fits to the Composite Spectra}
To simplify the application of our results as SFR indicators, each of the filter smoothed composite spectra is fit using a continuous function, $fit_{x}(\lambda)$. We perform Levenberg-Marquardt least-squares fits of a polynomial (up to 1st order) and Drude profiles (up to 5), $I_r(\lambda)$, to the smoothed composite spectra, using the IDL code \texttt{MPFITFUN},
\begin{equation}
fit_{x}(\lambda) = \sum_{i=1}^2 p_i\lambda^{(i-1)} + \sum_{r=1}^{5} I_r(\lambda) \,,
\end{equation}
where $x$ corresponds to the filter being considered, $p_i$ are constants, and $I_r(\lambda)$ are Drude profiles. Drude profiles, which are typically employed to characterize dust features, have the form
\begin{equation}
I_r(\lambda) = \frac{b_r\gamma_r^2}{(\lambda/\lambda_r-\lambda_r/\lambda)^2+\gamma_r^2} \,,
\end{equation}
where $\lambda_r$ is the central wavelength of the feature, $\gamma_r$ is the fractional FWHM, and $b_r$ is the central intensity, which is required to be non-negative. We emphasize that because these are smoothed spectra, the parameters of these fits are not of physical significance and are simply being employed for ease of application. 

\begin{figure*}
\begin{center}$
\begin{array}{cc}
\includegraphics[width=3.0in]{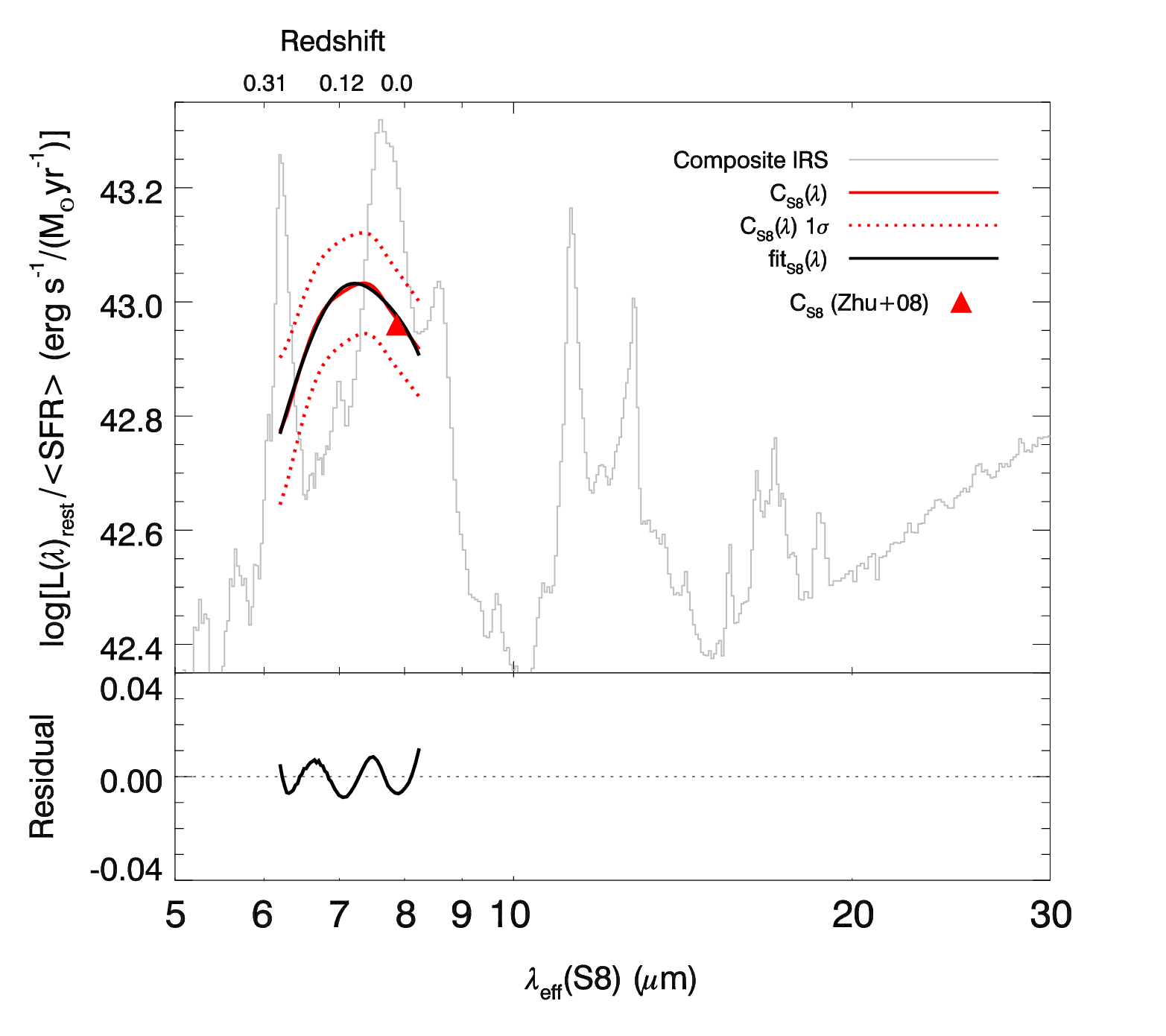} &
\includegraphics[width=3.0in]{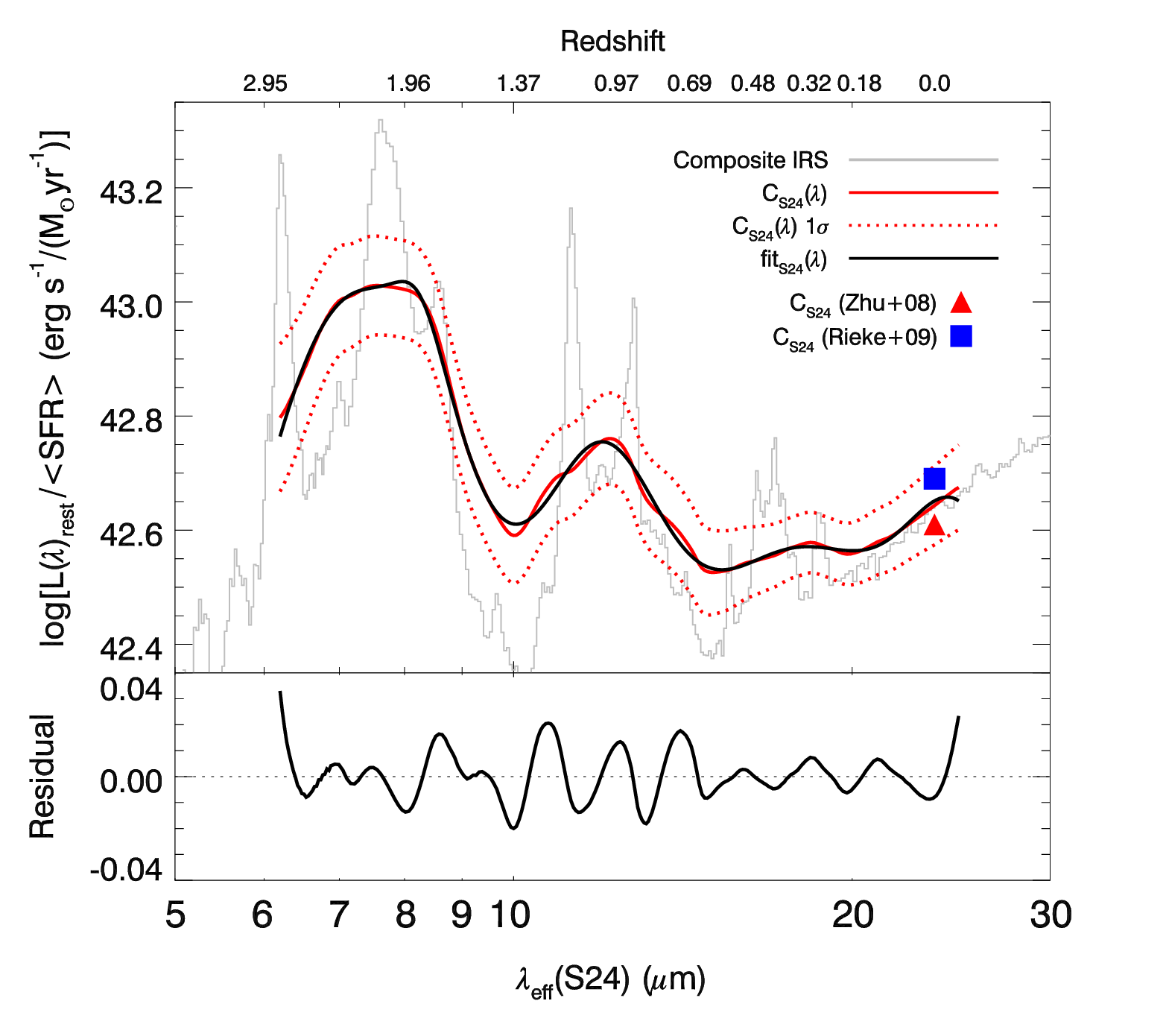} \\ 
\includegraphics[width=3.0in]{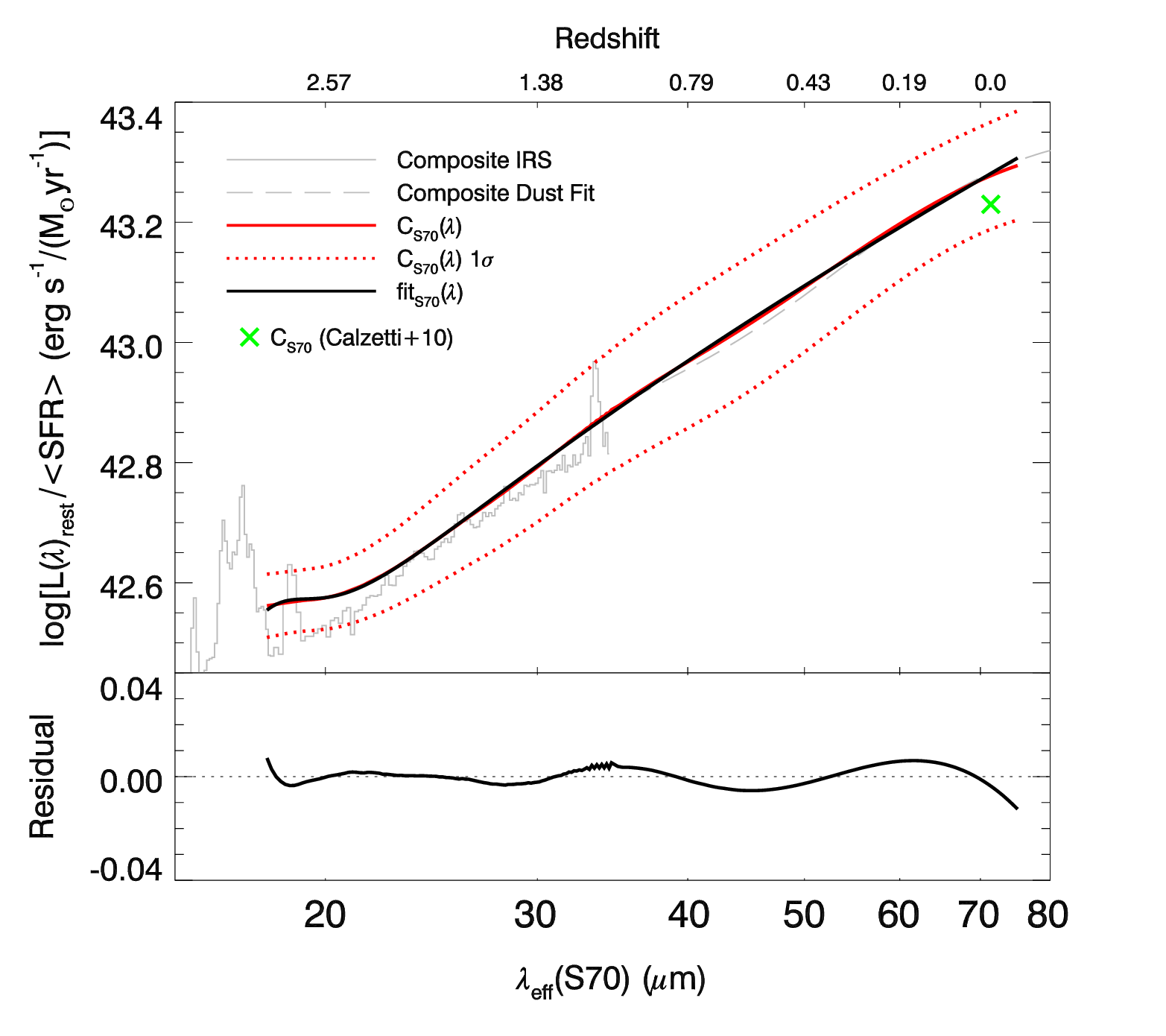} &
\includegraphics[width=3.0in]{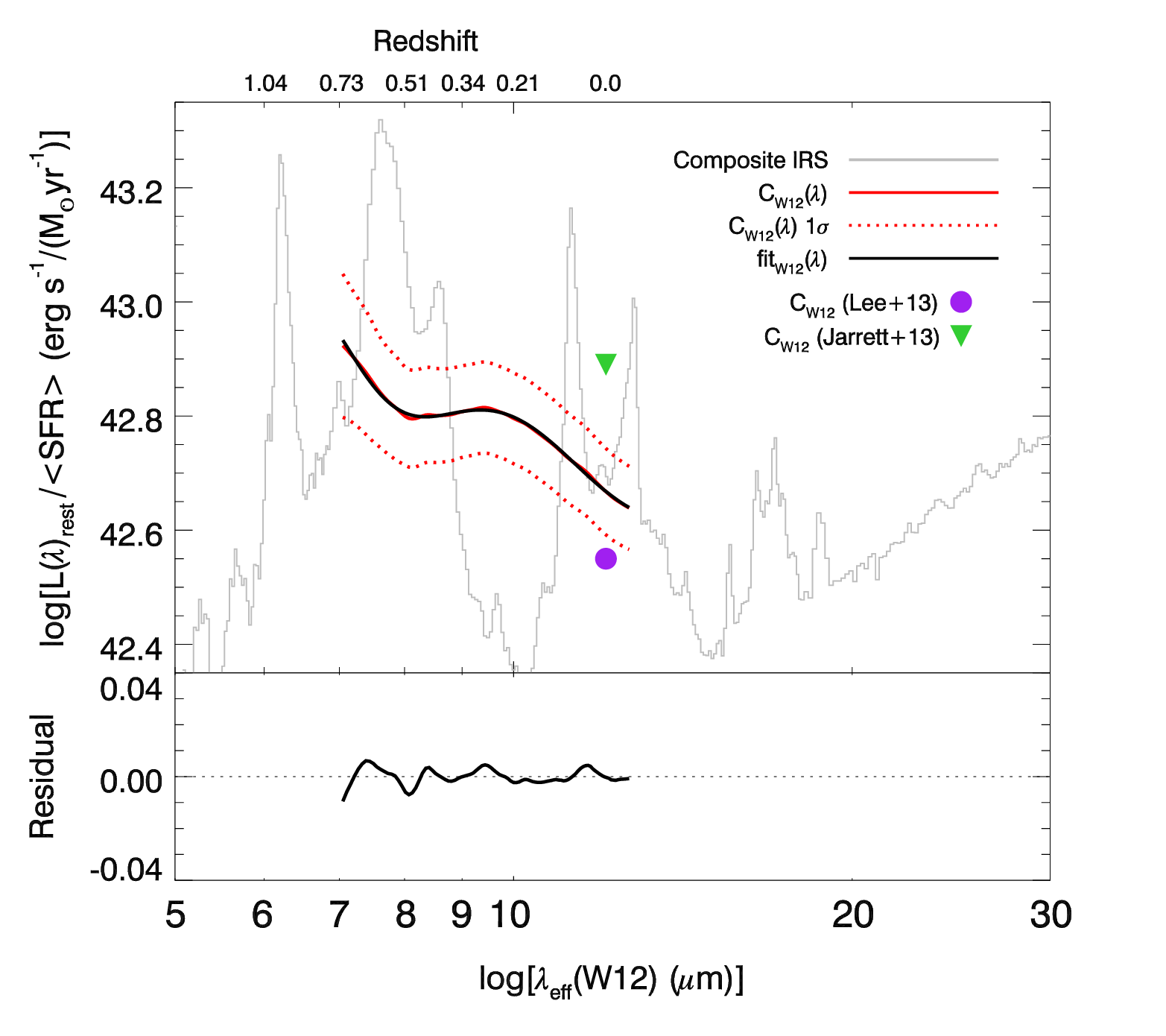} \\ 
\multicolumn{2}{c}{\includegraphics[width=3.0in]{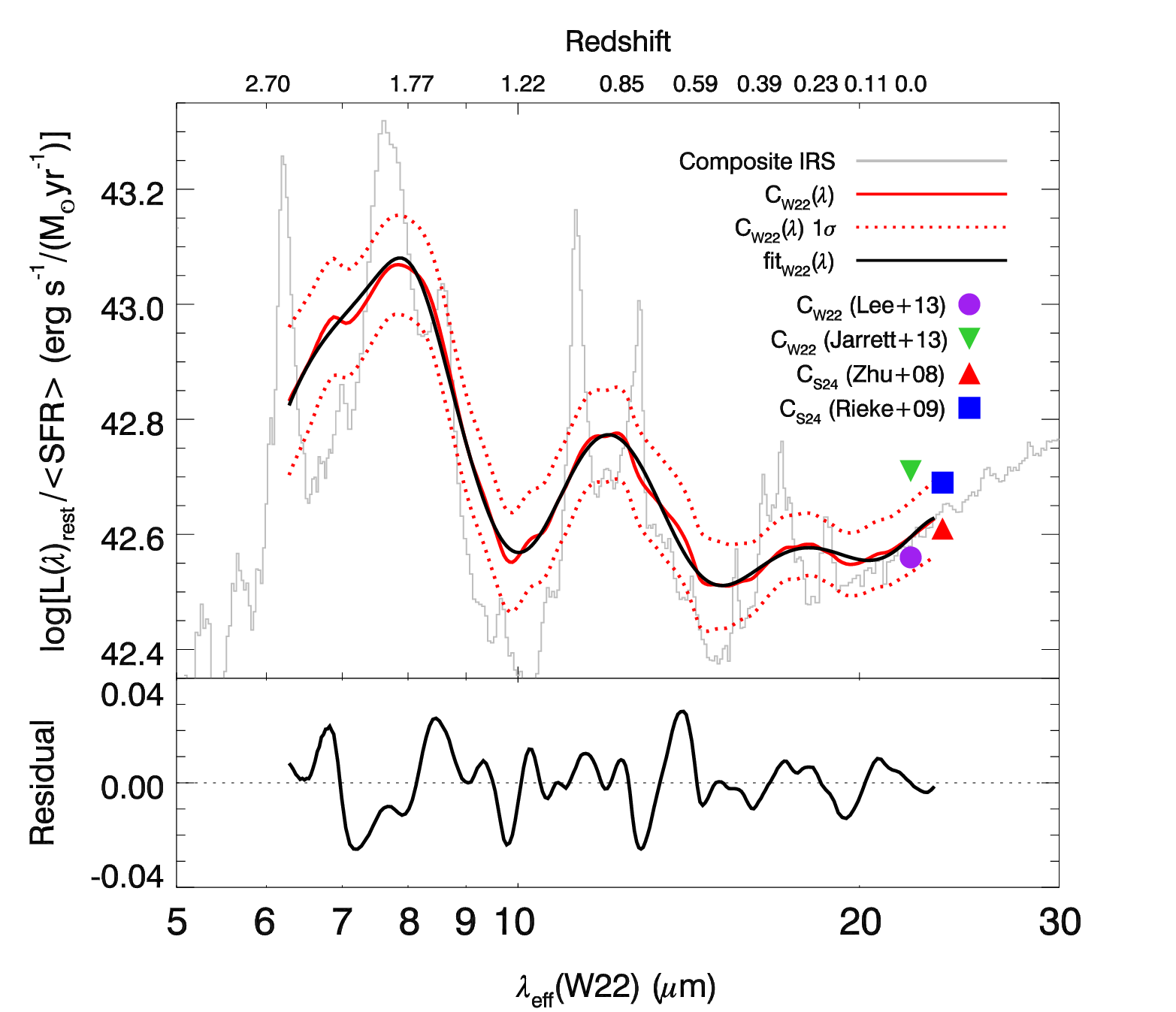}}
\end{array}$
\end{center}
\caption{The top of each panel shows the conversion factor for a \textit{Spitzer} or WISE band using all SFG galaxies (solid red lines), along with their uncertainty (dotted red lines), which for most cases is between $15-20\%$. The solid black line is a fit to the smoothed spectrum, $fit_{x}(\lambda)$. Local conversion factors from the literature are also shown for comparison (colored symbols). The region below $\sim$$6~\micron$ is excluded due to significantly increased uncertainty in the composite spectrum (see \S~\ref{filter_smooth}). The bottom of each panel shows the residuals between the conversion factor and a fit to the curve ($\log[C_{x}(\lambda)/fit_{x}(\lambda)]$). \label{fig:band_calibration_all_SFG}}
\end{figure*}

For the Drude profiles, the central wavelengths, $\lambda_r$, are fixed to wavelengths that roughly correspond to the peaks in the smoothed spectrum, while $\gamma_r$ and $b_r$ are left as free parameters. Therefore, there are up to 12 free parameters in total, two from the polynomial and ten from the Drude profiles. The values of $\lambda_r$ for each smoothed composite fit and all the other fit parameters are listed in Table~\ref{Tab:calibration_parameters}. The fits are shown for the individual bands in Figure~\ref{fig:band_calibration_all_SFG}. The fitting functions are typically accurate to within $\pm5\%$ (0.02 dex) of the true values and can be used in place of the templates [$C_x(\lambda) = fit_x(\lambda)$]. 

\subsection{Comparison to WISE SFR Calibrations}
The WISE All-Sky Survey \citep{wright10} provided photometry for over 563 million objects, and as such has great potential for future application of our calibrations. Recently, calibrations of the WISE bands as SFR indicators have emerged \citep{donoso12,shi12,jarrett13,lee13,cluver14}, some of which can be easily compared to our results. In particular, we focus on the results of \citet{jarrett13} and \citet{lee13} as these have linear calibrations of the WISE 12 and $22~\micron$ bands and can be directly compared to our values. The difference between the calibration values found in these works is rather large, corresponding to 0.34 dex ($\sim$120\%) and 0.15 dex ($\sim$40\%), for the WISE 12 and $22~\micron$ band, respectively. These large discrepancies are likely the result of the different approaches of the two works. \citet{jarrett13} rely of the previous calibrations of \citet{rieke09} at $24~\micron$, whereas \citet{lee13} attempt to determine SFRs from extinction-corrected H$\alpha$ emission. 

The composite of our sample of SFG spectra smoothed by the WISE 12 and $22~\micron$ filters is compared to these calibrations in Figure~\ref{fig:band_calibration_all_SFG}. We find that the results lie in-between the values found by \citet{jarrett13} and \citet{lee13}. Since the WISE $22~\micron$ band is so similar in shape and location to the \textit{Spitzer} $24~\micron$ band, the calibrations of \citet{zhu08} and \citet{rieke09} are also presented and show close agreement to our work.


\begin{sidewaystable}
\begin{center}
\caption{Continuous Star Formation Rate Calibration $fit_x(\lambda)$ Parameters \label{Tab:calibration_parameters}}
\begin{tabular}{cccccccccccccccccccc}
\hline\hline 
Band & $\lambda_{\rm{eff}}$ Range & $z$ limit & $p_1$ & $p_2$ & $\lambda_1$ & $b_1$ & $\gamma_1$ & $\lambda_2$ &  $b_2$ & $\gamma_2$ & $\lambda_3$ &  $b_3$ & $\gamma_3$ & $\lambda_4$ &  $b_4$ & $\gamma_4$ & $\lambda_5$ &  $b_5$ & $\gamma_5$\\
  & ($\micron$) & & & & ($\micron$) & & & ($\micron$) & & & ($\micron$) & & & ($\micron$) & & & ($\micron$) & & \\ \hline
$S8$ & $6.20-7.87$ & $0.27$ & -1.229e42 & \nodata & 7.1 & 1.075e43 & 0.349 & 8.2 & 3.102e42 & 0.224 & \nodata & \nodata & \nodata & \nodata & \nodata & \nodata & \nodata & \nodata & \nodata \\
$S24$ & $6.20-23.68$ & $2.82$ & -2.803e42 & \nodata & 7.1 & 1.107e43 & 0.413 & 8.2 & 4.496e42 & 0.165 & 12.0 & 5.300e42 & 0.341 & 17.9 & 4.154e42 & 0.576 & 25.0 & 5.093e42 & 0.366 \\ 
$S70$ & $17.85-71.42$& $3.00$ & -3.213e42 & 3.125e41 & 17.9 & 1.208e42 & 0.272 & \nodata & \nodata & \nodata & \nodata & \nodata & \nodata & \nodata & \nodata & \nodata & \nodata & \nodata & \nodata \\
$S70_{\rm{corr}}$ & $17.85-71.42$& $3.00$ & -1.929e42 & 2.825e41 & 35.0 & 4.065e42 & 0.461 & 47.0 & 2.029e42 & 0.732 & \nodata & \nodata & \nodata & \nodata & \nodata & \nodata & \nodata & \nodata & \nodata \\
$W12$ & $7.00-12.08$ & $0.73$ & -1.598e43 & 1.112e42 & 6.3 & 1.528e43 & 0.402 & 9.5 & 9.050e42 & 0.685 & \nodata & \nodata & \nodata & \nodata & \nodata & \nodata & \nodata & \nodata & \nodata \\
$W22$ & $6.28-22.19$ & $2.53$ & -1.791e43 & \nodata & 6.8 & 2.253e43 & 0.789 & 8.0 & 6.473e42 & 0.198 & 12.0 & 9.557e42 & 0.392 & 17.9 & 1.428e43 & 0.743 & 25.0 & 1.202e43 & 0.401 \\  
$F1000W$ & $6.48-10.00$ & $0.54$ & -2.463e42 & \nodata & 6.7 & 6.078e42 & 0.569 & 7.9 & 1.074e43 & 0.247 & 10.6 & 4.042e42 & 0.065 & \nodata & \nodata & \nodata & \nodata & \nodata & \nodata \\
$F1280W$ & $6.43-12.80$ & $0.99$ & -5.267e43 & 3.409e42 & 6.7 & 3.544e43 & 0.801 & 8.0 & 7.603e42 & 0.224 & 10.6 & 2.468e42 & 0.097 & 11.9 & 6.918e42 & 0.213 & \nodata & \nodata & \nodata \\
$F1500W$ & $6.37-15.00$ & $1.35$ & -2.105e43 & \nodata & 6.7 & 2.212e43 & 0.756 & 7.9 & 1.038e43 & 0.285 & 10.5 & 1.359e42 & 0.069 & 11.9 & 1.174e43 & 0.367 & 17.7 & 1.976e43 & 0.684 \\
$F1800W$ & $6.48-18.00$ & $1.78$ & -1.339e43 & 3.510e41 & 6.4 & 8.470e42 & 0.160 & 7.8 & 2.092e43 & 0.317 & 10.7 & 1.581e42 & 0.082 & 12.0 & 1.008e43 & 0.331 & 17.7 & 8.963e42 & 0.539 \\
$F2100W$ & $6.43-21.00$ & $2.27$ & -1.493e43 & 6.926e41 & 7.0 & 1.531e43 & 0.483 & 8.1 & 7.937e42 & 0.247 & 10.6 & 8.052e41 & 0.048 & 12.0 & 8.089e42 & 0.357 & 17.3 & 4.146e42 & 0.533 \\
$F2550W$ & $6.40-25.50$ & $2.99$ & -6.947e42 & 4.408e41 & 6.5 & 7.568e42 & 0.186 & 7.9 & 1.468e43 & 0.237 & 11.9 & 6.429e42 & 0.273 & 13.5 & 8.379e41 & 0.083 & 17.3 & 2.112e42 & 0.290 \\\hline
\end{tabular}
\end{center}
\textbf{Notes.} The second fit for $S70$ has a correction term that takes into account the FIR variation of SFG SEDs with redshift (see \S~\ref{C70z}). The functional form of these fits is $f_{\rm{IRS}}(\lambda) = \sum_{i=1}^2 p_i\lambda^{(i-1)} + \sum_{r=1}^{5} I_r(\lambda)$, where $I_r(\lambda) = \frac{b_r\gamma_r^2}{(\lambda/\lambda_r-\lambda_r/\lambda)^2+\gamma_r^2}$
\end{sidewaystable}

\section{Application to Higher Redshift Galaxies}
\subsection{Demonstration}
Here we demonstrate how to apply our calibrations to a SFG with a known redshift. Let us consider using the observed $24~\micron$ flux density for the galaxy SSGSS~1 to estimate the SFR of this galaxy. At $z=0.066$ for this particular galaxy, the $24~\micron$ band has an effective wavelength of $\lambda_{\rm{eff}}(24)=22.51~\micron$ and an observed flux density of $S_{obs}(24)=9.72\times10^{-26}$~erg~s$^{-1}$~cm$^{-2}$~Hz$^{-1}$, which corresponds to a luminosity of $\log[L_{\rm{obs}}(24)]=43.11$~erg~s$^{-1}$. Knowing the effective wavelength, we next want to use the composite $24~\micron$ band smoothed spectrum to determine the appropriate conversion factor, which is found to be $\log[C_{24}(22.51~\micron)]=42.62$~erg~s$^{-1}/(M_\odot \rm{yr}^{-1}$) using the smoothed composite template or the fitting function (see Figure~\ref{fig:band_calibration_all_SFG}). Finally making use of eq. (\ref{sfr}), we get that the SFR is simply the observed luminosity in this band divided by the conversion factor, $\mathrm{SFR}=10^{43.11}/10^{42.62}=3.09~M_\odot \rm{yr}^{-1}$. This value differs from the actual value of $\langle SFR \rangle=4.22~M_\odot \rm{yr}^{-1}$ by about $30\%$. 

Next we can consider the slightly more distant case of SSGSS 14, at $z=0.153$, and determine the SFR from its observed $24~\micron$ flux density of $S_{obs}(24)=4.44\times10^{-26}$~erg~s$^{-1}$~cm$^{-2}$~Hz$^{-1}$, which corresponds to a luminosity of $\log[L_{\rm{obs}}(24)]=43.55$~erg~s$^{-1}$. The $24~\micron$ band has an effective wavelength of $\lambda_{\rm{eff}}(24)=20.53~\micron$, which corresponds to $\log[C_{24}(20.53~\micron)]=42.57$~erg~s$^{-1}/(M_\odot \rm{yr}^{-1}$). Taking the ratio of these numbers gives $\mathrm{SFR}=9.63~M_\odot \rm{yr}^{-1}$. This value differs from the actual value of $\langle SFR \rangle=9.60~M_\odot \rm{yr}^{-1}$ by $<1\%$.

In the same manner, each calibration can be applied to any redshift that spans the $\lambda_{\rm{eff}}$ range covered by IRS. These examples highlight the importance of using large sample sizes in the application of these diagnostics, as a single case can have SED variations relative to the mean of our sample, which can give rise to slight inaccuracies in SFR estimates. It is important to emphasize that the accuracy of such an application is dependent on the shape of the SED of SFGs as a function of redshift. The extent to which this condition holds is examined in detail in \S~\ref{template_variation}.

\subsection{Limitations of this Sample} \label{limitations}
It is important to acknowledge the potential differences of this sample with respect to high-$z$ galaxies as well as the limitations for its use. As was mentioned, the selection criteria for the SSGSS sample limits it to relatively high metallicities, which may not be a well representative sample as one goes to high-$z$. In addition, if the dust content of high-$z$ galaxies is different, it is possible that the amount of UV light reprocessed by dust could change. For example, if high-$z$ galaxies had more dust, then our templates would overestimate the SFR, as it would be implicitly adding back in unobscured UV flux present in the SSGSS sample but that may not be there for the high-$z$ galaxies. Variations in the typical dust temperature of galaxies with redshift would also pose a problem, as this would result in variations in their FIR SED. The relative importance of some of these effects will be tested when we compare our SED to those at higher redshift (\S~\ref{template_variation}).   

Another area for concern is in the range of \lir\ values spanned by the SSGSS sample. Our template is made utilizing galaxies over a range of $9.53\le \log(L_{\rm{IR}}/L_{\odot})\le 11.37$, which is lower than the range that is currently accessible at high-$z$. However, the results of \citet{elbaz11} suggest that luminous infrared galaxies (LIRGs; $L_{\rm{IR}}>10^{11}L_{\odot}$) and ultra luminous infrared galaxies (ULIRGs; $L_{\rm{IR}}>10^{12}L_{\odot}$) identified in the GOODS-\textit{Herschel} sample at high-$z$ have similar SEDs to normal SFGs, in contrast to their starburst-like counterparts found locally \citep[e.g.,][]{rieke09}. They find that the entire population of IR-bright galaxies has a distribution with a median of $\mathrm{IR8}=L_{\rm{IR}}/L_{\rm{rest}}(8\mu\mathrm{m})=4.9~[-2.2,+2.9]$, where the term in brackets is the $1\sigma$ dispersion. Elbaz et al. suggest that this population can be separated into two groups: a main-sequence (MS) of normal SFGs for which $\mathrm{IR8}=4\pm2$ consisting of $\sim$80\% of the population, and starburst (SB) galaxies which occupy the region with $\mathrm{IR8}>8$ and represent about $\sim$20\% of the population. For reference, the IR8 value of our template is 4.8, which agrees with the median of the GOODS-\textit{Herschel} sample. The uniformity of IR8 values in MS galaxies, over the range $10^9<L_{\rm{IR}}/L_{\odot}<10^{13}$, suggests that the SED of normal SFGs do not change drastically with luminosity. This also indicates that our limited range in \lir\ coverage for the SSGSS sample should not drastically affect its utility towards higher luminosity MS galaxies.

Perhaps the biggest factor limiting the large scale application of this technique is in the ability to identify galaxy types at higher redshifts. The calibrations presented in this work are applicable to normal SFGs, typically referred to as being on the main-sequence of star formation, and not to cases undergoing starburst activity (different SED) or with AGN (significant IR emission not associated with star formation). This topic should be thoroughly addressed before widespread applications of these calibrations can be made to specific surveys.  

There are a several techniques that have been demonstrated to isolate out AGN and starburst galaxies, however, some of them rely on observations made outside the MIR. One of the most reliable techniques to identify AGN is thorough X-ray observations \citep[e.g.,][]{alexander03}, however these can miss obscured AGN and could be biased \citep{brandt&hasinger05}. In order to avoid obscuration effects, AGN selection techniques using the MIR and FIR have also been developed. These include \textit{Spitzer}+\textit{Herschel} color-cuts \citep{kirkpatrick12}, \textit{Spitzer}/IRAC color-cuts \citep{lacy04,stern05,donley12,kirkpatrick12}, and WISE color-cuts \citep{stern12,mateos12,assef13}. Emission line diagnostics, such as the BPT diagram \citep{kewley13} and the Mass-Excitation diagram \citep{juneau14}, are also effective techniques. It has been suggested that starburst galaxies can be identified as sources with $\mathrm{IR8}>8$ by \citet{elbaz11}.

\section{Discussion}
\subsection{Origins of the Scatter in SFR(\texorpdfstring{$\lambda$}{lambda})} \label{scatter}
In addition to grouping all of the SFGs together, we also examine grouping our galaxies based on their luminosity at rest-frame 3.6, 4.5, 8, 24, and $70~\micron$, as well as their \lir, \lir\ surface brightness, and $L($H$\alpha$)/$L(24~\micron$) ratios, in order to identify possible origins to the scatter within the SFR calibration of the entire group. These bands are chosen because $3.6~\micron$ and $4.5~\micron$ correlate with the underlying stellar population \citep[i.e., stellar mass;][]{meidt12,meidt14}, and the other bands correlate strongly with star formation \citep{zhu08,rieke09,calzetti10,hao11}. For each of these cases, the sample is divided into 6 bins with $9-10$ galaxies in each.

\begin{figure*}
\begin{center}$
\begin{array}{cc}
\includegraphics[width=3.5in]{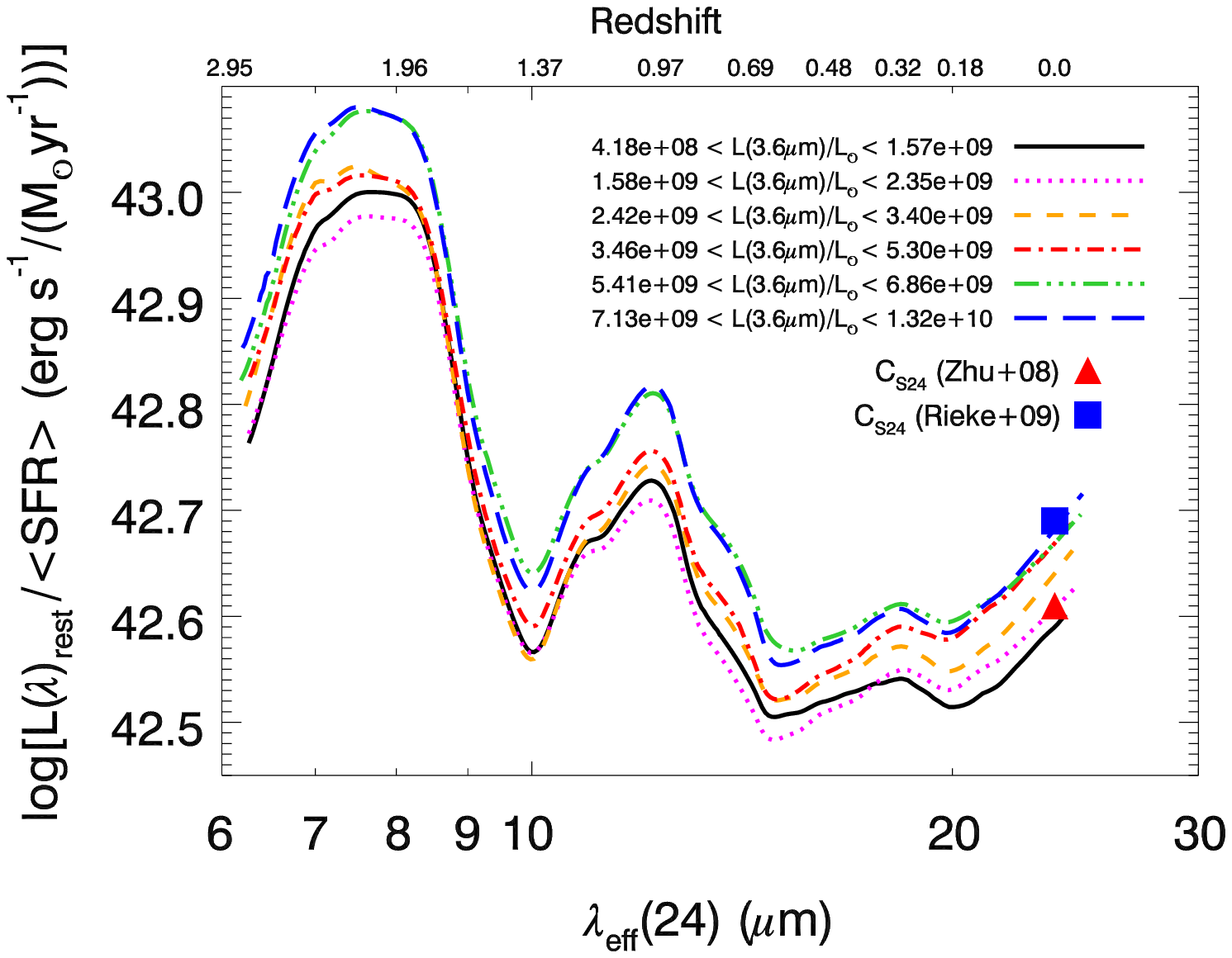} &
\includegraphics[width=3.5in]{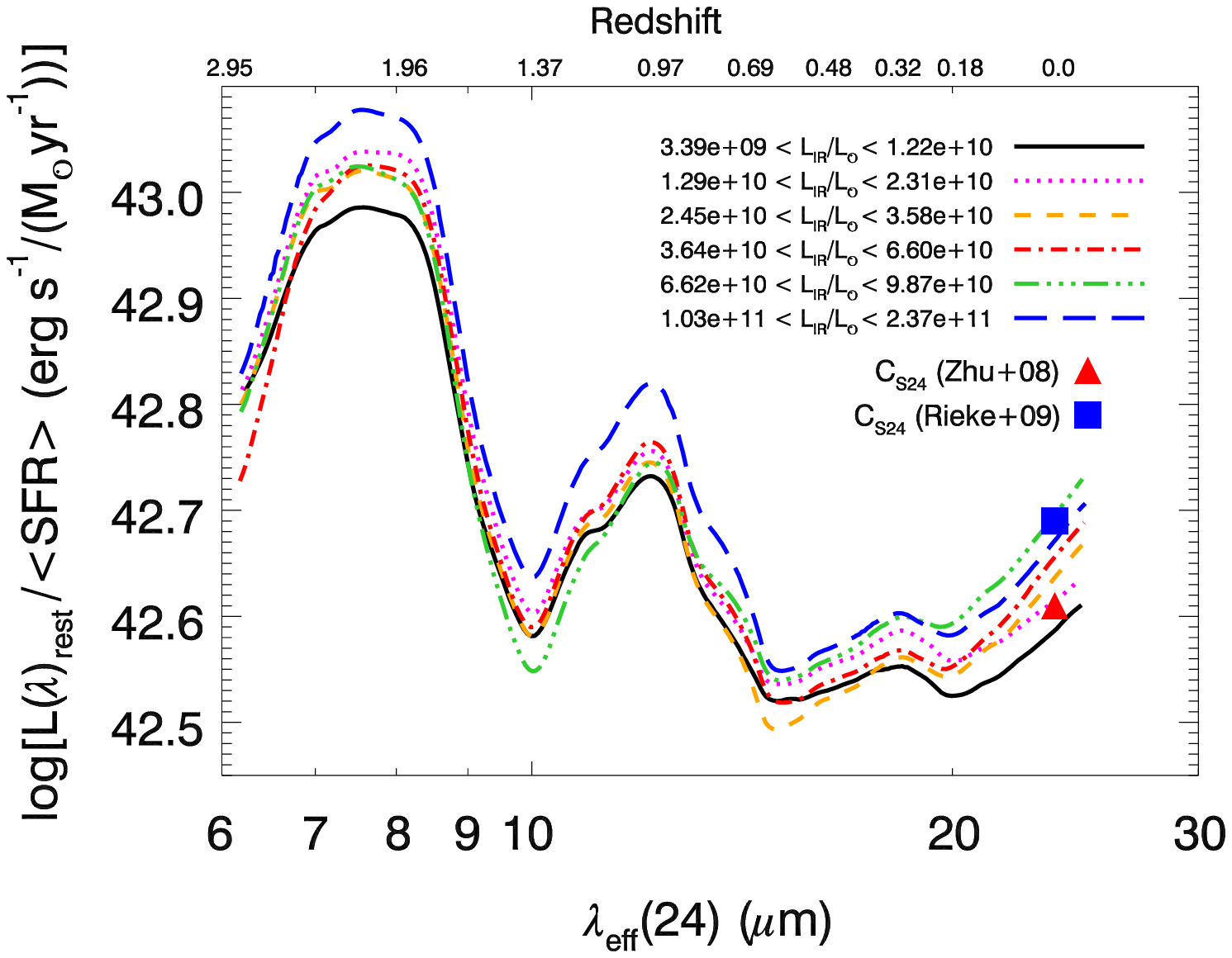} \\
\includegraphics[width=3.5in]{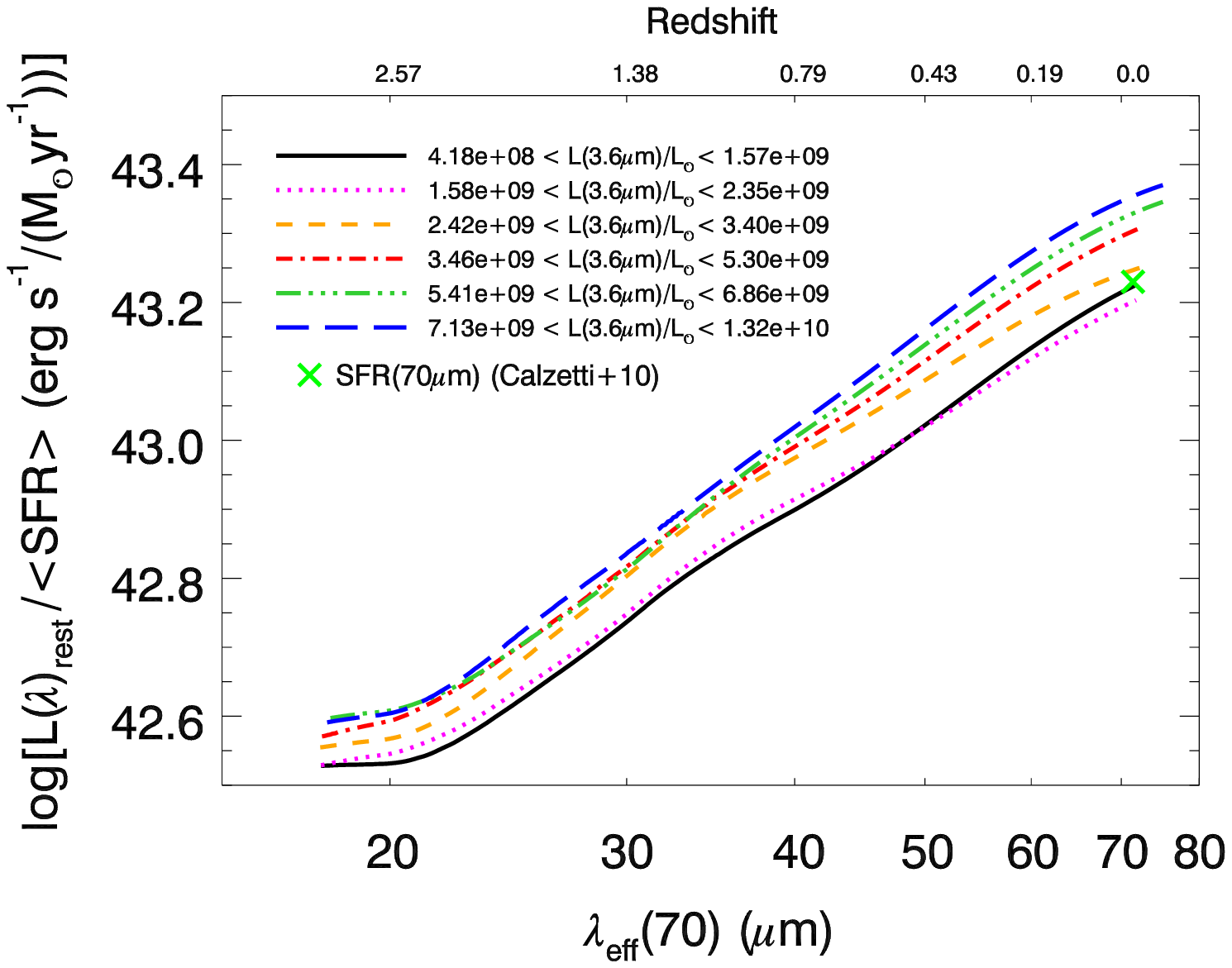} &
\includegraphics[width=3.5in]{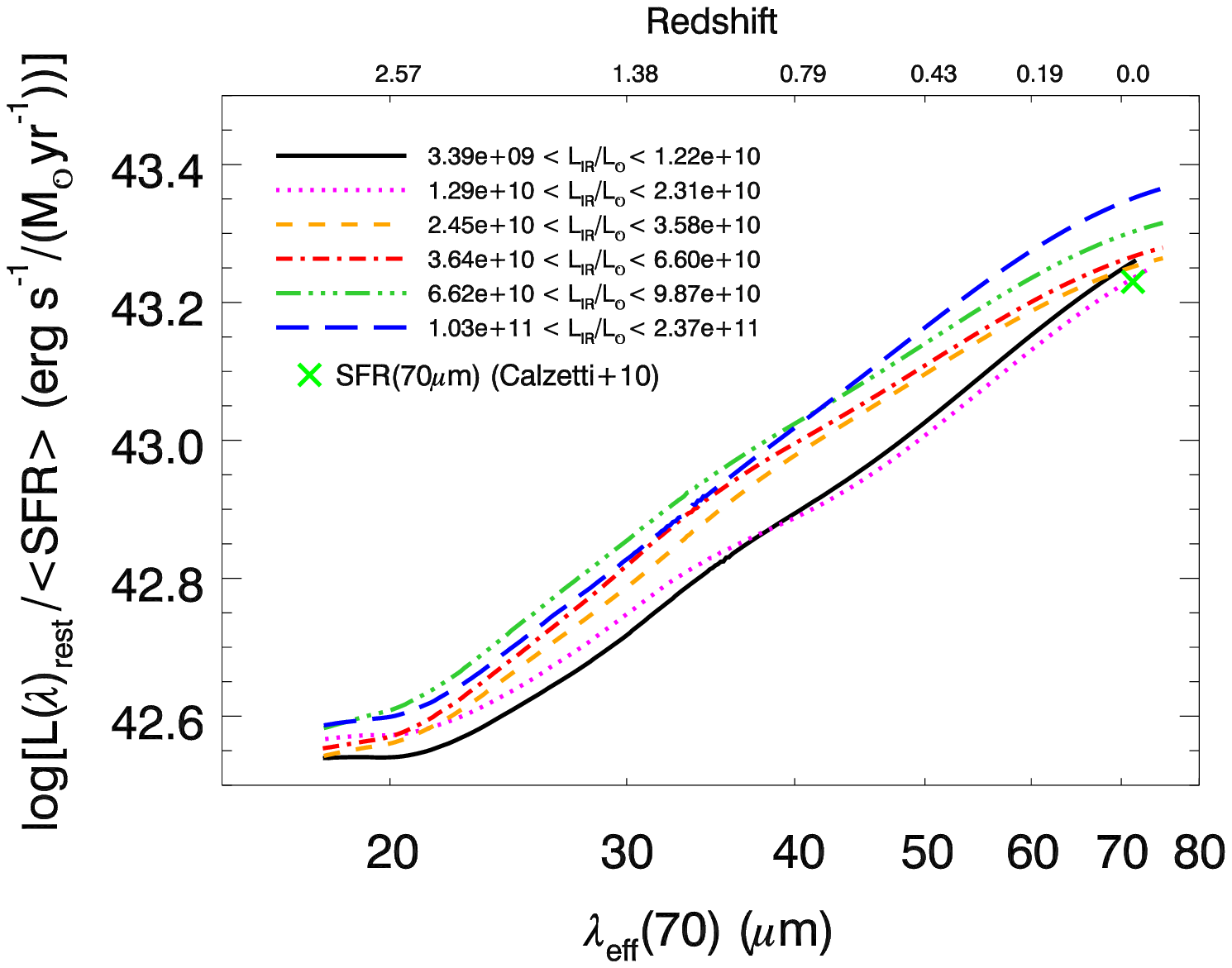}
\end{array}$
\end{center}
\caption{Top: $C_{24}(\lambda)$ conversion factor for galaxies when arranged into groups ($\sim$10 galaxies) according to $L(3.6\mu\rm{m})$, a proxy for stellar mass, and \lir, a proxy for the SFR. Bottom: $C_{70}(\lambda)$ conversion factor for galaxies when arranged according to $L(4.5\mu\rm{m})$ and \lir. Local conversion factors are also shown for comparison. The dispersion in each of the groups ($10-25\%$; see \S~\ref{scatter}) is not shown for clarity, but is comparable to the separation among the groups. Weak trends appear which would suggest larger conversion factors are needed for the higher luminosity galaxies. \label{fig:group_calibrations}}
\end{figure*}

Looking at each of the calibrations, weak trends are found suggesting larger conversion factors, at almost all MIR wavelengths, for galaxies with higher luminosities when arranged by any of the luminosities mentioned before. A few examples are shown in Figure~\ref{fig:group_calibrations}. These trends are very weak because the separation between the groups is comparable to the scatter within each of the groups, which is $10-25\%$ (the largest scatter at lowest luminosity galaxies), and similar to the uncertainty of the entire group average values. If real, these trends could suggest that (1) galaxies with a larger old stellar population require slightly larger conversion factors at all MIR wavelengths, as might be expected if light unassociated to star formation is contaminating the MIR; and/or (2) a slightly super-linear relationship exists between luminosity and SFRs in the MIR.

\begin{figure}
\plotone{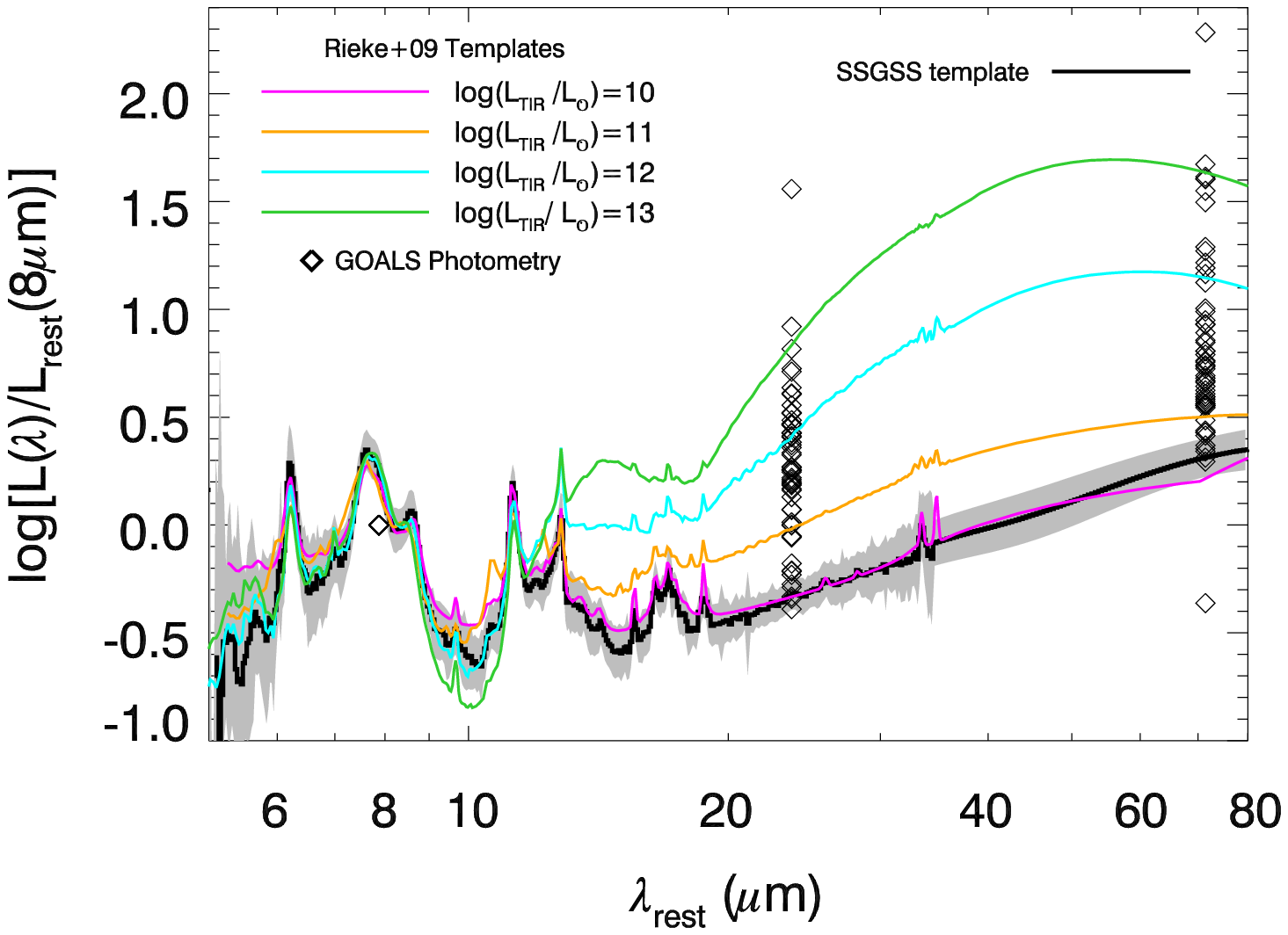}
\caption{Comparison of our SFG composite template to GOALS photometry \citep{u12} and \citet{rieke09} templates. The scatter of the SSGSS template is shown as the filled gray region. The values of $\mathrm{IR8}=L_{\rm{IR}}/L_{\rm{rest}}(8\mu\mathrm{m})$ for the \citet{rieke09} templates range from $\mathrm{IR8}=4.8$ at $L_{\rm{TIR}}=10^{10}L_{\odot}$ to $\mathrm{IR8}=59.6$ at $L_{\rm{TIR}}=10^{13}L_{\odot}$, whereas high-$z$ galaxies over this luminosity range have $\mathrm{IR8}=4.9~[-2.2,+2.9]$ \citep{elbaz11}. Therefore, there is significant FIR SED evolution that occurs for local (U)LIRGS that is absent at high-$z$.  \label{fig:GOALS_compare}}
\end{figure}

Our attempts to account these effects by introducing additional terms into the conversion factors do not appear to significantly reduce the overall scatter of the conversion factors. Such additional terms would also make application of this method to higher redshift galaxies more difficult, as more information would be needed (e.g., determination of rest-frame luminosities). Given our limited range in galaxy properties to determine the validity of any trends, we adopt the simplest approach and use the entire group average for our analysis. We refrain from using higher luminosity local galaxies, such as the (U)LIRGs in the GOALS sample \citep{armus09}, as an additional test of such claims because of the significant FIR SED evolution that occurs for $L_{\rm{IR}}\gtrsim10^{11}L_{\odot}$, which is absent from galaxies of these luminosities at high-$z$ \citep[see \S~\ref{limitations};][]{elbaz11}. In addition, a large fraction of these systems are likely to host AGN \citep{u12}, which would be excluded by our selection process. For reference to the reader, we illustrate the local SED evolution by comparing the GOALS photometry \citep{u12} and a few of the \citet{rieke09} templates to our own template, normalized by $L_{\rm{rest}}(8\micron)$, in Figure~\ref{fig:GOALS_compare}.

\subsection{Variation in SFG SEDs with Redshift} \label{template_variation}
Many studies have sought to characterize the SED of different galaxy types (e.g., SFG, AGN) as functions of redshift. In this section, the templates of \citet{elbaz11,kirkpatrick12,magdis12} and \citet{ciesla14} are compared to our own to thoroughly examine the extent to which the SED of SFGs change with redshift. Similar to the approach outlined in this work, these studies use large surveys to construct IR templates for different populations of galaxies at different redshifts. In general, the templates created in these studies suggest that the mean dust temperature of galaxies increases as one looks to higher redshifts. In addition to the change in dust temperature that is evident, \citet{magdis12} suggest that the value of IR8 increases mildly from $\mathrm{IR8}\sim4$ to $\mathrm{IR8}\sim6$ at $z>2$ for MS galaxies. 

If one considers the notion that both $L_{\rm{rest}}(8\micron$) and \lir\ are typically used for SFR indicators, such a change in IR8 would suggest that there is a change in the SFR converstion factor of one (or both) of these luminosities with redshift and this is important to keep in mind when comparing the templates. Changes in $L_{\rm{rest}}(8\micron$) could result from variations in PAH abundances relative to the total dust content, which has been found to correlate with metallicity \citep{engelbracht05,engelbracht08,marble10}, and also to the hardness of the radiation field \citep{madden06,gordon08,engelbracht08}. Given the sensitivity of \lir\ to the contributions from older stellar populations \citep{calzetti10}, it is also likely that the value of the conversion factor for SFR-\lir\ could also vary with redshift. 

\begin{figure*}
\begin{center}$
\begin{array}{cc}
\includegraphics[width=3.5in]{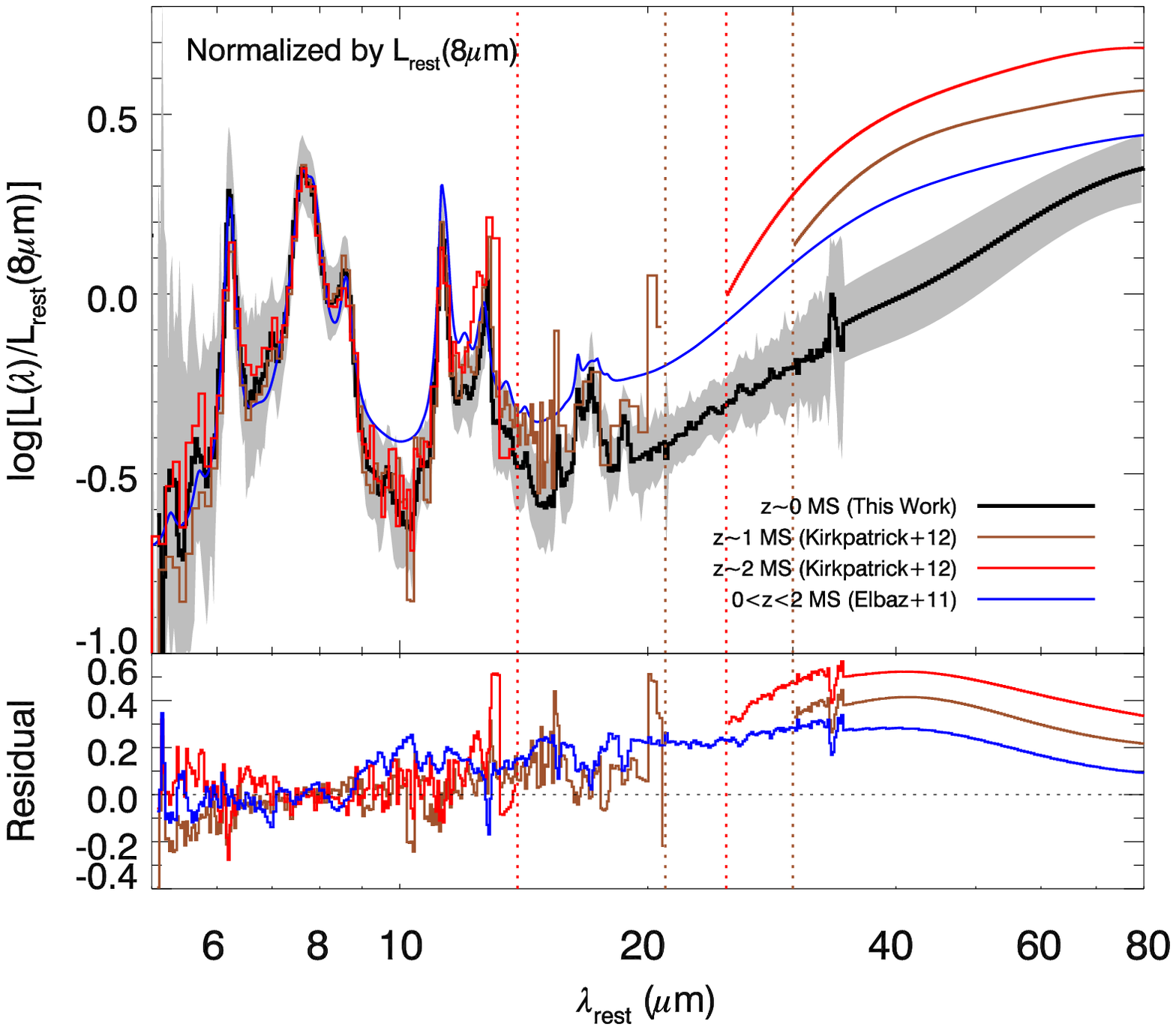} &
\includegraphics[width=3.5in]{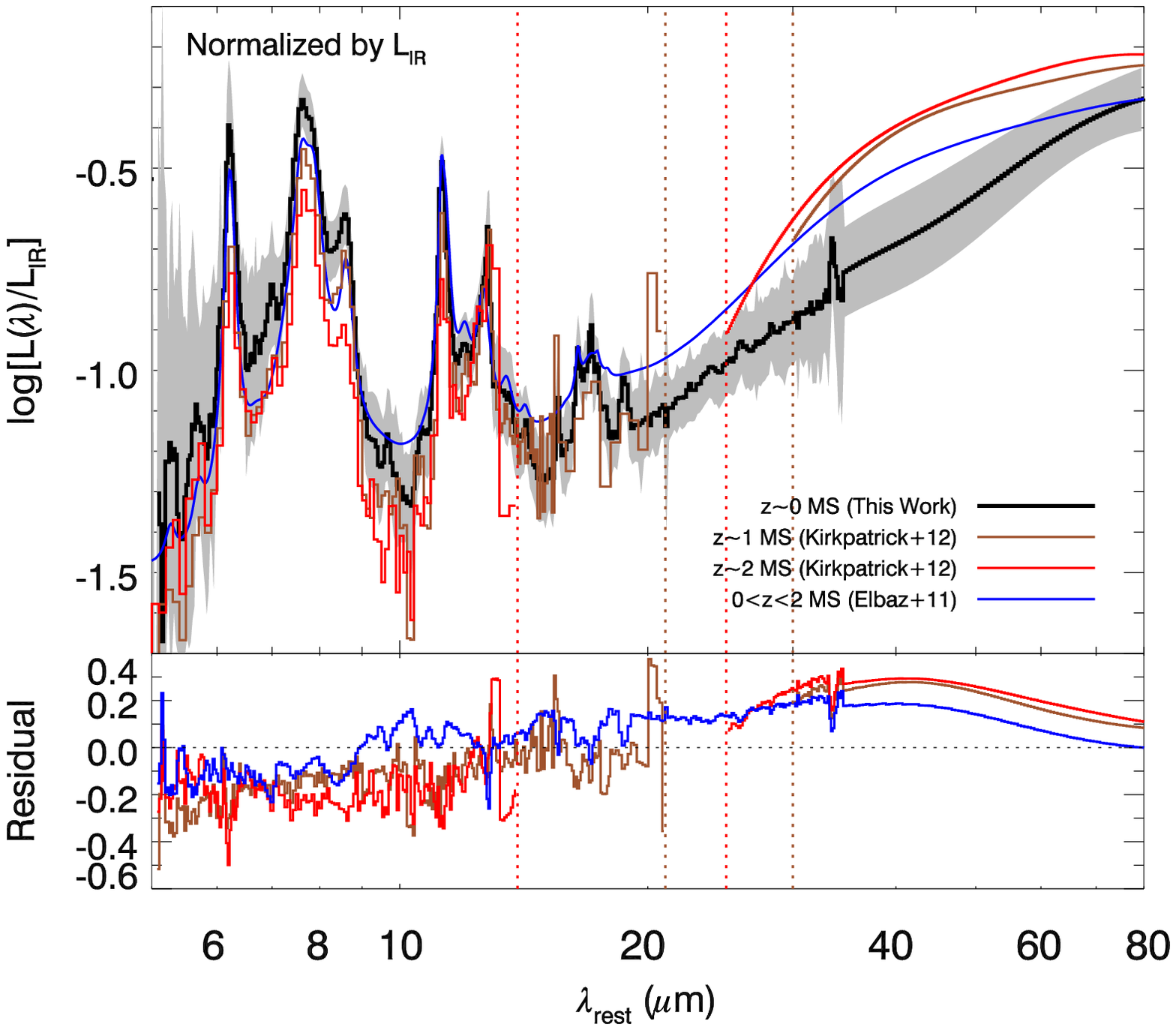} 
\end{array}$
\end{center}
\caption{The top of each panel shows the comparison of the composite SED of SFGs in our sample to those at higher redshifts for which spectroscopic information is available. The scatter of the SSGSS template is shown as the filled gray region. The SEDs have been normalized by $L_{\rm{rest}}(8\micron$) and \lir. The sections between the vertical dotted green and red lines, corresponding to the $z\sim1$ and $z\sim2$ templates from \citet{kirkpatrick12}, lack spectral data. The template of \citet{elbaz11} uses redshifted photometry to act as a spectroscopic analog. The bottom of each panel shows the residuals between our template and the other templates. The shape of the SED remains unchanged with redshift for $\lambda\lesssim20~\micron$, with only constant offsets occurring depending on the normalization. For $\lambda\gtrsim20~\micron$, significant SED evolution is present with increasing redshift. \label{fig:template_compare}}
\end{figure*}

\begin{figure*}
\begin{center}$
\begin{array}{cc}
\includegraphics[width=3.5in]{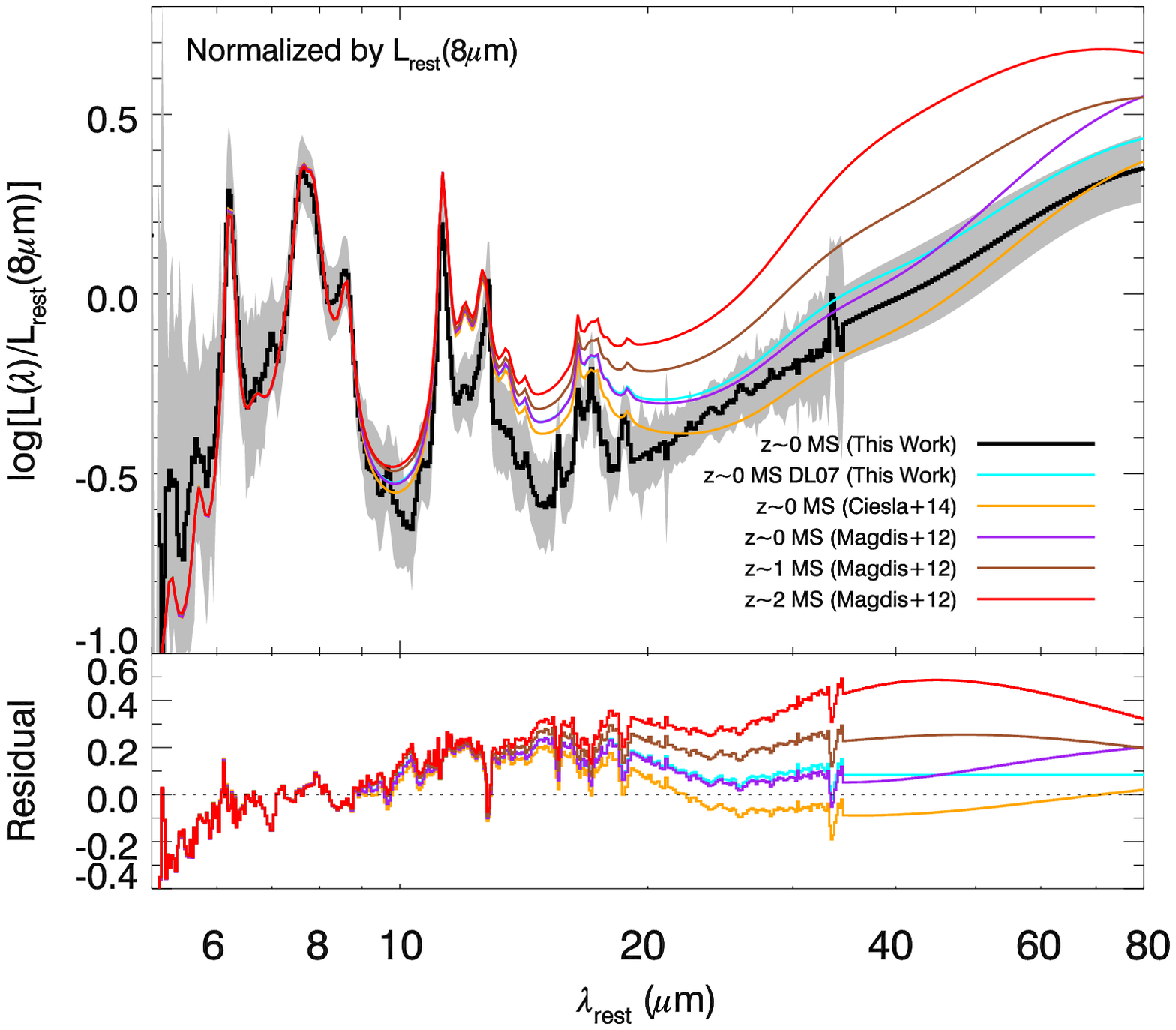} &
\includegraphics[width=3.5in]{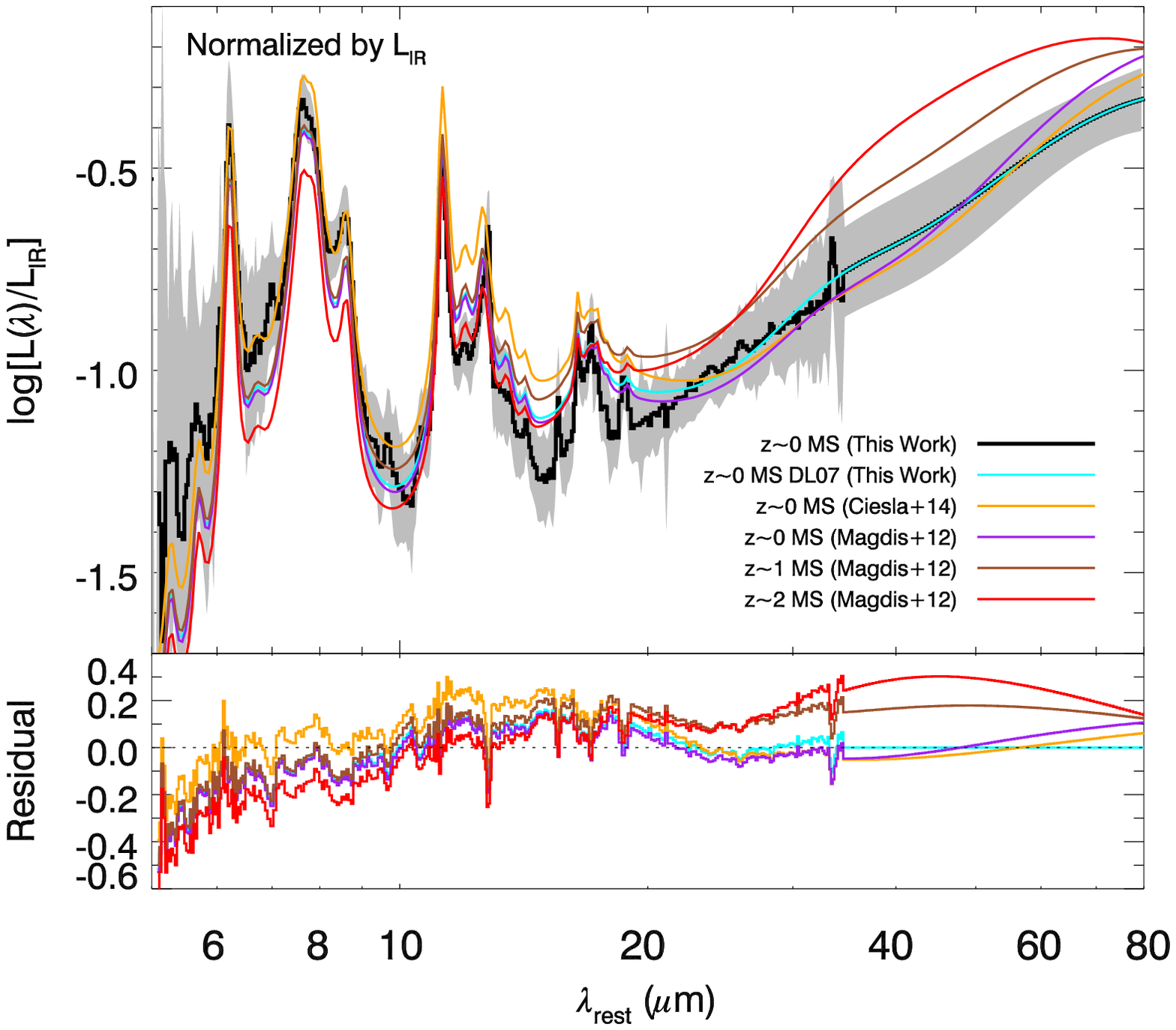} 
\end{array}$
\end{center}
\caption{The top of each panel shows the comparison of the composite SED of SFGs in the SSGSS sample to those at higher redshifts for which \citet{draine&li07} models have been used to fit the available photometry. The scatter of the SSGSS template is shown as the filled gray region. The SEDs have been normalized by $L_{\rm{rest}}(8\micron$) and \lir. Note that when normalized by $L_{\rm{rest}}(8\micron$) these model-based template have an excess in the $10-25~\micron$ region compared to the spectral data. This trend is seen in our own fits of \citet{draine&li07} models (cyan line), and indicates a limitation in the simple 3-component model typically adopted (see \S~\ref{DL07_method}). The bottom of each panel shows the residuals between our template and the other templates. \label{fig:template_compare_model}}
\end{figure*}

The comparison between the templates from the literature to our own is shown in Figures~\ref{fig:template_compare} and \ref{fig:template_compare_model}. We have chosen to normalize the templates in two ways, both of which correlate with star formation. Normalizing by a close proxy for star formation is crucial to compare how viable our continuous calibrations are at higher redshifts. The first method is to normalize by $L_{\rm{rest}}(8\micron$), which is chosen over use of the $24~\micron$ region because it is not available for the \citet{kirkpatrick12} templates. With this choice of normalization, it is also easier to directly compare the shape of the MIR SEDs. The second method is to normalize by \lir, as is traditionally done in many template comparisons. We reemphasize that the observed trend of IR8 increasing from 4 to 6 implies that these choices of normalization for the templates will give different results.

First, the templates of \citet{kirkpatrick12} are examined as these provide the best sample for comparison because they are based on direct spectral measurements of higher redshift galaxies. In addition, access to spectral data allowed them to accurately identify galaxies with significant AGN contribution and create separate templates for AGN and SFGs, the latter of which is considered here. The gap in spectral coverage of their templates, shown as the vertical dotted lines in Figure~\ref{fig:template_compare}, correspond to regions lacking spectral or photometric values with which to constrain the SED, and is ignored for our comparison. Looking at the templates normalized by $L_{\rm{rest}}(8\micron$), it is seen that the templates show remarkable agreement in SED shape for $\lambda<24~\micron$ and lie almost entirely within the scatter in our local SED template.  In contrast, there is clear disagreement in SED shape at $\lambda>24~\micron$ which becomes more drastic at higher redshift. This is mostly due to the larger IR8 values of these templates, which exceed the IR8 values observed in photometric samples at these redshifts \citep{magdis12}. For reference, the \citet{kirkpatrick12} $z\sim1$ template has IR$8=6.5$ and the $z\sim2$ template has IR$8=8.0$. We associate this difference to the selection criteria of this sample, which required bright sources at $24~\micron$ ($S_{24}>100~\mu$Jy) to obtain IRS spectroscopy and which corresponds to more \lir\ luminous galaxies at higher redshifts (see \S\ref{Cx_test} for more details). When instead normalized by \lir, slight offsets appear between the templates for $\lambda<24~\micron$ as a result of the larger \lir\ with increasing redshift. The fact that the shape remains fixed, regardless of possible offsets, gives credibility to this technique being applicable to up to $z\sim2$ for all bands at effective wavelengths below $24~\micron$. In contrast, even when normalized by \lir\ there is clear disagreement in SED shape at $\lambda>24~\micron$ which becomes more drastic at higher redshift. This effect, also observed by \citet{magdis12}, is argued to be due to the mean dust temperature of galaxies increasing with redshifts. 

There are multiple physical mechanisms that give rise to increased dust temperatures in galaxies. Locally, similar trends are seen in galaxies with increasing values of \lir\ \citep[e.g.,][]{rieke09}. However, for local LIRGs and ULIRGs there is also an associated decrease in the relative strength of the $8~\micron$ PAH feature relative to the FIR, corresponding to IR8 values more similar to SB galaxies, which deviates significantly from our template. Instead, \citet{magdis12} suggest that this could be the result of a hardening of the radiation field, $\langle U \rangle$, in MS galaxies with increasing redshift ($\langle U \rangle\propto(1+z)^{1.15}$). Adopting \citet{draine&li07} models to fit their galaxy SEDs, for which $\langle U \rangle\propto L_{\rm{IR}}/M_{\rm{dust}}$, they argue that this is explained by the redshift evolution of the $M_*-Z$ and $\mathrm{SFR}-M_*$ relations. Another physical mechanism that gives rise to this effect is compactness. More compact star formation in galaxies can give rise to elevated dust temperatures and appears to occur more frequently in MS galaxies at higher redshifts \citep{elbaz11,schreiber14}. Regardless of the origin of this effect, these results suggest that the $70~\micron$ band requires additional correction to be utilized as a SFR diagnostic as a function of redshift. We perform this analysis in the next section. 

Next, the templates of \citet{elbaz11} are considered. These templates make use of redshifted photometry of galaxies from $0<z<2$ to act as spectroscopic analog. The combination of all galaxies over this redshift range of results in an artificially broad FIR bump, due to the shifting of the FIR bump with $z$, and makes direct comparison of these templates tricky. In general there is good agreement in SED shape with their MS template and our own if this FIR broadening is taken into account.

Lastly, the templates based on \citet{draine&li07} model fitting of photometric data are considered, shown in Figure~\ref{fig:template_compare_model}. These include the templates of \citet{magdis12}, and \citet{ciesla14}. Considering first the $z\sim0$ cases normalized by $L_{\rm{rest}}(8\micron$), we note that all of these model-based templates show the same excess in the $10-25~\micron$ region compared to the spectral data that was seen in our own fits using \citet{draine&li07} models (cyan line in the Figure~\ref{fig:template_compare_model}). This suggests that these model-based templates may not be accurately representing the intrinsic SED over this region. The region beyond $25~\micron$ is likely to be more representative, as it is usually well fit a simple two component dust model. As with the spectral-based templates, the FIR bump peaks at shorter wavelengths with increasing redshift and also shows an increase in IR8. 
 
Taken together, it would appear that there is no strong evidence to suggest that the shape of the SED for SFGs varies significantly over the wavelength region of $6-30~\micron$. However, vertical offsets, corresponding to a constant factor offset, cannot be ruled out without direct comparison of SFR estimates for higher redshift galaxies.
 
 \subsection{Accounting for Dust Temperature Variation} \label{C70z}
The significant change in shape of the FIR bump makes the calibrations at the longer wavelengths (i.e., $C_{70}(\lambda)$) more difficult. Comparing the SED of the higher-$z$ galaxies at these wavelengths to the SSGSS SED, there is a significant difference (up to $0.5$~dex), which exceeds the scatter of SEDs for local SFGs. For this reason, a correction to $C_{70}(\lambda)$ does appear necessary if it is to be applied at higher redshifts. 

\begin{figure}
\plotone{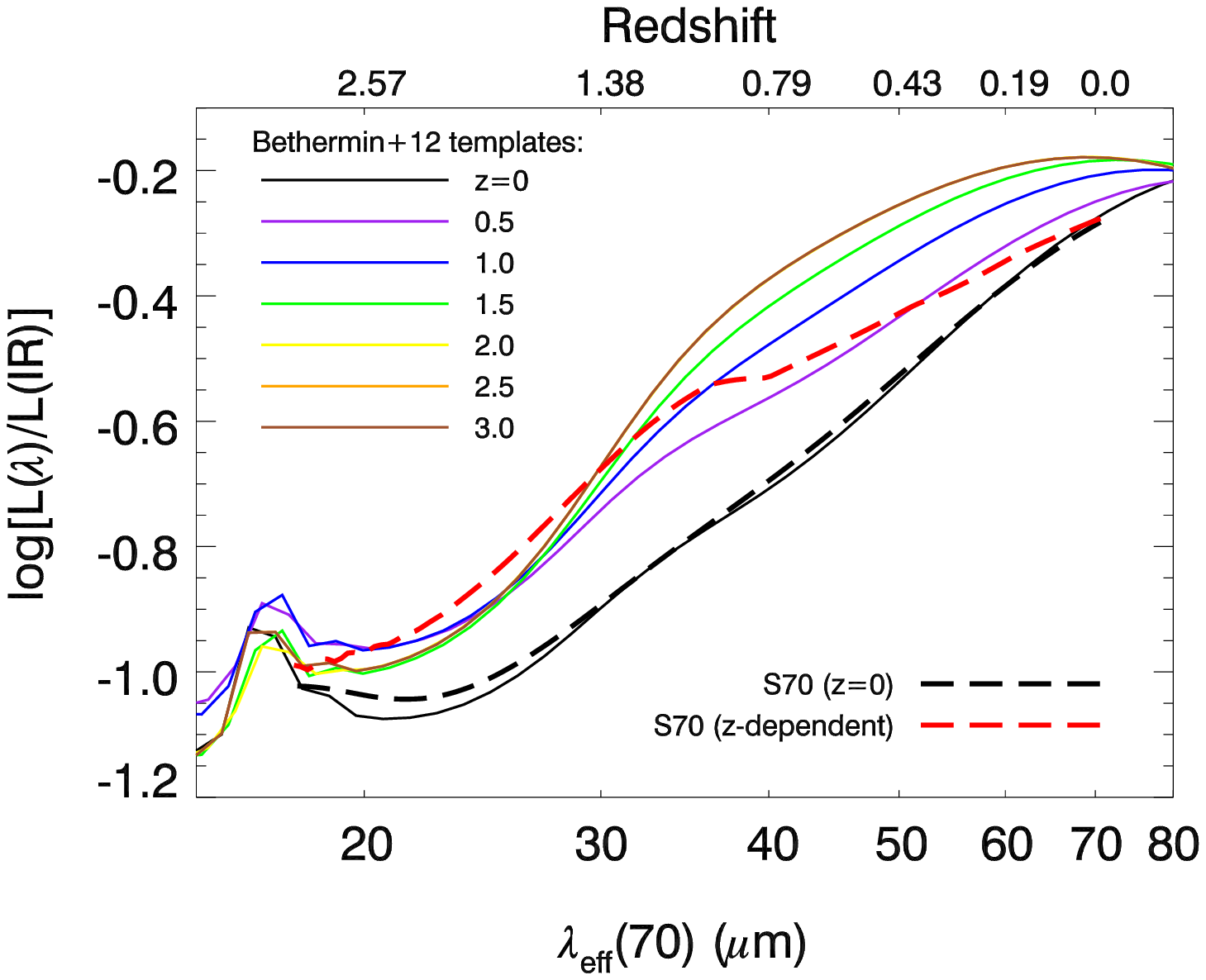}
\caption{The SED of SFGs changes with redshift, as demonstrated by the templates of \citet{bethermin12}. Taking this into account, the value of $C_{70}(\lambda)$ would change significantly from those derived from a $z\sim0$ SED (dashed black line). This change is demonstrated as the red dashed line, which shows the 70~$\micron$ filter convolution when accounting for the SED variations with redshift. We stress that the red dashed line is a z-dependent interpretation of the expected 70~\micron\ emission and cannot be considered a spectrum for an individual galaxy. \label{fig:C70_z}}
\end{figure}

We correct for the dust temperature variation using the the SED template grids of \citet{bethermin12}, which are built from the results of \citet{magdis12}. These templates have been normalized by \lir. By making use of this grid, the observed \textit{Spitzer} $70~\micron$ luminosity as function of redshift is estimated while accounting for the changing SEDs. A demonstration of how the observed luminosity changes is shown in Figure~\ref{fig:C70_z}. For example, at $z=0.5$ and $z=1$ the $70~\micron$ band measures rest-frame $46.7~\micron$ and $35.7~\micron$, respectively, and the observed band luminosity is derived from the $z=0.5$ (purple line) and $z=1$ (blue line) templates at those wavelengths. We perform a fit to this new conversion factor and present it in Table~\ref{Tab:calibration_parameters}. The accuracy of this correction will be tested in the following section.

\subsection{Testing the Calibrations} \label{Cx_test}
To test the utility of our calibrations, we compare SFRs of galaxies from other surveys to those found using our continuous, monochromatic values developed in this work. This requires a survey which has photometry available in one of the calibrated bands, as well as an independent technique to measure star formation from the those used to calibrate our conversion factors.  We choose to use SFRs based on \lir\ measurements as these are the most readily available diagnostic for deep IR surveys. For consistency with our local SSGSS sample, we adopt a conversion factor of $\log[C(L_{\mathrm{IR}})]=43.64$~erg~s$^{-1}/(M_\odot \rm{yr}^{-1}$), which corresponded to a $\tau\sim500$~Myr constant star formation (see \S~\ref{lir}). Furthermore, sources with significant AGN components need to be to identified and removed. It is worth noting that adopting different \lir\ conversion factors for this analysis will only lead a constant offset between these two SFRs at all redshifts and that we are most interested in assessing where breaks from a constant relation develop. 

\begin{figure*}
\begin{center}$
\begin{array}{cc}
\includegraphics[width=3.5in]{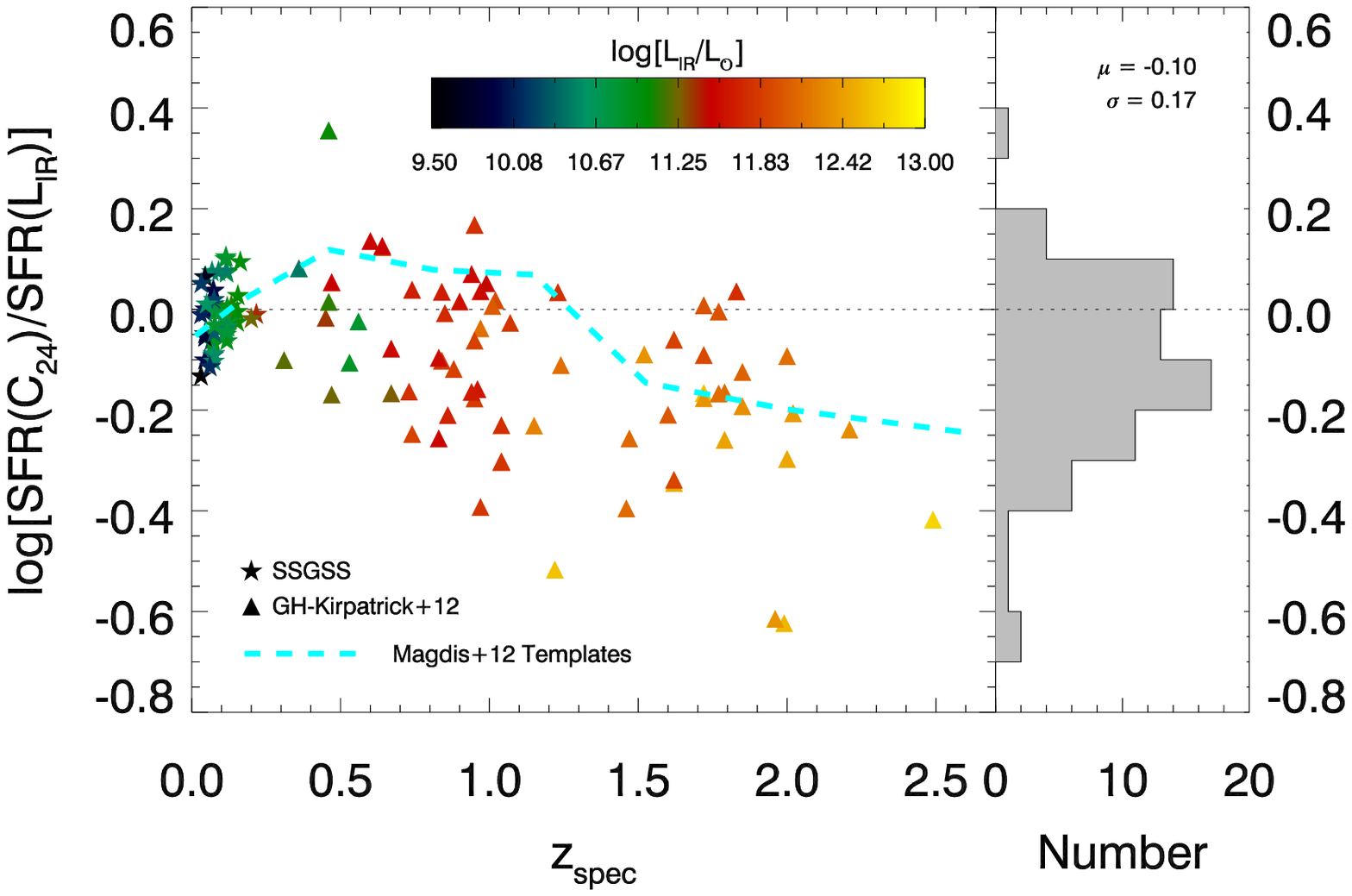} &
\includegraphics[width=3.5in]{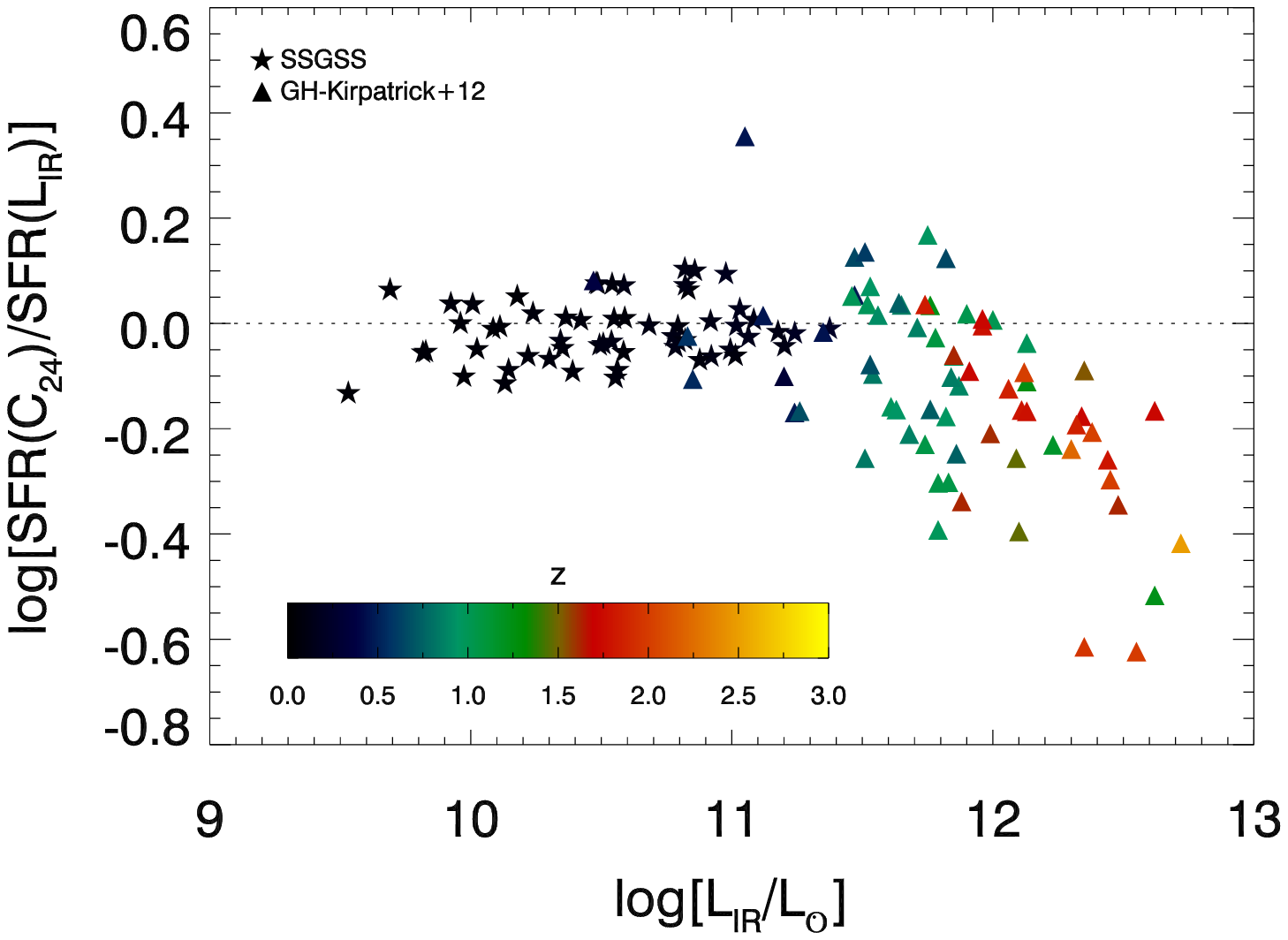} 
\end{array}$
\end{center}
\caption{Comparison of SFRs estimated from $C_{24}(\lambda)$ and \lir\ for the GOODS-\textit{Herschel} sample from \citet{kirkpatrick12}. Left: Comparison as a function of redshift. The values show agreement for $z\lesssim1$, beyond which the dataset is biased towards galaxies with log[\lir/$L_{\odot}]\gtrsim12$. The distribution of the \citet{kirkpatrick12} sources, along with the parameters of a best-fit Gaussian to this distribution, is also shown. Right: Comparison as a function of \lir. The values show agreement for cases with log[\lir/$L_{\odot}]\lesssim12$. For cases with log[\lir/$L_{\odot}]\gtrsim12$, which dominate $z\gtrsim1$ for this sample, the monochromatic SFR is lower than the \lir\ SFR. \label{fig:C24LIR_compare_K12}}
\end{figure*}

\begin{figure*}
\begin{center}$
\begin{array}{cc}
\includegraphics[width=3.5in]{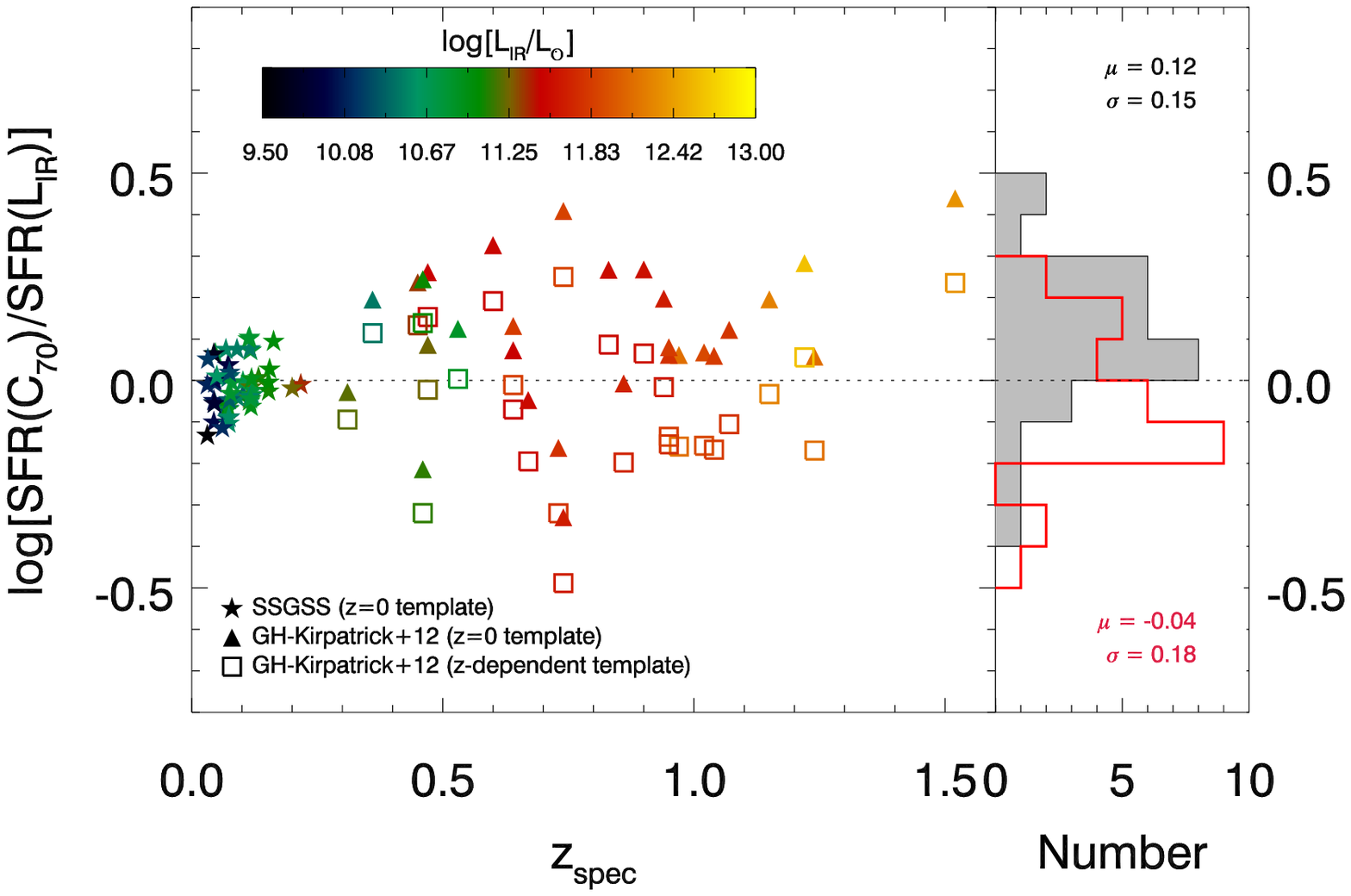} &
\includegraphics[width=3.5in]{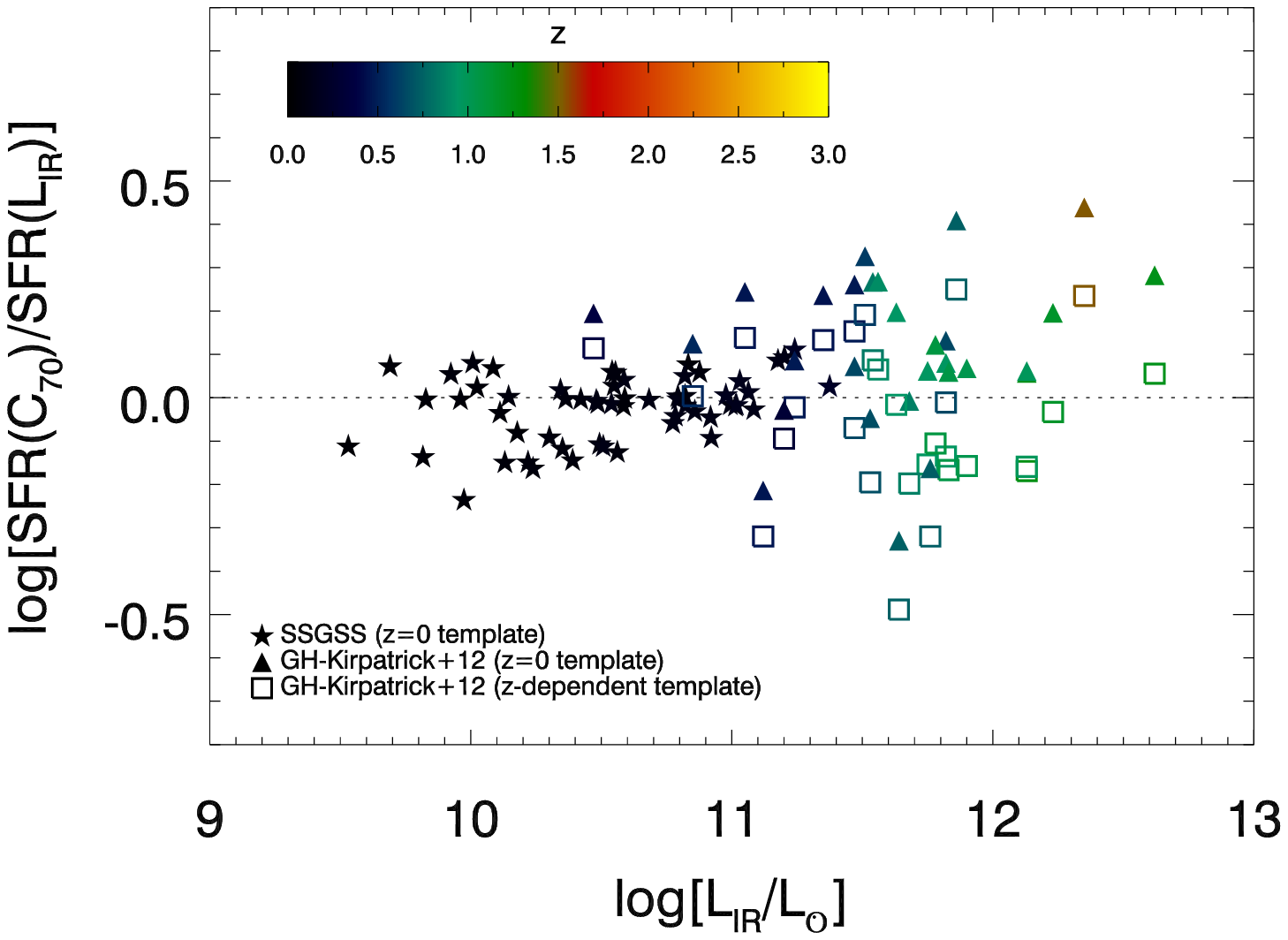} 
\end{array}$
\end{center}
\caption{Comparison of SFRs estimated from $C_{70}(\lambda)$ and \lir\ for the GOODS-\textit{Herschel} sample from \citet{kirkpatrick12}. Left: Comparison as a function of redshift when using our calibration based on a $z=0$ template (filled triangles) and the $z$-dependent template (open squares). The calibration derived from the $z$-dependent template appears to work better than the $z=0$ template and shows agreement for $z\lesssim1.2$, beyond which data is lacking. The distributions of the \citet{kirkpatrick12} sources when using the $z=0$ template (filled gray) and $z$-dependent template (open red), along with the parameters of a best-fit Gaussian to these distributions, are also shown. Right: Comparison as a function of \lir. \label{fig:C70LIR_compare_K12}}
\end{figure*}

First we use the sample of 70 sources identified as SFGs from \citet{kirkpatrick12}, corresponding to AGN contribution of less than $20\%$. These galaxies cover a redshift range of $0.3<z<2.5$ and have full \textit{Spitzer} and \textit{Herschel} photometry. We use the \lir\ measurements of these galaxies from \citet{kirkpatrick12} (private communication), determined from IRS measurements for the MIR and by fitting two modified blackbodies for the FIR. A comparison of SFR($C_{24}(\lambda)$) to SFR(\lir) is shown in Figure~\ref{fig:C24LIR_compare_K12}, both as a function of redshift and \lir. There is general agreement between the values up to redshifts of about $z\sim1$, which corresponds to galaxies with log[\lir/$L_\odot]<12$. Given that the sources of \citet{kirkpatrick12} were required to be very bright in the IR to obtain IRS spectral measurements at these redshifts, it is likely that their sources at $z>1$ are slightly biased to larger \lir\ luminosities (demonstrated by their larger IR8 values). These values deviate significantly from deeper photometric surveys of SFGs at these redshifts, which is shown in Figure~\ref{fig:C24LIR_compare_K12}, by the templates of \citet{magdis12} (dashed cyan line). We remind the reader that the \citet{magdis12} templates are based on \citet{draine&li07} models, which was found to show significant offsets compared to the observed spectra of the SSGSS galaxies, and is only shown for reference. We also examine the comparison of SFR($C_{70}(\lambda)$) to SFR(\lir), shown in Figure~\ref{fig:C70LIR_compare_K12}. The redshift-dependent correction of $C_{70}(\lambda)$ seems to work well for galaxies of $z\lesssim1.2$, for which data is available for this band.

As a second test we use the sample from \citet{elbaz11}. This sample covers a redshift range of $0.03<z<2.85$ and has \textit{Spitzer} and \textit{Herschel} photometry. The \lir\ values for these galaxies have been determined by \citet{schreiber14}, and are estimated from the \citet{chary&elbaz01} template that provides the best fit to the Herschel data. For our analysis we only consider sources for which at least one photometric band covers wavelengths greater than 30$~\micron$, as these cases achieve better accuracy of the FIR region. The photometric redshifts of these sources are obtained from \citet{pannella14} \citep[using the EAZY code;][]{brammer08}, and we require the sources to have suitable quality flags. These photometric redshifts achieve a relative accuracy ($\Delta z=(z_{\mathrm{phot}}-z_{\mathrm{spec}})/(1+z_{\mathrm{spec}})$) of 3\%, with less than 3\% of cases suffering from catastrophic failures \citep[$\Delta z>0.2;$][]{pannella14}. These sources lack spectroscopic measurements to identify AGN or starburst sources and we rely on the color-cut  outlined by \citet{kirkpatrick12} for AGN and also remove sources with IR8$>$8 from our sample, which are believed to be predominately starburst galaxies \citep{elbaz11}. We follow the method of \citet{elbaz11} to determine rest-frame $8~\micron$ from $k$-correcting the $8~\micron$ band ($z<0.5$), $16~\micron$ band ($0.5<z<1.5$), and $24~\micron$ band ($1.5\le z\le 2.5$) assuming these galaxies follow the IR SED of M82. These constraints leaves us with a SFG sample of $825$ sources with $24~\micron$ observations and $66$ with $70~\micron$ observations.

\begin{figure*}
\begin{center}$
\begin{array}{cc}
\includegraphics[width=3.5in]{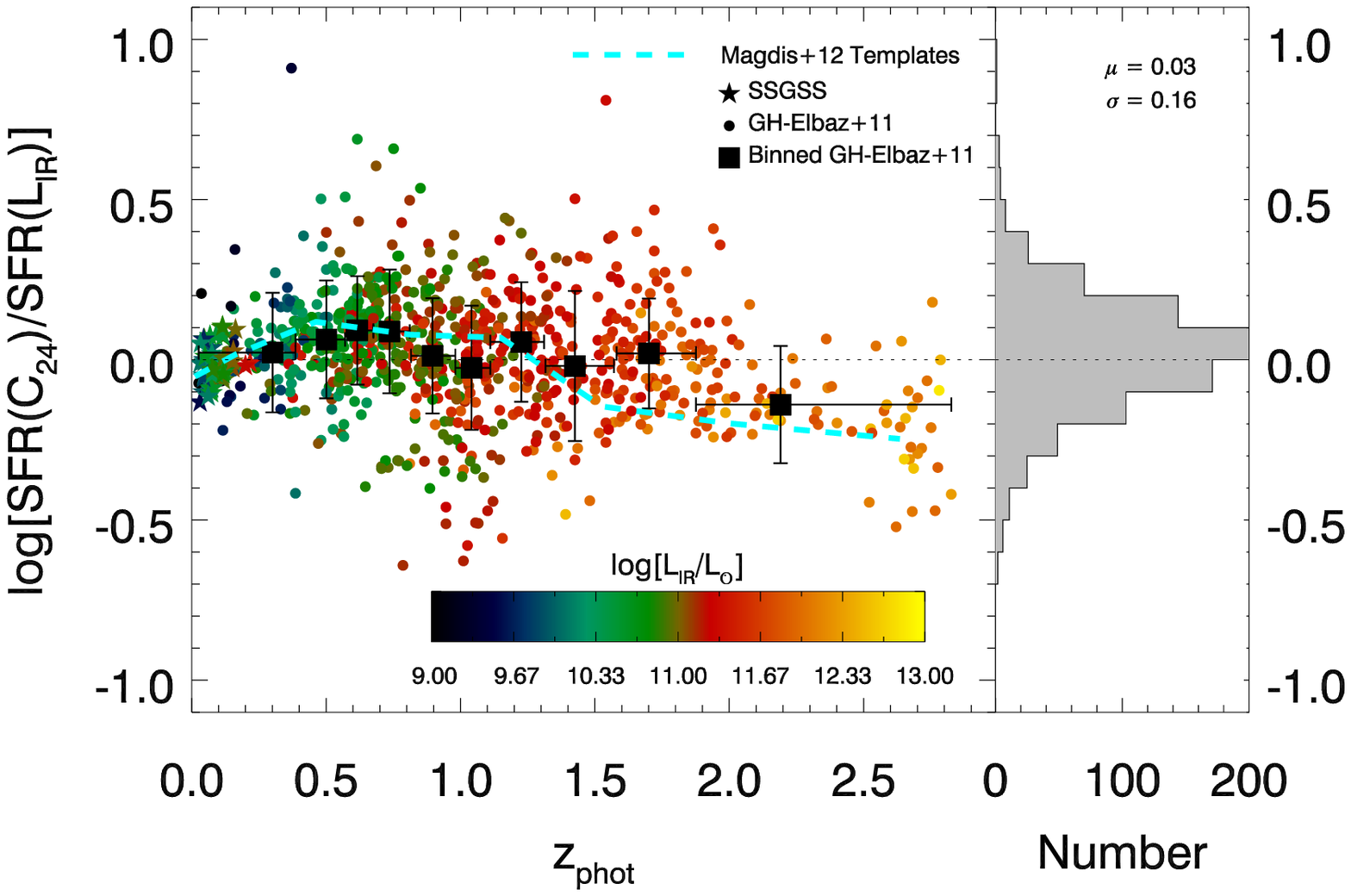} &
\includegraphics[width=3.5in]{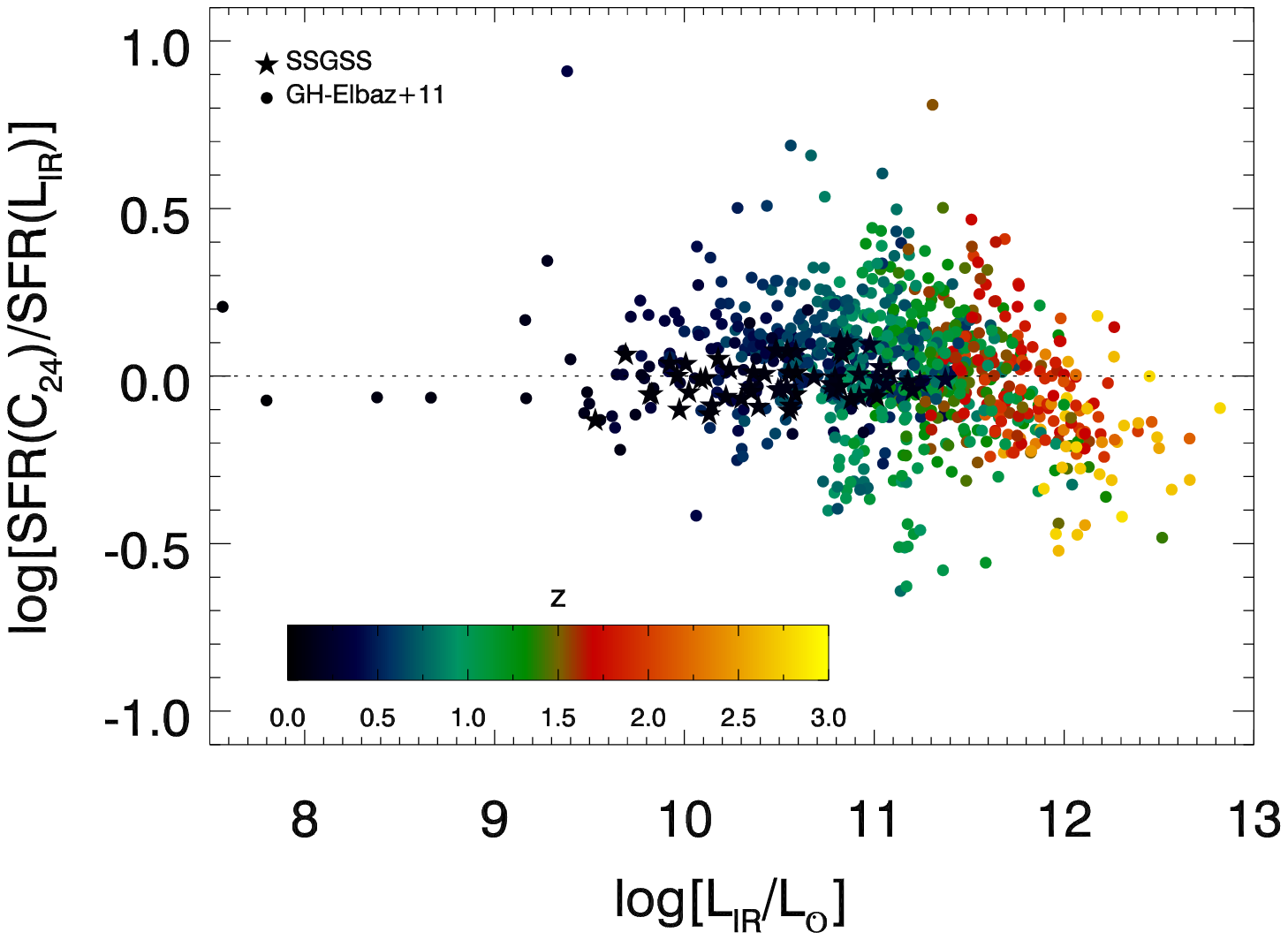} 
\end{array}$
\end{center}
\caption{Comparison of SFRs estimated from $C_{24}(\lambda)$ and \lir\ for the GOODS-\textit{Herschel} sample from \citet{elbaz11}. Left: Comparison as a function of redshift. The values show agreement for $z\lesssim2$, beyond which the dataset is biased towards galaxies with log[\lir/$L_{\odot}]\gtrsim12$. The distribution of the \citet{elbaz11} sources, along with the parameters of a best-fit Gaussian to this distribution, is also shown. Right: Comparison as a function of \lir. The values show reasonable agreement for cases with log[\lir/$L_{\odot}]\lesssim12$. For cases with log[\lir/$L_{\odot}]\gtrsim12$, which dominate $z\gtrsim2$ for this sample, the monochromatic SFR is lower than the \lir\ SFR. \label{fig:C24LIR_compare}}
\end{figure*}

\begin{figure*}
\begin{center}$
\begin{array}{cc}
\includegraphics[width=3.5in]{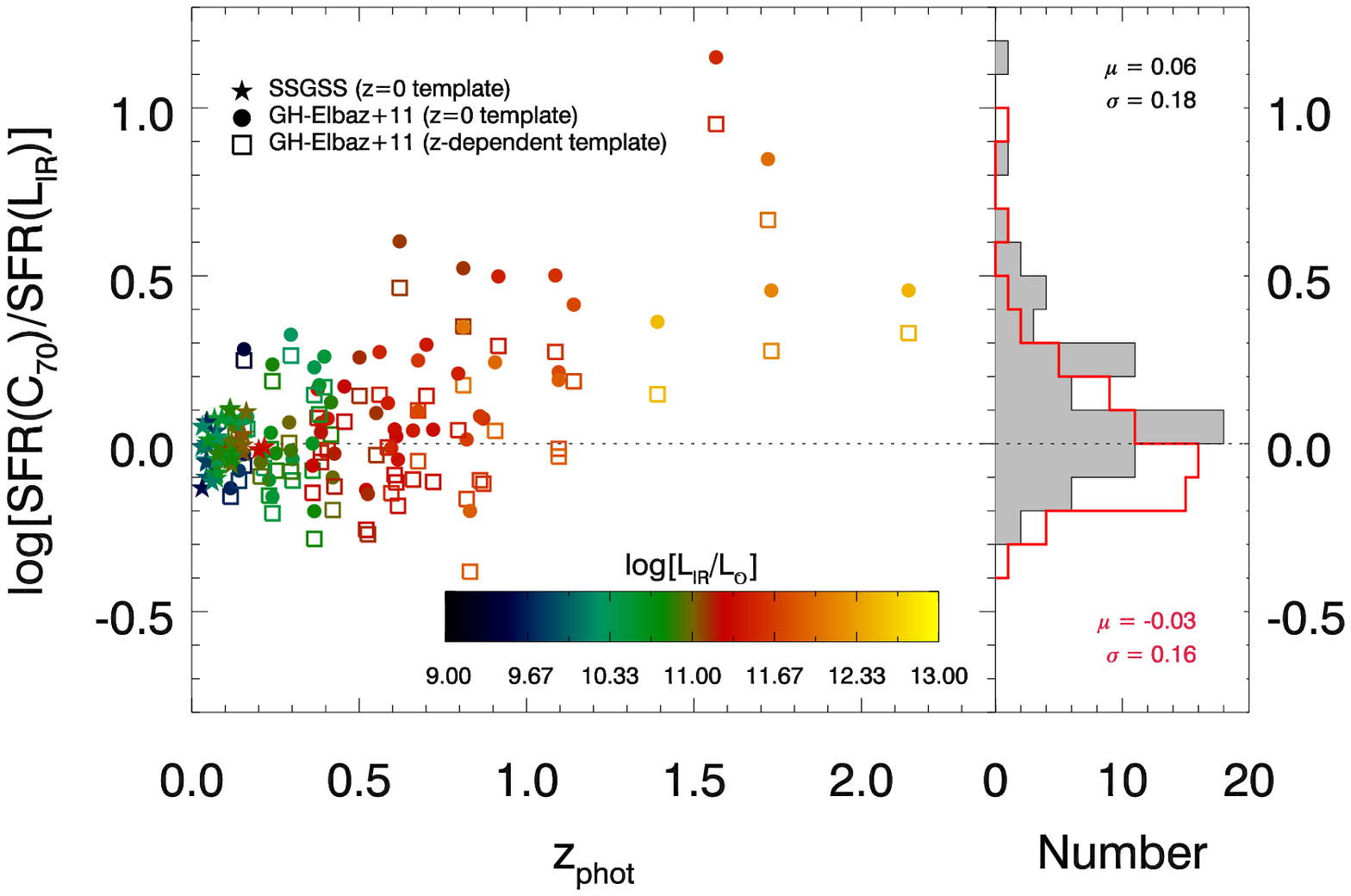} &
\includegraphics[width=3.5in]{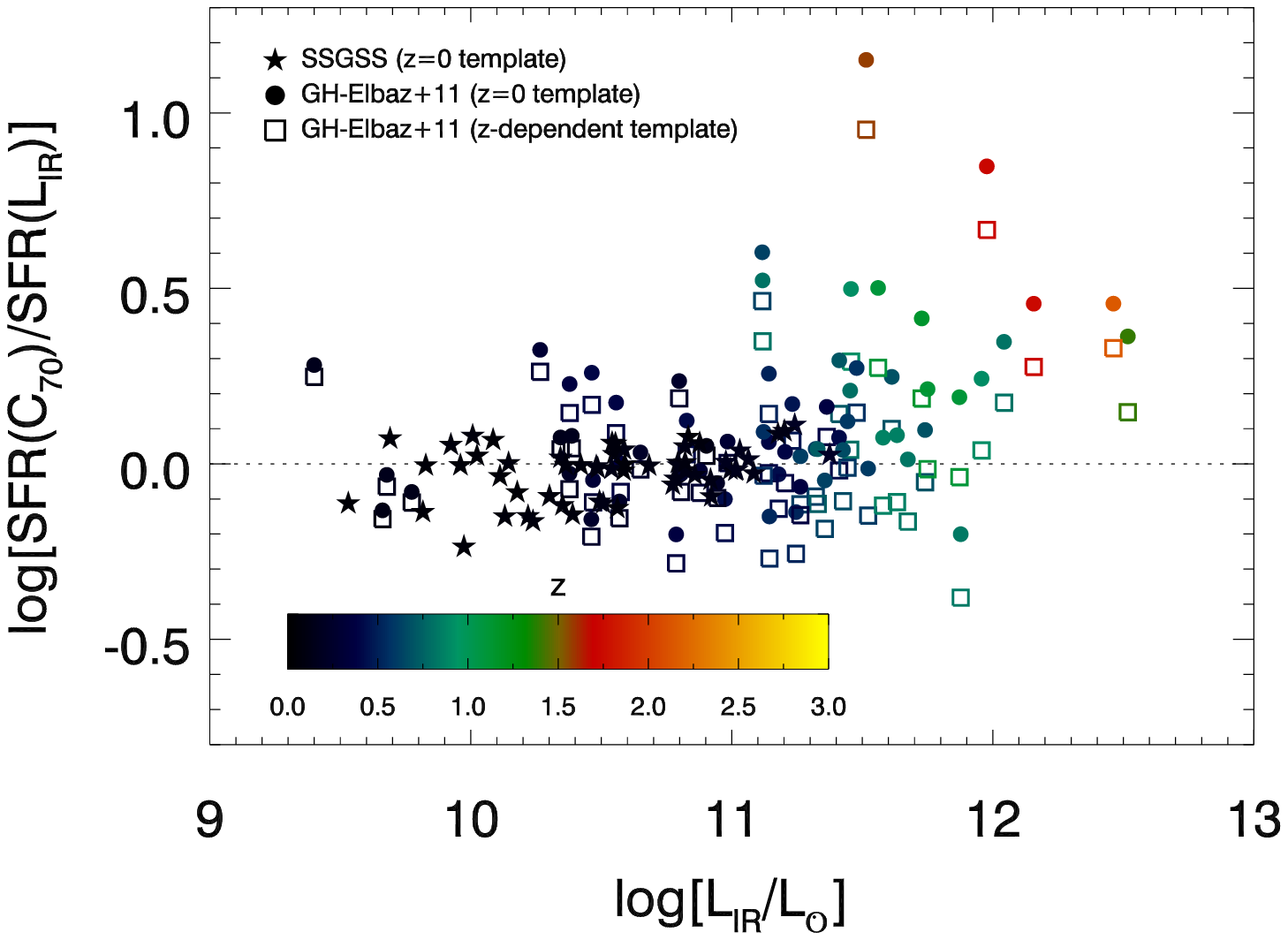} 
\end{array}$
\end{center}
\caption{Comparison of SFRs estimated from $C_{70}(\lambda)$ and \lir\ for the GOODS-\textit{Herschel} sample from \citet{elbaz11}. Left: Comparison as a function of redshift when using our calibration based on a $z=0$ template (filled triangles) and the $z$-dependent template (open squares). The calibration derived from the $z$-dependent template appears to work better than the $z=0$ template and shows agreement for $z\lesssim1$, beyond which data is lacking. The distributions of the \citet{elbaz11} sources when using the $z=0$ template (filled gray) and $z$-dependent template (open red), along with the parameters of a best-fit Gaussian to these distributions, are also shown. Right: Comparison as a function of \lir. \label{fig:C70LIR_compare}}
\end{figure*}

Using the measured \lir\ values from these sources, the same comparison is made as before and is shown in Figure~\ref{fig:C24LIR_compare}. In this case there appears to be more agreement among the diagnostics out to redshifts of $z\lesssim2$, with a $1\sigma$ dispersion of 0.16 dex (45\%). Small changes appear to develop beyond $z>2$, which is consistent with the observed trend of IR8 going 4 to 6 by $z=3$ (a difference of $\sim$$0.2$ dex). This trend is apparent in the ratios of observed $24~\micron$ luminosity to \lir\ in the templates of \citet{magdis12} (dashed cyan line). Similar to the \citet{kirkpatrick12} sample, the largest discrepancies occur in galaxies with log[\lir/$L_\odot]>12$. Next, we examine the comparison of SFR($C_{70}(\lambda)$ to SFR(\lir) shown in Figure~\ref{fig:C70LIR_compare}. As before, the redshift-dependent correction of $C_{70}(\lambda)$ seems to work well for galaxies of $z\lesssim1$, with a $1\sigma$ dispersion of 0.18 dex (50\%). For $z\gtrsim 1$, significant differences appear but this is likely due to the poor sensitivity of the 70$~\micron$ band detecting only the most luminous galaxies in these bands at high redshifts. With the limited number of sources available at $z>1$, the reliability of our corrections cannot be determined for this range.

Common to the $C_{24}(\lambda)-$\lir\ comparisons for the two samples considered is the trend that as one goes to $z\gtrsim2$ and/or $L_{\rm{IR}}\gtrsim10^{12}L_{\odot}$ the SFRs predict from \lir\ will be larger than those inferred from the MIR. This may demonstrate that $C_{24}(\lambda)$ is unsuitable when considering galaxies with log[\lir/$L_\odot]>12$ or $z\gtrsim2$, but it could also indicate a change occurs in \lir conversion factor at these luminosities/redshifts. Unfortunately, it is unclear whether what we are observing is a redshift effect or a luminosity effect, as there is a degeneracy between these variables that cannot be resolved with our current data. In other words, we may simply be seeing a selection effect. Given the higher sensitivity of JWST, would including sources with $L_{\rm{IR}}\lesssim12$ at $z\gtrsim2$ follow the same trend of increasing IR8 when considering the entire population? Prior to the JWST mission, a technique that does not rely on \lir\ to determine SFRs for higher redshift galaxies will be necessary to state confidently what effect is occurring.

\section{Conclusions}
We have presented continuous, monochromatic star formation rate (SFR) indicators over the mid-infrared wavelength range of $6-70~\micron$, using a sample of 58 star forming galaxies in the SSGSS at $z<0.2$.  The continuous wavelength coverage granted with this sample has allowed for the calibration of \textit{Spitzer}, WISE, and JWST bands as SFR diagnostics covering, continuously, redshifts from $0<z<3$. We find that these diagnostics are consistent with monochromatic calibrations of SFGs in the local universe, and achieve accuracies of 30\% or better. They also appear consistent with templates of high-$z$ SFGs, with no significant evidence of variations in the shape of the SED over $6-30~\micron$ region. Subtle changes of $\mathrm{IR8}=L_{\rm{IR}}/L_{\rm{rest}}(8\mu\mathrm{m})$ with redshift appear to cause variations at $z\gtrsim2$, but currently it is unclear whether this could be due to a selection bias at these redshifts. Due to the significant changes in the FIR region beyond $30~\micron$ with redshift, the use of this region as a SFR diagnostic requires correction to our local template, however, this has been demonstrated to work well up to redshifts of at least $z\sim1$. 

These powerful diagnostics are critical for future studies of galaxy evolution and allows for much easier application to large survey programs with a limited number of MIR wavelength bands. This technique is only valid for SFGs, and therefore methods are required to remove AGN and starburst from any sample before use. With the upcoming JWST mission, we hope that these diagnostics will provide important contributions as we begin to examine more typical main-sequence galaxies up to $z\sim3$.

\section*{Acknowledgments}
The authors thank the referee whose suggestions helped to clarify and improve the content of this work. We also thank the MIRI instrument team for providing the MIRI filter curves prior to their publication. AJB thanks K. Grasha, A. Kirkpatrick, A. Pope, and D. Marchesini for comments and discussion that improved the content of this paper. AJB and DC gratefully acknowledge partial support from the NASA ROSES Astrophysics Data Analysis Program, under program number NNX13AF19G.  

This research has made use of the NASA/IPAC Infrared Science Archive, which is operated by the Jet Propulsion Laboratory, California Institute of Technology, under contract with the National Aeronautics and Space Administration (NASA). 

This work is based on observations made with the \textit{Spitzer} Space Telescope, which is operated by the Jet Propulsion Laboratory, California Institute of Technology under a contract with NASA.

This work has made use of SDSS data. Funding for the SDSS and SDSS-II has been provided by the Alfred P. Sloan Foundation, the Participating Institutions, the National Science Foundation, the US Department of Energy, the National Aeronautics and Space Administration, the Japanese Monbukagakusho, the Max Planck Society and the Higher Education Funding Council for England. The SDSS website is http://www.sdss.org/. 

The SDSS is managed by the Astrophysical Research Consortium for the Participating Institutions. The Participating Institutions are the American Museum of Natural History, Astrophysical Institute Potsdam, University of Basel, University of Cambridge, Case Western Reserve University, University of Chicago, Drexel University, Fermilab, the Institute for Advanced Study, the Japan Participation Group, Johns Hopkins University, the Joint Institute for Nuclear Astrophysics, the Kavli Institute for Particle Astrophysics and Cosmology, the Korean Scientist Group, the Chinese Academy of Sciences (LAMOST), Los Alamos National Laboratory, the Max-Planck-Institute for Astronomy (MPIA), the Max-Planck-Institute for Astrophysics (MPA), New Mexico State University, Ohio State University, University of Pittsburgh, University of Portsmouth, Princeton University, the United States Naval Observatory and the University of Washington.

This work is based on observations made with the NASA Galaxy Evolution Explorer. \textit{GALEX} is operated for NASA by the California Institute of Technology under NASA contract NAS5-98034.

This publication makes use of data products from the Wide-field Infrared Survey Explorer, which is a joint project of the University of California, Los Angeles, and the Jet Propulsion Laboratory/California Institute of Technology, funded by the NASA.



\end{document}